\documentstyle[12pt]{article}
\def\PRL #1 #2 #3{{\em Phys. Rev. Lett. \/} {\bf#1} (#2) #3}
\def\NPB #1 #2 #3{{\em Nucl. Phys. \/} {\bf B#1} (#2) #3}
\def\NPBFS #1 #2 #3 #4{{\em Nucl. Phys. \/} {\bf B#2} [FS#1] (#3) #4}
\def\CMP #1 #2 #3{{\em Commun. Math. Phys. \/} {\bf #1} (#2) #3}
\def\PRD #1 #2 #3{{\em Phys. Rev. \/} {\bf D#1} (#2) #3}
\def\PLA #1 #2 #3{{\em Phys. Lett. \/} {\bf #1A} (#2) #3}
\def\PLB #1 #2 #3{{\em Phys. Lett. \/} {\bf B#1} (#2) #3}
\def\JMP #1 #2 #3{{\em J. Math. Phys. \/} {\bf #1} (#2) #3}
\def\PTP #1 #2 #3{{\em Prog. Theor. Phys. \/} {\bf #1} (#2) #3}
\def\SPTP #1 #2 #3{{\em Suppl. Prog. Theor. Phys. \/} {\bf #1} (#2) #3}
\def\AoP #1 #2 #3{{\em Ann. of Phys. \/} {\bf #1} (#2) #3}
\def\PNAS #1 #2 #3{{\em Proc. Natl. Acad. Sci. USA} {\bf #1} (#2) #3}
\def\RMP #1 #2 #3{{\em Rev. Mod. Phys. \/} {\bf #1} (#2) #3}
\def\PR #1 #2 #3{{\em Phys. Reports \/} {\bf #1} (#2) #3}
\def\AoM #1 #2 #3{{\em Ann. of Math. \/} {\bf #1} (#2) #3}
\def\UMN #1 #2 #3{{\em Usp. Mat. Nauk \/} {\bf #1} (#2) #3}
\def\FAP #1 #2 #3{{\em Funkt. Anal. Prilozheniya \/} {\bf #1} (#2) #3}
\def\FAaIA #1 #2 #3{{\em Functional Analysis and Its Application \/}
{\bf
#1} (#2) #3}
\def\BAMS #1 #2 #3{{\em Bull. Am. Math. Soc. \/} {\bf #1} (#2)
#3} \def\TAMS #1 #2 #3{{\em Trans. Am. Math. Soc. \/} {\bf #1} (#2) #3}
\def\InvM #1 #2 #3{{\em Invent. Math. \/} {\bf #1} (#2) #3}
\def\LMP #1 #2 #3{{\em Letters in Math. Phys. \/} {\bf #1} (#2) #3}
\def\IJMPA #1 #2 #3{{\em Int. J. Mod. Phys. \/} {\bf A#1} (#2) #3}
\def\AdM #1 #2 #3{{\em Advances in Math. \/} {\bf #1} (#2) #3}
\def\RMaP #1 #2 #3{{\em Reports on Math. Phys. \/} {\bf #1} (#2) #3}
\def\IJM #1 #2 #3{{\em Ill. J. Math. \/} {\bf #1} (#2) #3}
\def\APP #1 #2 #3{{\em Acta Phys. Polon. \/} {\bf #1} (#2) #3}
\def\TMP #1 #2 #3{{\em Theor. Mat. Phys. \/} {\bf #1} (#2) #3}
\def\JPA #1 #2 #3{{\em J. Physics \/} {\bf A#1} (#2) #3}
\def\JSM #1 #2 #3{{\em J. Soviet Math. \/} {\bf #1} (#2) #3}
\def\MPLA #1 #2 #3{{\em Mod. Phys. Lett. \/} {\bf A #1} (#2) #3}
\def\JETP #1 #2 #3{{\em Sov. Phys. JETP \/} {\bf #1} (#2) #3}
\def\JETPL #1 #2 #3{{\em  Sov. Phys. JETP Lett. \/} {\bf #1} (#2) #3}
\def\PHSA #1 #2 #3{{\em Physica} {\bf A#1} (#2) #3}
\def\CQG #1 #2 #3{{\em Class. Quantum Grav. \/} {\bf #1} (#2) #3}
\def\SJNP #1 #2 #3{{\em Sov. J. Nucl. Phys. (Yadern.Fiz.) \/} {\bf #1}
(#2) #3}

\begin{document}
\thispagestyle{empty}
\begin{flushright}
HUB-EP-99/26\\
DFPD 99/TH/25\\
hep--th/9906142
\end{flushright}

\bigskip
\begin{center}
{\large \bf Superbranes and Superembeddings}

\bigskip
Dmitri Sorokin\footnote{Alexander von Humboldt Fellow.
On leave from Kharkov Institute of Physics and
Technology, Kharkov, 310108, Ukraine}\\

\bigskip
{\it Humboldt-Universit\"at zu Berlin,
Mathematisch-Naturwissenshaftliche Fakultat,\\
Institut f\"ur Physik,
Invalidenstrasse 110, D-10115 Berlin\\
}
and\\
{\it  Universit\`a Degli Studi Di Padova,
Dipartimento Di Fisica ``Galileo Galilei''\\
ed INFN, Sezione Di Padova\\
Via F. Marzolo, 8, 35131 Padova, Italia}
\end{center}
\bigskip

\begin{abstract}
We review the geometrical approach to the description of the dynamics
of superparticles, superstrings and, in general, of
super--p--branes, Dirichlet
branes and the M5-brane, which is based on a
generalization of the  elements of surface theory
to the description of the embedding of supersurfaces into target
superspaces.

Being manifestly supersymmetric in both, the superworldvolume of the
brane
and the target superspace, this approach unifies the
Neveu--Schwarz--Ramond
and the Green--Schwarz formulation and provides the
fermionic ${\kappa}$--symmetry of the Green--Schwarz--type superbrane
actions
with a clear geometrical meaning of standard worldvolume local
supersymmetry.

The dynamics of superbranes is encoded in a generic superembedding
condition.

Depending on the superbrane and the target--space dimension,
the superembedding condition produces either only off--shell constraints
(as in the case of $N=1$
superparticles and $N=1$ superstrings),
or also results in the full set of the
superbrane equations of motion
(as, for example, in the case of the M--theory branes).
In the first case worldvolume superspace actions for the superbranes can
be
constructed, while in the second case
only component or generalized superfield actions are known.

We describe the properties of the
doubly supersymmetric brane actions and show
how they are related to the standard Green--Schwarz formulation.

In the second part of the article basic geometrical grounds of
the (super)embedding
approach are considered and applied to the description
of the M2--brane and the
M5--brane. Various applications of the superembedding approach are
reviewed.
\end{abstract}

\newpage
\tableofcontents
\newpage
\section{Introduction}
String theory to much more extent than any other theory requires for
its description various important developments of classical and
quantum field theory, group theory, geometry and topology.

To comprehend different properties of string theory and to uncover
underlying relations between them one should use for the analysis
 different (often intertwined) mathematical tools and methods.

One of such basic geometrical tools is surface theory which
describes the embedding of surfaces into higher dimensional
manifolds. The application of surface theory to the description of
string theory is quite natural, since the string is a one--dimensional
relativistic object which sweeps a two--dimensional surface
(worldsheet) when it propagates in a (target) space--time. The
dynamics of the string completely determines the geometrical
properties of the worldsheet describing its embedding into the target
space, and vice versa, specifying geometrical properties of the
embedding of a surface into a target space one can, in principle, get
the full information about the details of the dynamics of a string
whose worldsheet is associated with this surface.

For instance, a metric $g_{mn}(\xi)$ $(m,n=0,1)$ of a string
worldsheet locally parametrized by coordinates $\xi^m=(\tau,\sigma)$ is
an
induced metric which is related to a D--dimensional target--space
metric $g_{\underline{mn}}(X)$ $(\underline{m},\underline{n}=
0,1,...,D-1)$ through the condition
\footnote{In what follows the underlined indices will correspond to
target--space objects (coordinates, fields, etc.), while not
underlined indices will correspond to the worldsurface.
This will allow us to escape, to some extent, the proliferation of
indices.}
\begin{equation}\label{1}
g_{mn}(\xi)=\partial_mX^{\underline{m}}g_{\underline{mn}}(X(\xi))
\partial_nX^{\underline{n}},
\end{equation}
which is the simplest example of the embedding condition.

Note that with taking into account worldsheet reparametrization
symmetry the induced metric condition (\ref{1}) amounts to Virasoro
constraints on string dynamics (see \cite{gsw,lt,pol} for details on
superstring theory).

The classical trajectories of the string are surfaces of a minimal
(or more generally, extremal) area. This follows from the Nambu--Goto
string action
\begin{equation}\label{2}
S=-T\int d^2\xi\sqrt{-\det{g_{mn}(X(\xi))}},
\end{equation}
whose geometrical meaning is to be the total area of the worldsheet
of a string with the tension $T$. The variation of the action (\ref{2})
vanishes for minimal area surfaces, which produces the string
equations of motion. Therefore, from the geometrical point of view
the string equations describe minimal (area) embedding of the
worldsheet into target space.

A direct application of the geometrical methods of surface theory
in string theory was initiated in \cite{lr,om} and revealed a
connection of the string
equations of motion with two--dimensional (exactly solvable)
non--linear equations, such as the sin--Gordon and Liouville
equation.

Though, of course, all string formulations imply that string worldsheet
is a surface embedded into a target space--time, the geometrical
embedding
approach explores this in the most direct way.
It deals with such objects as induced vielbeins
of the surface, extrinsic
curvature and torsion of the surface, and reduces the string
equations to the system of the Codazzi, Gauss and Ricci equations
completely determining the embedding of the surface (see \cite{kn}
for the details on surface theory and \cite{barnes}
on its applications to strings).

The embedding approach was also used in connection with the
problem of formulating a consistent quantum string theory in
non--critical
space--time dimensions and has been developed in a number
of papers  (see \cite{barnes,zhelt} and references therein).

In addition to particles and strings a variety of other extended
relativistic objects have been discovered in
$D=10$ superstring theories and $D=11$ supergravity. This includes a
$D=11$ membrane \cite{bst1}, a $D=11$ five--brane \cite{guv} and
Dirichlet branes \cite{le,po}. Collectively all these extended objects
are called superbranes or super--p--branes, where $p$ denotes the number
of
spatial dimensions of a given brane.

The existence of the superbranes reflects and causes important
duality chains which connect $D=11$ supergravity with five basic
$D=10$ superstring theories and string theories among themselves
(see \cite{duality,pol} for a review on dualities). This gives rise to a
conjecture that $D=11$ supergravity and the $D=10$ string theories
can be associated with different vacua of a single underling
quantum theory called U--theory \cite{sen1}. This theory is also
often called M--theory \cite{m}, but following Sen \cite{sen1} and
Schwarz
\cite{jhs} we shall reserve the latter
name for a $D=11$ sector of U--theory whose low energy effective
field theory limit is $D=11$ supergravity, and which also contains
a membrane (M2--brane) and a five--brane (M5--brane) as part of its
physical spectrum.

Since the superbranes are surfaces in target superspaces it is
natural to apply for studying their properties the same geometrical
methods as for strings, i.e. to describe the propagation
of superbranes by specifying the embedding of brane worldvolumes
in target superspaces. For instance, a bosonic part of a
super--p--brane worldvolume action is a $(p+1)$--dimensional analog
of the Nambu--Goto action (\ref{2}). It produces bosonic equations
of motion which are equivalent to minimal (volume) embedding of
the corresponding worldvolume into space--time.

As we have already mentioned, at the classical level the use of
geometrical embedding methods allow one to find a connection between
equations of motion of bosonic branes and integrable \cite{hoppe}, and
in the
case of strings, with exactly solvable nonlinear equations
\cite{lr,om,barnes,zhelt,b1}. Analogous relations were also
found for superstrings \cite{lio,wz}.
As a result, one can relate, for example, (super)strings to
exactly--solvable
Wess--Zumino--Novikov--Witten models \cite{group,costr,wz}.
( for a review on WZNW models and related topics
see \cite{f,halpern} and references therein).
This provides us with useful information about the details of
brane dynamics.

In the case of supersymmetric extended objects the use of the
embedding approach has turned out to be even more fruitful, since
it has allowed one to clarify the geometrical origin of local symmetries
of superbrane worldvolumes \cite{stv}--\cite{bpstv},
to relate different formulations
of superbrane dynamics \cite{vz}--\cite{berk}, and to get equations of
motion
for those superbranes for which the construction of worldvolume
actions encountered problems, such as
D(irichlet)-branes \cite{hs1} and the M5-brane \cite{hs2}.
It has also allowed one to make a progress towards a solution of
the problem of covariant quantization of superstrings
\cite{banquant,to,apt,berquant,tor,dpolyak}.
Let us also mention that supergravity can also be described as a
theory of supersurfaces \cite{rs}.

To apply the embedding approach to the description of superbranes one
should first generalize the method itself and find supersymmetric
analogs of bosonic embedding conditions such as the induced metric
condition (\ref{1}) and the minimal embedding of the superbrane
worldvolume
which would be equivalent to the superbrane equations of motion
\cite{bpstv}.
A reasonable generalization is to consider superbrane worldvolumes as
{\it super\/}surfaces locally
parametrized by $(p+1)$ bosonic coordinates $\xi^m$
and $n$ fermionic coordinates $\eta^\mu$ ($\mu=1,...,n$) embedded
(in a specific way) into target {\it super\/}spaces locally
parametrized by $D$
bosonic coordinates $X^{\underline{m}}$ and $2^{[{D\over 2}]}$
Grassmann spinor coordinates $\Theta^{\underline\mu}$
(where brackets denote the integer part of ${D\over 2}$). Thus we get
a superembedding.

The assumption that such a superembedding should underlie the
worldvolume
dynamics of superbranes is prompted by a well known fact that the $D=10$
superstring theories can be formulated in two different ways.

In the Neveu--Schwarz--Ramond (or spinning string) formulation
\cite{nsr}
superstring propagation is described by a supersurface, possessing
$n=1$ local worldsheet supersymmetry, embedded into bosonic space--time.
Space--time supersymmetry appears in this model only upon quantization
as a symmetry of quantum string physical states singled out by the
Gliozzi--Scherk--Olive projection \cite{gso}.

On the other hand in the Green--Schwarz formulation \cite{gs} the
superstring worldsheet is a bosonic surface embedded into a target
superspace. This formulation is manifestly space--time supersymmetric
and,
in addition, possesses a local worldsheet fermionic symmetry called
kappa--symmetry. The number of independent $\kappa$--symmetry
transformations is half the number of target superspace Grassmann
coordinates. Kappa--symmetry was first observed in the case of
superparticles \cite{al,sig} and is inherent to all superbranes in the
Green--Schwarz formulation \cite{gsw,bst1,dbrane,m5,s1}.
It plays an important role ensuring
that superbranes form stable, so called, Bogomol'nyi--Prasad--Sommerfeld
configurations whose presence in a superspace background preserves half
of the space--time supersymmetries of a background vacuum. For $D=10$
superstrings this means, in particular, that their quantization results
in consistent quantum supersymmetric theories.

At the same time local fermionic
$\kappa$--symmetry causes problems with performing the covariant
Hamiltonian
analysis and
quantization of superbrane theories. This is due to the fact that
the first--class constraints corresponding to $\kappa$--symmetry form an
infinite reducible set, and in a conventional formulation of
superparticles and
superstrings it turned out impossible
to single out an irreducible set of the fermionic first--class
constraints in a
Lorentz covariant way.
(Note however that it is possible to do this in the case of D--branes
\cite{kallosh}).

It is therefore desirable to replace $\kappa$--symmetry with something
more natural and simple. To do this one can notice that
$\kappa$--symmetry
transformations resemble supersymmetry transformations. Space--time
supersymmetry transformations of $X^{\underline{m}}$ and
$\Theta^{\underline{\mu}}$ with a supersymmetry parameter
$\epsilon^{\underline{\mu}}$ are
\begin{equation}\label{3}
\delta\Theta^{\underline{\mu}}=\epsilon^{\underline{\mu}}, \qquad
\delta X^{\underline{m}}=
i\bar\Theta\Gamma^{\underline{m}}\delta\Theta,
\end{equation}
while $\kappa$--symmetry transformations of $X^{\underline{m}}$ and
$\Theta^{\underline{\mu}}$ on the superbrane worldvolume
have the following form
\begin{equation}\label{4}
\delta\Theta^{\underline{\mu}}(\xi)=
{1\over 2}(1+\bar \Gamma)^{\underline{\mu}}_{~~\underline{\nu}}
\kappa^{\underline{\nu}}(\xi), \qquad
\delta X^{\underline{m}}(\xi)=
-i\bar\Theta\Gamma^{\underline{m}}\delta\Theta,\quad (\bar\Gamma)^2=1
\end{equation}
where ${1\over 2}(1+\bar \Gamma)$ is a projection operator constructed
from the
$\Gamma^{\underline m}$ matrices. The projector
is specific for each type of the superbranes, and its presence implies
that
the number
of independent $\kappa$--symmetry transformations is twice less than
the number of the space--time supersymmetries (\ref{3}).

Notice the difference in the sign of the $X^{\underline m}$--variations
in (\ref{3}) and (\ref{4}). This difference implies that if ordinary
supersymmetry is associated with left boosts in the target superspace,
the $\kappa$--symmetry is associated with right boosts.

It is instructive to understand how the presence of
the branes breaks symmetries of target superspace vacua. For this
consider an infinite flat membrane in a
$D=11$ Minkowski superspace.

i) $D=11$ Lorenz symmetry $SO(1,10)$
is broken down to $d=3$ Lorentz symmetry $SO(1,2)$
of the worldvolume
times $SO(8)$ rotations in directions orthogonal to the membrane;

ii) $D=11$ Poincare invariance is broken down to a translational
symmetry along the membrane worldvolume (generated by momenta $P_m$,
$m$=0,1,2),

iii) therefore, unbroken supersymmetries are those generated by
supercharges
whose anticommutators close on the unbroken translations along the
membrane
worldvolume.
There are 16 such generators $Q_\alpha$ ($\alpha$=1,...,16) among 32.
$$
\{Q,Q\}=2iP_m\Gamma^m.
$$
16 target--superspace Grassmann coordinates (let us call them
$\eta^\alpha$)
  which are shifted by the action of
the 16 unbroken supercharges can be regarded as ones
`parallel' to the membrane. It is important to stress that it is the
{\it unbroken} supersymmetry transformations of $\eta^\alpha$ which can
be
compensated by appropriate worldvolume $\kappa$--symmetry
transformations,
and, hence, these 16 Grassmann coordinates can be associated with
fermionic
directions which extend the bosonic worldvolume of the membrane
to a superspace embedded into the target superspace.
Other 16 target--space Grassmann coordinates (let us call them
$\theta^{\alpha'}$) correspond to the spontaneously broken
supersymmetries.
They are Goldstone fermion physical modes of the brane.

From the above analogy we see that, since $\kappa$--symmetry  is
a local worldvolume fermionic symmetry, it would be natural to endow it
with the
direct meaning of being a manifestation of conventional $n$--extended
local
supersymmetry of the superbrane worldvolume. This would allow one
to get rid of `unpleasant' features of $\kappa$--symmetry by dealing
directly with well understood properties of conventional linearly
realized supersymmetry. Such an interpretation of $\kappa$--symmetry was
proposed in \cite{stv}.

For this one should  construct a formulation of superbrane dynamics
which would be manifestly supersymmetric on both the worldvolume and
the target superspace. Such a doubly supersymmetric formulation would
then
unify the NSR and GS formulations. Note that the NSR formulation of
super--branes with $p> 1$ is unknown and the doubly supersymmetric
formulation may provide one with an idea what ``spinning'' branes might
look like.

To have the worldvolume and target space supersymmetry manifest the
doubly supersymmetric formulation should be constructed as a superfield
theory on both the worldvolume superspace and the target superspace, the
former being (super)embedded into the latter. We thus again arrive at
the superembedding approach.

The idea to use doubly supersymmetric models
for a combined treatment of the NSR and the GS superstrings
first appeared in \cite{gates1}. It has then been applied to the
construction of an interesting class of so called spinning
superparticle \cite{ssp} and  spinning superstring \cite{sss} models.
In general, however, these models describe
objects with twice larger number of physical states
than, for example, corresponding NSR spinning strings or
GS superstrings have themselves. They also do not resolve the problem
of $\kappa$--symmetry, since the latter is still present in the
spinning superbranes as an independent symmetry in addition to
double supersymmetry.

A main reason why spinning
superbranes have `redundant' symmetries and physical states is that
embedding of their superworldvolumes into target superspaces is too
general. To describe conventional (i.e. not spinning) superbranes in the
superembedding
approach (with kappa--symmetry replaced by local worldvolume
supersymmetry)
one should find an appropriate superembedding condition
which would amount to conditions required
for the description of the dynamics of the `standard' superbranes.

The basic superembedding condition
was first found in \cite{stv}
for superparticles and then, as a result of the development of the
superembedding approach \cite{stv}--\cite{hs2},
was proved to be generic to all known
types of superbranes. From the geometrical point of view this condition
is quite natural and simple.

Consider the pullback onto the superworldvolume (with
$n=2^{[{D\over 2}]-1}$ Grassmann directions)
of the vector component $E^{\underline{a}}(Z)$ of a target space
supervielbein $E^{\underline{M}}(Z)=(E^{\underline{a}},
E^{\underline{\alpha}}$) (where $Z^{\underline{M}}=(X^{\underline{m}},
\Theta^{\underline{\mu}})$, and the indices from the beginning of the
alphabets correspond to the (co)tangent space of the target superspace
which has $2n=2^{[{D\over 2}]}$ Grassmann directions).
The pullback of $E^{\underline{a}}$ is a one--superform
on the superworldvolume and,
therefore, can be expended in a superworldvolume supervielbein basis
$e^A(\xi,\eta)=(e^a,e^\alpha)$
\begin{equation}\label{5}
E^{\underline{a}}(Z(\xi,\eta))=e^aE_a^{\underline{a}}
+e^\alpha E_\alpha^{\underline{a}}.
\end{equation}

The superembedding condition reads that the pullback components
of the vector supervielbein $E^{\underline{a}}(Z)$ along the Grassmann
directions of the superworldvolume are zero
\begin{equation}\label{6}
E_\alpha^{\underline{a}}(Z(\xi,\eta))=0.
\end{equation}
The geometrical meaning of this condition is that at any point
on the superworldvolume the worldvolume tangent space Grassmann
directions
form a subspace of the
Grassmann tangent space of the target superspace.

In a certain sense eq. (\ref{6}) is analogous to the induced metric
condition
(\ref{1}) (actually, eq. (\ref{6}) implies a superspace generalization
of
(\ref{1})).
As we shall see below, the dynamical meaning of the superembedding
condition (\ref{6}) is that it produces Virasoro--like constraints
on the dynamics of the superbranes, and in many cases its integrability
conditions lead to superbrane equations of motion. It also provides
a link between $\kappa$--symmetry of the standard GS formulation
and local worldvolume supersymmetry of the
doubly supersymmetric formulation of superbranes \cite{stv},
as well as a classical relation between the NSR and GS formulation
in the case of superparticles \cite{vz,stvz} and superstrings
\cite{apt,berk}.

In the cases when the superembedding condition does not produce the
superbrane equations of motion one can construct worldvolume superfield
actions for corresponding superbranes. Such actions have been
constructed
for $N=1$, \footnote{The capital $N$ stands for the number of
the spinorial supercharges of target--space supersymmetry.}
$D=3,4,6,10$ massless superparticles \cite{stv,sor,chern,ds,ps92,gs92},
$N=1$, $D=3,4,6,10$
superstrings \cite{hsstr,hsstr1,to,ik,dis,dghs92},
$N=2$, $D=2,3,4,6$ massless and massive
superparticles \cite{gaunt,ps92,gs2,bms}, an
$N=2$, $D=3$ superstring \cite{chp,gs2} and an $N=1$, $D=4$
supermembrane \cite{hrs}.

In the case where worldvolume superembedding corresponds to the
type II $D=10$ superstrings and D--branes, the $D=11$ supermembrane
and the super--five--brane the superembedding condition produces
superbrane equations of motion \cite{gs2,bpstv,hs1,hs2}.
It is remarkable that in such cases the superembedding condition
contains full information about the physical contents of supersymmetric
theories on the worldvolumes of the superbranes. For instance, in the
case of the D-branes one derives from (\ref{6}) that the D--branes
carry on their
worldvolumes vector gauge fields of the Dirac--Born--Infeld--type,
and the M5--brane carries a self--dual second--rank tensor
gauge field. The superembedding approach allowed one to get
the full set of nonlinear equations of motion of D--branes \cite{hs1}
and
the M5--brane \cite{hs2,hsw1} before the Green--Schwarz--type
worldvolume actions
for these objects were constructed in \cite{dbrane,m5,s1}
\footnote{A gauge fixed action for a space--filling D3--brane was
first constructed in \cite{bg2} as a result of studying partial
breaking of $N=2$, $D=4$ supersymmetry.}.

When the superembedding condition produces superbrane equations of
motion one cannot construct worldvolume superfield actions for these
branes. The problem is the same as in the case of extended
supersymmetric field theories and supergravities when
constraints imposed on superfields put the theory on the mass shell.
In these cases one should either consider component actions (such as
Green--Schwarz--type actions), or apply a generalized action principle
\cite{bsv} based on a group--manifold approach to the description of
supersymmetric theories \cite{gma}. In the latter case one gets an
action
which produces the superfield superembedding condition as an equation
of motion, though the action itself is not a fully fledged superfield
action in the sense that the integral is taken only over a bosonic
submanifold of the superworldvolume.

In some cases the superembedding condition is not enough even for
the off--shell descriptions of superbranes \cite{hs1}. This happens,
for instance,
with D6-- and D8--branes of the type IIA superstring theory \cite{open}
or with space--filling branes, such as
the D9--brane of the type IIB superstring \cite{akub}.
In these cases an additional superworldvolume condition is required to
constrain superfields describing gauge fields propagating in the
worldvolume
of the branes.

As another important and profound feature of the superembedding approach
it is worth mentioning that it provides a natural link between
space--time and twistor \cite{twistor,fer} description of
relativistic systems.
Actually, in the first series of papers on the doubly supersymmetric
description of superbranes the approach was called twistor--like, and
only quite recently it has acquired the present name of
the geometrical \cite{bpstv,bpst}
or simply the superembedding \cite{aspects} approach.

In the superembedding approach auxiliary commuting spinors
appear as superpartners of
the target superspace Grassmann coordinates and have properties of
twistors \cite{twistor,fer,shir,stv} and/or Lorentz harmonics
\cite{gikos},\cite{sok}--\cite{bzm,banquant}.

The use of twistor--like variables to formulate the theory of
supersymmetric relativistic objects has the following deep grounds.

When a superbrane propagates in a nontrivial gauge superfield or
supergravity background
$\kappa$--symmetry requires background superfields to
satisfy superfield constraints, which (in most of the cases)
are equivalent to background superfield equations of motion.

For instance, the $\kappa$--symmetry of a massless superparticle
propagating in a super--Yang--Mills background requires
geometrical integrability of the background
along light--like lines of the
superparticle trajectories, which implies constraints on background
superfields.
In the case of $N=3,4$, $D=4$ or
$N=1$, $D=10$ super--Yang--Mills background, for example, this is
equivalent to the SYM field equations \cite{twi,chau}. Analogous results
were obtained for supergravity backgrounds \cite{twi,shat}. And it was
understood that the correspondence between geometrical integrability
along massless superparticle orbits and super--Yang--Mills and
supergravity equations is based on a twistor interpretation of these
theories. In this connection twistor formulations of
superparticles and superstrings have been studied
in a number of papers
\cite{shir,twbc}--\cite{bmrs,shat,d10p,ps92,berquant,preit,claus1,claus2}.

As we shall see, the superembedding condition (\ref{6}) contains a
Cartan--Penrose relation between vectors and commuting spinors. And
the Cartan--Penrose relation turns out to be the twistor--like
solution of the Virasoro--type constraints which govern the dynamics
of any brane.

Therefore, through the superembedding condition one establishes the
relationship between various space--time, twistor--like and
harmonic formulations of superparticles, superstrings and
superbranes.

In this review we shall consider basic features of the superembedding
approach. We shall show how one arrives at the superembedding condition
by constructing an appropriate worldline superfield action for a
massless superparticle. We shall demonstrate how local worldvolume
supersymmetry transformations reduce to $\kappa$--symmetry
transformations
upon the elimination of the auxiliary components of worldvolume
superfields
which made local supersymmetry manifest. For those superbranes
for which
the superembedding condition does not put the theory on the mass
shell we
shall present doubly supersymmetric (worldvolume superfield) actions
which produce the superembedding condition, and will discuss
their symmetry properties.

We shall perform the analysis of the superembedding condition in cases
when it puts the theory on--shell producing superbrane equations of
motion.
In these cases, such as the
$D=11$ M2--brane and the M5--brane,
we shall demonstrate how superbrane equations of
motion
arise as a requirement of the integrability of the superembedding
condition,
and will discuss the problem of constructing doubly supersymmetric
actions for such branes.

Since the (super)embedding approach has a rather wide range of
applications,
it is not possible in one review to cover in detail all points where
this
approach has been found to be useful. Some of them we shall just sketch
referring
the reader to corresponding original literature.
For instance, we leave aside the relation of the method to integrable
models, and
an important and interesting problem of covariant quantization
of Green--Schwarz
superstrings, which itself requires a separate review article.
The details
on solving this problem with methods related to the superembedding
approach the reader may find in
\cite{berquant,to,apt,tor,banquant,dpolyak}.

The article is organized as follows.

It can be conventionally divided into two parts. The first part consists
of Sections 2 and 3, and Sections 4 and 5 constitute the second part.

In Section 2 we introduce the basic ideas of the
superembedding approach with a simple example of superparticles.
This instructive example should facilitate the reader's understanding of
all the features, ingredients and techniques of the approach,
which will be applied (in exactly the same way) in subsequent Sections
to the description of more complicated models of superstrings
and superbranes.

In Section 3 we discuss the superembedding formulation of
$N=1$, $D=3,4,6,10$ superstrings.
As in the case of $N=1$, $D=3,4,6,10$ superparticles,
the superembedding condition does not produce dynamical
equations of motion of the $N=1$ superstrings,
and one can construct worldsheet superfield actions, from which the
dynamical
equations are derived. An interesting feature of these actions is that
the
string tension appears there through an auxiliary dynamical variable
\cite{dghs92}.
This is a realization of the idea of a dynamical generation of the brane
tension, various aspects of which have been discussed in references
\cite{zh2,t,t1}.
We then describe worldsheet superfield actions for $N=1$ superstrings in
curved target superspaces and show that the latter should obey
supergravity
constraints. Finally, we introduce into the doubly supersymmetric
construction
chiral (heterotic) fermions which extend the $N=1$, $D=10$ closed
superstring
to a $D=10$ heterotic string.

In Section 4 we present basic geometrical ideas which underlie the
theory
of bosonic surfaces and then extend them to the description of
superembeddings.

In Section 5 we apply general tools of superembeddings to the
description of classical dynamics of the supermembrane and the
super--5--brane of M--theory.

In  Section 6 we briefly review other applications of the superembedding
approach.

\section{Doubly supersymmetric particles}
\setcounter{equation}0
\subsection{The bosonic particle}
We start by recalling the form of actions for a massless
bosonic particle which we will then subject to
various kinds of supersymmetrization. This will allow us to describe
the dynamics of particles having spin degrees of freedom.

A well known quadratic action for the massless bosonic particle
propagating in D--dimensional Minkowski space is
\begin{equation}\label{2.1.1}
S=\int d\tau L(x,\dot x, e)=\int d\tau {1\over {2e(\tau)}}
\dot x^{\underline m}
\dot x^{\underline n}\eta_{\underline{mn}},
\end{equation}
where $\eta_{\underline{mn}}$ is the diagonal Minkowski metric
$(-,+,...,+)$,
$\tau$ is the time variable parametrizing a particle worldline
$x^{\underline m}(\tau)$ $({\underline m}=0,1...,D-1)$
in the target space--time,
$\dot x^{\underline m}\equiv\partial_\tau x^{\underline m}$
and $e(\tau)$
is an auxiliary field which can be regarded as a (nondynamical)
gravitational field on the worldline of the particle.
The latter ensures the invariance of the
action (\ref{2.1.1}) under worldline reparametrizations
(diffeomorphisms)
$\tau'=f(\tau)$, the worldline fields $x^{\underline m}(\tau)$
and $e(\tau)$ transforming as scalars and a one--dimensional
covariant vector, respectively,
\begin{equation}\label{2.1.2}
{x'}^{\underline m}(\tau')=
x^{\underline m}(\tau), \qquad e'(\tau')={d\tau\over{d\tau'}}e(\tau)
={1\over{\dot f(\tau)}}e(\tau).
\end{equation}
The consequence of the reparametrization invariance is the
constraint on the dynamics of the particle
$$
{1\over {e^2}}\dot x^{\underline m}\dot x_{\underline m}=0,
$$
or
\begin{equation}\label{2.1.3}
p^{\underline m}p_{\underline m}=0,
\end{equation}
which is obtained by varying (\ref{2.1.1}) with respect to $e(\tau)$.
In eq. (\ref{2.1.3}) $p_{\underline m}
={{\delta L}\over{\delta\dot x^{\underline m}}}={1\over e}
\dot x^{\underline m}$ is the canonical momentum of the particle.

The constraint (\ref{2.1.3}) implies that the particle is massless.

Using the canonical momentum variable $p_{\underline m}(\tau)$
we can rewrite
the action (\ref{2.1.1}) in the first order form
\begin{equation}\label{2.1.4}
S=\int d\tau(p_{\underline m}\dot x^{\underline m}-
{e\over{2}}p_{\underline m}p^{\underline m}).
\end{equation}

We will now generalize both versions of the massless particle
action to describe various supersymmetric particles.

\subsection{Spinning particles}

Let us assume that the trajectory of a particle in a bosonic
space--time is not a line but a supersurface parametrized by one
bosonic variable $\tau$ and one fermionic variable $\eta$ (which
can be regarded as a one-dimensional worldline spinor). The
functions $X^{\underline m}(\tau,\eta)$ describing the embedding
of this
supersurface into D-dimensional bosonic target space become worldline
superfields
\begin{equation}\label{2.2.1}
X^{\underline m}(\tau,\eta)=x^{\underline m}(\tau)
+i\eta\chi^{\underline m}(\tau),
\end{equation}
where, as above, $x^{\underline m}(\tau)$ are the coordinates of the
particle in
the target space and $\chi^{\underline m}$
are their Grassmann--odd superpartners.\footnote{In (\ref{2.2.1})
and below
capital letters stand for worldvolume superfields,
while corresponding small letters denote leading components of
these superfields which are associated with conventional
dynamical variables of the superbranes.}

The one--dimensional graviton $e(\tau)$ becomes a member of
a (supergravity) superfield
\begin{equation}\label{2.2.2}
E(\tau,\eta)=e(\tau)+2i\eta\psi(\tau)
\end{equation}
and thus acquires a gravitino field $\psi(\tau)$ as its Grassmann--odd
superpartner.

The particle momentum $p_{\underline m}$ enters a superfield
\begin{equation}\label{2.2.02}
P_{\underline m}(\tau,\eta)=p_{\underline m}(\tau)
+i\eta\rho_{\underline m}(\tau).
\end{equation}

To get an action for a spinning superparticle one can
supersymmetrize both the action (\ref{2.1.1})
or its first--order form (\ref{2.1.4}). Since the generalization
of the latter is more sophisticated \cite{solom} we shall not
describe it below.

The action (\ref{2.1.1}) is generalized to a superworldline
integral as follows \cite{spin1/2}
\begin{equation}\label{2.2.3}
S=-\int d\tau d\eta{i\over{2E}}D X^{\underline m}
\partial_\tau X_{\underline m}=
-\int d\tau d\eta{1\over{2E}}DX^{\underline m}
D(DX_{\underline m}),
\end{equation}
where
\begin{equation}\label{2.2.04}
D={\partial\over{\partial\eta}}+i\eta{\partial\over{\partial\tau}},
\qquad D^2={1\over 2}\{D,D\}=i\partial_\tau
\end{equation}
is a Grassmann covariant derivative.

The action (\ref{2.2.3}) is invariant under local worldline
superreparametrizations which include
bosonic worldline reparametrizations and local supersymmetry
transformations. The infinitesimal superreparametrizations are
\begin{equation}\label{2.2.4}
\begin{array}{rl}
\tau'-\tau&=\delta\tau=\Lambda(\tau,\eta)-{1\over 2}\eta D\Lambda,
\\
\eta'-\eta&=\delta \eta = -{i\over 2}D\Lambda,\\
D'-D&=\delta D=-{{1}\over 2}\dot\Lambda D,
\end{array}
\end{equation}
where $\Lambda(\tau,\eta)=a(\tau)+i\eta\alpha(\tau)$ is the
superreparametrization
parameter which contains the worldline
bosonic reparametrization parameter $a(\tau)$ and the local
supersymmetry
parameter $\alpha(\tau)$.
Under (\ref{2.2.4}) the superfield
$X^{\underline m}(\tau,\eta)$ (\ref{2.2.1})
transforms as a scalar
$X^{\prime{\underline m}}(\tau',\eta')=X^{\underline m}(\tau,\eta)$,
its components having the following variation properties
($\delta X^{{\underline m}}\equiv X^{\prime{\underline m}}(\tau,\eta)
-X^{\underline m}(\tau,\eta)$)
$$
\delta x^{\underline m}=-a(\tau)\dot x^{\underline m}
-{i\over 2}\alpha(\tau)\chi^{\underline m},
$$
\begin{equation}\label{2.2.5}
\delta \chi^{\underline m}=
-a(\tau)\dot\chi^{\underline m}-{1\over 2}\dot a(\tau)\chi^{\underline
m}
-{1\over 2}\alpha(\tau)\dot x^{\underline m}.
\end{equation}
The superfield $P_{\underline m}(\tau,\eta)$ also transforms as
a scalar $P'_{\underline m}(\tau',\eta')=P_{\underline m}(\tau,\eta)$
$$
\delta p_{\underline m}=-a(\tau)\dot p_{\underline m}
-{i\over 2}\alpha(\tau)\rho_{\underline m},
$$
\begin{equation}\label{2.2.05}
\delta \rho_{\underline m}=
-a(\tau)\dot\rho_{\underline m}-{1\over 2}\dot a(\tau)\rho_{\underline
m}
-{1\over 2}\alpha(\tau)\dot p_{\underline m}.
\end{equation}
Finally, the transformation properties of the
superfield $E(\tau,\eta)$ are
\begin{equation}\label{2.2.6}
\delta E=-\partial_\tau(\Lambda E) + {i\over 2}D\Lambda DE,
\end{equation}
from which one derives the variation of the worldline graviton
$e(\tau)$ and the gravitino $\psi(\tau)$
\begin{equation}\label{2.2.7}
\begin{array}{rl}
\delta e& =-\partial_\tau(ae)-{i}\alpha(\tau)\psi,
\\
\delta\psi&=-a(\tau)\dot\psi-{3\over 2}\dot a \psi
-{1\over4}\alpha(\tau)
\dot e-{1\over 2}\dot\alpha(\tau)e.
\end{array}
\end{equation}

Let us note that the geometry of the superworldline associated with
the superreparametrizations defined in eq. (\ref{2.2.4}) is conformally
flat. This means that worldline supervielbeins
$e^A(\tau,\eta)=(e^\tau,e^\eta)$
describing such a
geometry differ from flat supervielbeins
$e^A_0(\tau,\eta)=(e^\tau_0,e^\eta_0=d\eta)$ by a conformal factor
(the subscript 0 indicates the flat basis). Namely,
$$
e^\tau=E(\tau,\eta)e^\tau_0=E(\tau,\eta)(d\tau+i\eta d\eta),
$$
\begin{equation}\label{2.2.07}
e^\eta=E^{1\over 2}(\tau,\eta)d\eta
-ie^\tau_0DE^{1\over 2}.
\end{equation}
(Note that as in the flat case
such a form of the supervielbein satisfies the worldline
torsion constraint $T^\tau\equiv de^\tau=-ie^\eta\wedge e^\eta$.)

In the case of the conformally flat worldline supergeometry
(and a corresponding  worldsheet
supergeometry of superstrings)
it is convenient to use flat supercovariant derivatives
(\ref{2.2.4}) and supervielbeins rather than curved
ones and just take into account conformal factors where they are
required
for the covariance.

Integrating (\ref{2.2.3}) over $\eta$ (using
the Grassmann integration rules $\int d\eta=0$ and
$\int d\eta\eta=1$) we get the component action
\begin{equation}\label{2.2.8}
S=\int d\tau\left[{1\over{2e}}(\dot x^{\underline m}
\dot x_{\underline m}+i\dot\chi^{\underline m}\chi_{\underline m})-
{i\over{e^2}}\psi\chi_{\underline m}\dot x^{\underline m}\right].
\end{equation}
In addition to the `masslessness' constraint (\ref{2.1.3}) the action
(\ref{2.2.8}) yields the fermionic constraint which comes from the
variation of the last term in (\ref{2.2.8}) with respect to
$\psi(\tau)$
\begin{equation}\label{2.2.9}
\chi^{\underline m}p_{\underline m}=0,
\end{equation}
where now the canonical momentum is
\begin{equation}\label{2.2.pm}
p_{\underline m}={1\over e}
(\dot x_{\underline m}-{i\over e}\psi\chi_{\underline m}).
\end{equation}
(Note that since $\chi_{\underline m}$ are Grassmann--odd,
$\chi_{\underline m}\chi^{\underline m}\equiv 0$).

Upon quantization the constraint (\ref{2.2.9}) becomes the Dirac
equation imposed on a first--quantized spinorial wave function.
Therefore, the action (\ref{2.1.3}) or (\ref{2.2.8}) describes
a relativistic particle with spin 1/2, and the Grassmann vector
$\chi^{\underline m}$ describes the spinning degrees of freedom of
the particle \cite{berma,spin1/2}.

To get an action for relativistic
particles of an arbitrary spin ${n\over 2}$ (where $n$ is a natural
number) one should consider the worldline of a particle to be
a supersurface parametrized by $n$ fermionic variables $\eta^q$
$(q=1,...,n)$ \cite{gert,spin>}.

\subsection{Superparticles, $\kappa$-symmetry}
Let us now do an opposite thing. Instead of considering the embedding
of a worldline supersurface into a bosonic D--dimensional space--time
consider an embedding of a bosonic worldline into a flat
target superspace parametrized by bosonic coordinates
$x^{\underline m}$ $({\underline m}=0,1,...,D-1)$ and
Grassmann spinor coordinates
$\theta^{\underline\mu}$ $({\underline \mu}=1,...,2^{[{D\over 2}]})$.
Depending on the dimension $D$ the spinors $\theta$ can be chosen
to be Dirac, Majorana or Majorana--Weyl ones.

The global supersymmetry transformations of the target superspace
coordinates are
\begin{equation}\label{2.3.1}
\delta\theta^{\underline{\mu}}=\epsilon^{\underline{\mu}}, \qquad
\delta x^{\underline{m}}=
i\bar\theta\Gamma^{\underline{m}}\delta\theta,
\end{equation}
and we would like to construct an action for a particle which
would be invariant under these transformations.

For the construction of such an action one uses supercovariant
one--forms
${\cal E}^{\underline A}=\left({\cal E}^{\underline a},
{\cal E}^{\underline\alpha}\right)$
which form an orthogonal supervielbein basis in
the flat target superspace
\begin{equation}\label{2.3.2}
{\cal E}^{\underline a}= \left(dx^{\underline m}-
id\bar\theta\Gamma^{\underline m}\theta\right)
\delta_{\underline m}^{\underline a}, \quad
{\cal
E}^{\underline\alpha}=d\theta^{\underline\mu}\delta_{\underline\mu}^
{\underline\alpha},
\end{equation}
The target superspace geometry
is flat, i.e. it has zero curvature, but nonzero torsion (the external
differential acts from the right)
\begin{equation}\label{2.3.3}
{\cal T}^{\underline a}=d{\cal E}^{\underline a}
=-id\bar\theta\Gamma^{\underline a}d\theta.
\end{equation}

[When superspaces are flat we shall usually not make a distinction
between
tangent--space `flat' indices (from the beginning of the alphabets) and
coordinate indices (from the middle of the alphabets), since in this
case the two types of indices are simply related by the unit matrix,
as in the definition of the supervielbeins (\ref{2.3.2}).]

The pullback of the supervielbeins (\ref{2.3.2}) onto the particle
worldline with the image
$Z^{\underline M}(\tau)=\left(x^{\underline m}(\tau),
\theta^{\underline\mu}(\tau)\right)$ is
\begin{equation}\label{2.3.4}
{\cal E}^{\underline a}(Z(\tau))=d\tau{\cal E}_{\tau}^{\underline
a}(Z(\tau))
= d\tau\left(\partial_\tau x^{\underline a}-
i\partial_\tau\bar\theta\Gamma^{\underline a}\theta \right),
\end{equation}
$$
{\cal E}^{\underline\alpha}(Z(\tau))=
d\tau\partial_\tau\theta^{\underline\alpha}(\tau).
$$
The superparticle action is obtained by replacing
$\dot x^{\underline m}$ in (\ref{2.1.1}) or (\ref{2.1.4}) with the
vector component
${\cal E}_{\tau}^{\underline a}$
of the supervielbein pullback (\ref{2.3.4}). Thus we get
\cite{casal,pv,BS}
\begin{equation}\label{2.3.5}
S=\int d\tau{1\over {2e(\tau)}}
{\cal E}_{\tau}^{\underline a}
{\cal E}_{\tau}^{\underline b}\eta_{\underline{ab}},
\end{equation}
or in the first order form
\begin{equation}\label{2.3.6}
S=\int d\tau(p_{\underline a}{\cal E}_{\tau}^{\underline a}-
{e\over{2}}p_{\underline a}p^{\underline a}).
\end{equation}

The superparticle equations of motion derived from (\ref{2.3.5}) or
(\ref{2.3.6}) are
\begin{equation}\label{2.3.06}
\partial_\tau({1\over {e(\tau)}}
{\cal E}_{\tau}^{\underline a})=\partial_\tau p^{\underline a}=0, \quad
({\cal E}_{\tau}^{\underline a}\Gamma_{\underline a})^{\underline\alpha}
_{~~\underline\beta}\partial_\tau\theta^{\underline\beta}=
(p^{\underline a}\Gamma_{\underline a})^{\underline\alpha}
_{~~\underline\beta}\partial_\tau\theta^{\underline\beta}=0,
\end{equation}
$$
{\cal E}_{\tau}^{\underline a}=\partial_\tau x^{\underline a}-
i\partial_\tau\bar\theta\Gamma^{\underline a}\theta=e(\tau)p_{\underline
a}.
$$.

By construction the actions (\ref{2.3.5}) and (\ref{2.3.6}) are
invariant under the target--space supersymmetry transformations
(\ref{2.3.1}), and under the worldline reparametrization of
$x^{\underline m}$, $e(\tau)$ (\ref{2.1.2}), $p'_{\underline m}(\tau')=
p_{\underline m}(\tau)$ and $\theta'(\tau')=\theta(\tau)$,
which is responsible for
the mass--shell condition (\ref{2.1.3}).
In addition, these actions possess
a hidden worldline fermionic $\kappa$--symmetry \cite{al,sig}.
The $\kappa$--transformations of the worldline fields are
\begin{equation}\label{2.3.7}
\begin{array}{rl}
\delta_\kappa\theta^{\underline{\mu}}&=
i(p_{\underline m}\Gamma^{\underline m}
)^{\underline{\mu}}_{~~\underline{\nu}}
\kappa^{\underline{\nu}}(\tau), \\
\delta_\kappa x^{\underline{m}}&=
-i\bar\theta\Gamma^{\underline{m}}\delta_\kappa\theta~~~\Rightarrow~~~
i_\kappa {\cal E}^{\underline a}\equiv \delta Z^{\underline M}
{\cal E}_{\underline M}^{\underline a}=0,\\
\delta_\kappa e&=4\dot{\bar\theta}\kappa,\\
\delta_\kappa p_{\underline m}&=0,
\end{array}
\end{equation}
where $\kappa^{\underline\mu}(\tau)$ is a Grassmann spinor parameter
of the $\kappa$--transformations.

One can notice that not all $\kappa^{\underline\mu}(\tau)$
contribute to the $\kappa$--transformations, and that, in fact,
the transformations (\ref{2.3.7}) are infinitely
reducible.

Consider a particular choice of
$\kappa^{\underline\mu}(\tau)$ when
\begin{equation}\label{2.3.8}
\kappa^{\underline\mu}(\tau)=i
(p_{\underline m}\Gamma^{\underline m}
)^{\underline{\mu}}_{~~\underline{\nu}}
\kappa^{\prime\underline{\nu}}(\tau).
\end{equation}
Then, due to the defining relations for the Dirac matrices
\begin{equation}\label{2.3.08}
\{\Gamma^{\underline m},\Gamma^{\underline n}\}=
\Gamma^{\underline m}\Gamma^{\underline n} +
\Gamma^{\underline m}\Gamma^{\underline n}=2\eta^{\underline{mn}},
\end{equation}
the transformation of $\theta$ takes the form
\begin{equation}\label{2.3.9}
\delta_\kappa\theta^{\underline{\mu}}(\tau)=
-p_{\underline m}p^{\underline m}\kappa^{\prime\underline{\mu}}.
\end{equation}
This transformation vanishes on the mass shell $p^2=0$
(\ref{2.1.3}).
Therefore, the parameters of the form (\ref{2.3.8}) do not eliminate
any on--shell gauge degrees of freedom of the superparticle.

If in (\ref{2.3.8}) we choose
$\kappa^\prime=
p_{\underline m}\Gamma^{\underline m}\kappa^{\prime\prime}$
we will see that
on the mass shell (\ref{2.1.3}) the parameter (\ref{2.3.8})
turns to zero. We can continue such a substitution an infinite number
of times and find that at any stage there are $\kappa$--parameters
for which the $\kappa$--transformations are trivial on the
mass shell. The reason for this (infinite) reducibility is that
the spinorial matrix $p_{\underline m}\Gamma^{\underline m}$ has
the rank which on the mass shell (\ref{2.1.3}) is half the
maximum rank, i.e. half the dimension $2^{[{D\over 2}]}$ of the
spinor. This means that among the $2^{[{D\over 2}]}$
components of $\kappa^{\underline\mu}$ only half of the components
are independent and effectively contribute to the $\kappa$--symmetry
transformations. This also concerns fermionic constraints on the
superparticle dynamics whose appearance is the consequence of
$\kappa$--symmetry. In the Dirac terminology \cite{dirac}
these constraints belong to the first class, and in the canonical
Hamiltonian formulation they are regarded as the generators of the local
$\kappa$--symmetry. The constraints in question are
\begin{equation}\label{2.3.10}
\pi_{\underline{\mu}}(\tau)(p_{\underline m}\Gamma^{\underline m}
)^{\underline{\mu}}_{~~\underline{\nu}}=0,
\end{equation}
where $\pi_{\underline\mu}=
{{\delta L}\over{\delta\dot\theta^{\underline\mu}}}
=i\bar\theta_{\underline{\nu}}(p_{\underline m}\Gamma^{\underline m}
)^{\underline{\nu}}_{~~\underline{\mu}}$ is
the momentum canonically conjugate to the Grassmann coordinate
$\theta$. Since the expression for $\pi$ does not contain
time--derivatives it is a fermionic spinorial
constraint
\begin{equation}\label{2.3.11}
D_{\underline\mu}=\pi_{\underline\mu}-
i\bar\theta_{\underline{\nu}}(p_{\underline m}\Gamma^{\underline m}
)^{\underline{\nu}}_{~~\underline{\mu}}=0.
\end{equation}
Under the canonical Poisson brackets $[p_{\underline m},x^{\underline
n}]=
\delta_{\underline m}^{\underline n}$,
$\{\pi_{\underline\mu},\theta^{\underline\nu}\}=
\delta_{\underline\mu}^{\underline\nu}$\footnote{The brackets $[,]$
and $\{,\}$ stand for commutation and anticommutation relations,
respectively.}  the constraints
(\ref{2.3.11}) have the anticommutation properties of the
super--Poincare
algebra
\begin{equation}\label{2.3.12}
\{\bar D^{\underline\mu}, D_{\underline\nu}\}
=-2i(p_{\underline m}\Gamma^{\underline m}
)^{\underline{\mu}}_{~~\underline{\nu}}.
\end{equation}
As we have seen, the matrix on the right hand side of (\ref{2.3.12})
is degenerate and has the rank one half of the maximum rank.
Thus half the constraints (\ref{2.3.12}) are of the second class
and another half are of the first class, the anticommutator of the
latter
being (weakly) zero.

Without introducing auxiliary fields
it turns out to be impossible to split the fermionic spinor constraints
(\ref{2.3.11}) into irreducible Lorentz covariant sets
of first and second class
constraints because the spinor
representation of the Lorentz group is the fundamental one and
cannot be decomposed into any other Lorentz group representations.

The maximum we can do in the present situation is to get an infinite
reducible covariant  set of the first class fermionic constraints
(\ref{2.3.10})
multiplying (\ref{2.3.11}) by $p_{\underline m}\Gamma^{\underline m}$,
or to break manifest Lorentz covariance.

If we try to covariantly quantize the superparticle model, the infinite
reducibility of $\kappa$-symmetry and of the corresponding constraints
will require the introduction of an infinite set of ghosts for ghosts
\cite{infini} and, in addition, we should also manage with the second
class
constraints contained in (\ref{2.3.11}). All this makes the problem
of the covariant quantization of superparticles and superstrings
a difficult one.

To be able to split the fermionic constraints (\ref{2.3.11}) into
irreducible sets of first and second class constraints it has been
proposed to enlarge the space of superparticle (and superstring)
variables with auxiliary bosonic spinor variables (twistors
\cite{fer,shir}--\cite{gums,stv},
or Lorentz harmonics \cite{gikos},\cite{sok}--\cite{bzm,banquant})
and to use these variables to carry
out a covariant split of the constraints.

Below we shall see that these auxiliary variables naturally appear
in the doubly supersymmetric formulations of the super--p--branes.

\subsection{Spinning superparticles}
Let us now make the next step and to construct particle models
which would be invariant under both the worldline (\ref{2.2.4})
and the target--space (\ref{2.3.1}) supersymmetry transformations.
The worldline of such a particle is a supersurface $z^M=(\tau,\eta)$
embedded into the target superspace $Z^{\underline M}
=\left(X^{\underline m},
\Theta^{\underline\mu}\right)$. I.e. the particle trajectory is
\begin{equation}\label{2.4.1}
Z^{\underline M}(z^M)=Z^{\underline M}(\tau,\eta)
=\left(X^{\underline m}(\tau,\eta),
\Theta^{\underline\mu}(\tau,\eta)\right),
\end{equation}
where the worldvolume superfield $X^{\underline m}(\tau,\eta)$ is
the same as in (\ref{2.2.1}) and
\begin{equation}\label{2.4.2}
\Theta^{\underline\mu}(\tau,\eta)=\theta^{\underline\mu}(\tau)+
\eta\lambda^{\underline\mu}(\tau).
\end{equation}
We see that the Grassmann spinor coordinate $\theta(\tau)$
of the particle acquires a commuting spinor superpartner
$\lambda(\tau)$.

Under worldline superreparametrizations (\ref{2.2.4})
$\Theta(\tau,\eta)$
transforms as a scalar superfield
$$
\delta \theta^{\underline\mu}=-a(\tau)\dot \theta^{\underline\mu}
-{1\over 2}\alpha(\tau)\lambda^{\underline\mu},
$$
\begin{equation}\label{2.4.02}
\delta \lambda^{\underline\mu}=
-a(\tau)\dot\lambda^{\underline\mu}-{1\over 2}\dot a(\tau)
\lambda^{\underline\mu}-{i\over
2}\alpha(\tau)\dot\theta^{\underline\mu}.
\end{equation}

The pullback onto the superworldline of the superinvariant forms
${\cal E}^{\underline A}=\left({\cal E}^{\underline a},
{\cal E}^{\underline\alpha}\right)$
(\ref{2.3.2}) takes the form
\begin{equation}\label{2.4.3}
\begin{array}{rl}
{\cal E}^{\underline\alpha}(Z(z^M))&=dz^M{\cal
E}_{M}^{\underline\alpha}(Z(z))
=dz^M\partial_M\Theta^{\underline\alpha}(z)=
e^\tau_0\partial_\tau\Theta^{\underline\alpha}+
e^\eta_0D\Theta^{\underline\alpha}, \\
{\cal E}^{\underline a}(Z(z))&=dz^M {\cal E}_{M}^{\underline a}(Z(z)) \\
&=  e^\tau_0\left(\partial_\tau X^{\underline a}-
i\partial_\tau\bar\Theta\Gamma^{\underline a}\Theta \right)+
e^\eta_0\left(D X^{\underline a}-
iD\bar\Theta\Gamma^{\underline a}\Theta \right),
\end{array}
\end{equation}
where we have expanded the target
superforms in the flat supervielbein
basis (\ref{2.2.07}) of the superworldline.

Note that the worldline spinor components of the superforms
(\ref{2.4.3})
are basic ones in the sense that the `$\tau$'--components can
be constructed from the former
by applying to the `$\eta$'--components the covariant derivative $D$.
We thus have
\begin{equation}\label{2.4.4}
\begin{array}{rl}
\partial_\tau\Theta&=-iD\left(D\Theta\right),\\
{\cal E}_{\tau}^{\underline a}&=\partial_\tau X^{\underline a}-
i\partial_\tau\bar\Theta\Gamma^{\underline a}\Theta\\
&= -iD\left(D X^{\underline a}-
iD\bar\Theta\Gamma^{\underline a}\Theta \right)
-D\bar\Theta\Gamma^{\underline a}D\Theta.
\end{array}
\end{equation}
Therefore, it is sufficient to use the basic `$\eta$'--components
to construct doubly supersymmetric actions.

It turns out that depending on which action (\ref{2.2.3}), (\ref{2.3.5})
or (\ref{2.3.6})
is generalized to acquire the second supersymmetry we get different
doubly supersymmetric particle models.

The second supersymmetrization of the actions (\ref{2.2.3})
and (\ref{2.3.5}) results in a same doubly--supersymmetric action which
describes the dynamics of so called spinning superparticles
\cite{ssp,stvss}.

To get this action we should, for example, simply
replace $DX^{\underline m}$
in (\ref{2.2.4}) with the basic `$\eta$'--component of the
target--space vector supervielbein pullback (\ref{2.4.3})
\begin{equation}\label{2.4.05}
{\cal E}_{\eta}^{\underline a}=D X^{\underline a}-
iD\bar\Theta\Gamma^{\underline a}\Theta.
\end{equation}
The resulting action is \cite{ssp,stvss}
\begin{equation}\label{2.4.5}
S=-\int d\tau d\eta{1\over{2E}}{\cal E}_{\eta}^{\underline a}
D{\cal E}_{\eta\underline a}.
\end{equation}
We shall not discuss the properties of the particle model
described by the action (\ref{2.4.5}). The reader
may find the details in references \cite{ssp,stvss,sor}.
We only mention that the first--quantized
spectrum of this particle
is the direct product of the spectra of the corresponding
spinning particle (\ref{2.2.3}) and superparticle (\ref{2.3.5}).
And this is the reason for the name `spinning superparticle'.

For instance, in D=4 the first--quantized states of
the spinning particle (\ref{2.2.3}) have spin ${1\over 2}$.
The quantization of
the $N=1, D=4$
massless superparticle results in states described by a chiral
supermultiplet $(0,{1\over 2})$ which contains one complex scalar
and one Weyl spinor. The direct product of these spectra fits into an
$N=1$, $D=4$ chiral scalar $(0,{1\over 2})$ and vector $({1\over 2},1)$
supermultiplet. Therefore, the first--quantized states of the
$D=4$ spinning superparticle have two spins 0, two spins ${1\over 2}$
and
spin 1. This spin content is the same as one described by an $N=2, D=4$
vector
supermultiplet. The latter arises as a result of the quantization of
an $N=2, D=4$ superparticle \cite{abs}. The equivalence of the two
models was demonstrated in \cite{stvss,sor}.

We now turn to the construction of a doubly supersymmetric action
which would describe a particle model with the same physical content
as the standard superparticle model based on the actions (\ref{2.3.5})
and
(\ref{2.3.6}), the role of the local worldline supersymmetry being to
substitute the $\kappa$--symmetry.

\subsection{Worldline superfield actions for standard superparticles.
The superembedding condition.}
We have seen that the number $n$ of the independent $\kappa$--symmetry
transformations is half the number of components of the Grassmann
spinor coordinates of
the superparticle. Therefore, if a superparticle propagates in
superspace
with D bosonic and $2^{[{D\over 2}]}$ fermionic directions, one should
construct an $n=2^{[{D\over 2}]-1}$ worldline supersymmetric action for
being able to completely replace $n$ $\kappa$--symmetries
with $n$--extended worldline supersymmetry. We shall see that such
actions
can be constructed for $N=1$, $D=3,4,6$ and 10 massless
superparticles. A particular property of these critical bosonic
dimensions is that there exists a twistor--like (Cartan--Penrose)
representation of a light--like vector in terms of a commuting spinor.
Note that standard classical Green--Schwarz superstrings also exist only
in these dimensions \cite{gs,gsw},
which is implicitly related to the same fact.
Superparticles (including massive ones as \cite{ik,gaunt})
and superstring models in other space--time dimensions can be
obtained from these basic models by a dimensional reduction
\cite{ps92,bms}.

Let us first consider the simplest case of an

\subsubsection{$N=1, D=3$ superparticle and twistors}
In this case the Majorana spinor has two components and
$\kappa$--symmetry
has only one independent parameter. Our goal is to identify the
independent
$\kappa$--symmetry transformations of (\ref{2.3.7}) with the
local worldline supersymmetry transformations (\ref{2.2.4}). To this end
we write down a worldline supersymmetric version of the first--order
action
(\ref{2.3.6})
\begin{equation}\label{2.5.1}
S=\int d\tau d\eta\left[P_{\underline a}{\cal E}_{\eta}^{\underline a}-
{{\cal L}\over{2}}P_{\underline a}P^{\underline a}\right],
\end{equation}
where the worldline even superfield $P_{\underline a}$ and the odd
superfield
${\cal E}_{\eta}^{\underline a}$ are defined in eqs. (\ref{2.2.02}) and
(\ref{2.4.05}), respectively, and the Lagrange multiplier superfield
${\cal L}(\tau,\eta)$
is Grassmann odd
for the second term of (\ref{2.5.1}) to have the right Grassmann parity.

By construction the action (\ref{2.5.1}) is invariant under
the worldline superrepara\-me\-t\-rizations (\ref{2.2.4})
and target--space supersymmetry transformations (\ref{2.3.1}),
and it is also invariant under
fermionic transformations (\ref{2.3.7}), where all variables are
replaced by
corresponding superfields,
and ${d\over{d\tau}}$ is
replaced by the supercovariant derivative $D$.
At the first glance it seems
that we have not got rid of the $\kappa$--symmetry since it appeared
again
at the worldline superfield level.
However, as we shall see, the second term of (\ref{2.5.1}) is
unnecessary \cite{ps92}.
Its role would be to produce a superfield generalization of the mass
shell
condition (\ref{2.1.3}), but, it turns out that eq. (\ref{2.1.3})
follows
already from the first term of the action (\ref{2.5.1}). We therefore
skip the second term. Then the resulting action \cite{stv}
\begin{equation}\label{2.5.2}
S=-i\int d\tau d\eta P_{\underline a}{\cal E}_{\eta}^{\underline a}=
-i\int d\tau d\eta P_{\underline a}\left[D X^{\underline a}-
iD\bar\Theta\Gamma^{\underline a}\Theta\right]
\end{equation}
does not have any redundant $\kappa$--symmetry, and,
as we shall now demonstrate,
is equivalent to the standard superparticle action (\ref{2.3.6}) in
$D=3$.

As we shall see in Subsection 2.5.4. in  $N=1$, $D=4,6$ and 10
target superspaces
the $n=1$ worldsheet superfield action (\ref{2.5.2}) also
describes standard superparticles, since then in addition to $n=1$
worldsheet supersymmetry it has $D-3$ hidden local
fermionic $\kappa$--symmetries.

Note that the action (\ref{2.5.2}) is of a `topological' Chern--Simons
nature \cite{chern} since it is invariant under the local worldline
superreparametrizations but does not contain the superworldline metric.

Performing $\eta$--integration in (\ref{2.5.2}) we get the following
action for the components of the superfields (\ref{2.2.1}),
(\ref{2.2.02})
and (\ref{2.4.2})
\begin{equation}\label{2.5.3}
S=\int d\tau p_{\underline a}(\dot x^{\underline a}-
i\dot{\bar\theta}\Gamma^{\underline a}\theta-\bar\lambda
\Gamma^{\underline a}\lambda)+
i\int d\tau\rho_{\underline a}(\chi^{\underline a}-
\bar\lambda\Gamma^{\underline a}\theta).
\end{equation}
The second term in (\ref{2.5.3}) reads that $\rho_{\underline a}$ and
$\chi^{\underline a}$ are auxiliary fields satisfying algebraic
equations
\begin{equation}\label{2.5.4}
\rho_{\underline a}=0, \quad \chi^{\underline a}=
\bar\lambda\Gamma^{\underline a}\theta.
\end{equation}
The equations of motion of $x^{\underline a}$ and
$\theta^{\underline\alpha}$ are, respectively,
\begin{equation}\label{2.5.5}
\partial_\tau p_{\underline a}=0, \quad
(p_{\underline a}\Gamma^{\underline a})^{\underline\alpha}
_{~~\underline\beta}\partial_\tau\theta^{\underline\beta}=0.
\end{equation}
They are the same as eqs. (\ref{2.3.06}) yielded by the action
(\ref{2.3.6}). We should now show that the mass shell condition
(\ref{2.1.3}) also follows from the action (\ref{2.5.3}). This can
be derived in two ways.

The variation of (\ref{2.5.3}) with respect
to $p_{\underline a}$ gives
\begin{equation}\label{2.5.6}
{\cal E}_{\tau}^{\underline a}|_{\eta=0}\equiv\dot x^{\underline a}-
i\dot{\bar\theta}\Gamma^{\underline a}\theta=\bar\lambda
\Gamma^{\underline a}\lambda,
\end{equation}
from which it follows that
\begin{equation}\label{2.5.7}
\eta_{\underline{ab}}{\cal E}_{\tau}^{\underline a}
{\cal E}_{\tau}^{\underline b}|_{\eta=0}= 0.
\end{equation}

This is because the square of the r.h.s. of (\ref{2.5.6}) is identically
zero
\begin{equation}\label{2.5.07}
(\bar\lambda
\Gamma^{\underline a}\lambda)(\bar\lambda
\Gamma_{\underline a}\lambda)\equiv 0
\end{equation}
due to a property of the $\Gamma$--matrices in $D=3$
\begin{equation}\label{2.5.8}
(\Gamma^{\underline m})^{\underline\alpha}_{~~\underline\beta}
(C\Gamma_{\underline m})_{\underline{\gamma\delta}}+
(\Gamma^{\underline m})^{\underline\alpha}_{~~\underline\gamma}
(C\Gamma_{\underline m})_{\underline{\delta\beta}}+
(\Gamma^{\underline m})^{\underline\alpha}_{~~\underline\delta}
(C\Gamma_{\underline m})_{\underline{\delta\beta}}=0,
\end{equation}
where $C_{\underline{\alpha}\underline{\beta}}=
\epsilon_{\underline{\alpha}\underline{\beta}}$
is the charge conjugation matrix which can be used to raise and
lower the spinor indices.

The identities (\ref{2.5.07}) and (\ref{2.5.8}) can be easily checked
in the Majorana representation of the Dirac matrices where the Majorana
spinor is real, and
\begin{equation}\label{2.5.9}
C=\Gamma_0=\left(
\begin{array}{cc}
0  &  1\\
-1 &  0
\end{array}
\right)\,,
\qquad
\Gamma_1=\left(
\begin{array}{cc}
0  &  1\\
1   &  0
\end{array}
\right)\,,
\qquad
\Gamma_2=\left(
\begin{array}{cc}
1   &  0\\
0   &   -1
\end{array}
\right)
\,.
\end{equation}

The identity (\ref{2.5.8}) also holds in
$D=4, 6$ and 10 space--time dimensions, and the eq. (\ref{2.5.07}) is
valid in these dimensions for the Majorana (or Weyl), simplectic
$SU(2)$ Majorana--Weyl \cite{kugo,west} and the Majorana--Weyl spinors,
respectively. As a consequence of these identities any light--like
vector in these dimensions can be represented as the
bilinear combination
of commuting spinors (\ref{2.5.6}). Such a representation is called
the Cartan--Penrose (or twistor)  representation of the light--like
vector. Note that its existence in the special dimensions
$D=3,4, 6$ and 10 is related to the existence of four division algebras
associated with real, complex, quaternionic and octonionic numbers,
respectively, (see \cite{jac}, and  \cite{kugo,gg,gt} as nice reviews
for physicists) which form an algebraic basis
for corresponding twistor constructions \cite{twistor,twbc}.
In application to
the superembedding formulation of superparticles and superstrings
these structures were exploited, for example, in references
\cite{shir,twbc,stv,hsstr1,twes,twplyus,ceder}.

Let us now derive the Cartan--Penrose relation as a solution
of the equation of motion of $\lambda$
\begin{equation}\label{2.5.10}
(p_{\underline a}\Gamma^{\underline a})^{\underline\alpha}
_{~~\underline\beta}\lambda^{\underline\beta}=0.
\end{equation}
If the matrix $p_{\underline a}\Gamma^{\underline a}$ is nondegenerate
the equations (\ref{2.5.5}) and (\ref{2.5.10}) are satisfied only when
\begin{equation}\label{2.5.11}
\partial_\tau\theta=0, \qquad \lambda=0,
\end{equation}
which (in virtue of (\ref{2.5.6})) leads to
$$
\partial_\tau x^{\underline m}=0.
$$
Such a solution describes a particle which is frozen in
a single point of super space--time (it even does not evolve
along the time direction $x^0$). Since this solution is physically
trivial it can be discarded by requiring that
the components of $\lambda$ do not turn to zero simultaneously
\footnote{Note that in the standard formulation of relativistic
particles
this requirement is analogous to excluding the point $e(\tau)=0$ from
the
solutions of the particle equations of motion (\ref{2.3.06}).}. This
requirement is in agreement with the definition of twistor variables
\cite{twistor} which are commuting spinor variables parametrizing a
projective
space. Then the eq. (\ref{2.5.10}) has nontrivial solutions if the
matrix  $p_{\underline a}\Gamma^{\underline a}$ is degenerate,
the general solution of (\ref{2.5.10}) being
\begin{equation}\label{2.5.12}
p_{\underline a}={1\over{e(\tau)}}\bar\lambda
\Gamma_{\underline a}\lambda,\quad \Rightarrow \quad
p_{\underline a}p^{\underline a}=0,
\end{equation}
or, because of the $\Gamma$--matrix identity (\ref{2.5.8}),
\begin{equation}\label{2.5.13}
p_{\underline a}(C\Gamma^{\underline a})_{\underline{\alpha}
\underline{\beta}}=
{2\over{e(\tau)}}\lambda_{\underline\alpha}\lambda_{\underline\beta}.
\end{equation}
Thus we have again arrived at the Cartan--Penrose representation of
the light--like vector.

Comparing (\ref{2.5.12}) with (\ref{2.5.6}) we see that
\begin{equation}\label{2.5.14}
p^{\underline a}={1\over{e(\tau)}}{\cal E}_{\tau}^{\underline
a}|_{\eta=0}=
{1\over{e(\tau)}}(\dot x^{\underline a}-
i\dot{\bar\theta}\Gamma^{\underline a}\theta)
\end{equation}
has the meaning of the superparticle canonical momentum as in the
standard formulation of Subsection 2.3, and the multiplier $e(\tau)$
can be identified with the worldline gravitation field.

One can notice that the Cartan--Penrose relation (\ref{2.5.12})
establishes a one--to--one correspondence between the two--component
commuting spinor $\lambda^{\underline\alpha}$
and the $D=3$ light--like vector
$p^{\underline a}$ which also has two independent components.
[The graviton $e(\tau)$ is completely auxiliary and can be gauge fixed
to a constant by the worldline reparametrizations (\ref{2.1.2}).]

Therefore, either $p^{\underline a}$ or $\lambda^{\underline\alpha}$
can be considered as independent dynamical variables in the phase space,
and the number of physical degrees of
freedom of the doubly supersymmetric particle is the same as that
of the standard superparticle.
If we choose $p^{\underline a}$ as
an independent
variable then the independent equations of motion (\ref{2.5.5}),
(\ref{2.5.12}) and (\ref{2.5.14}) of the
dynamical variables  $p^{\underline a}$, $x^{\underline a}$ and
$\theta^{\underline\alpha}$
derived from the action (\ref{2.5.3}) coincide with the massless
superparticle equations (\ref{2.3.06}) and (\ref{2.1.3})
yielded by the action (\ref{2.3.6}).
The local worldline supersymmetry can be used to eliminate one (pure
gauge) degree of freedom of $\theta^{\underline\alpha}$.
All this testifies
to the classical equivalence of the two actions.

The only thing which remains to establish is the relationship of the
worldline supersymmetry and $\kappa$--symmetry. For this consider
the worldline supersymmetry transformations (\ref{2.2.5}),
(\ref{2.2.05})
and (\ref{2.4.02}) of
$\theta^{\underline\mu}$, $\lambda^{\underline\mu}$,
$x^{\underline a}$ and $p^{\underline a}$, when the relations
(\ref{2.5.4}) for the auxiliary fields $\rho_{\underline a}$ and
$\chi^{\underline a}$ hold,
\begin{equation}\label{2.5.15}
\delta\theta^{\underline\mu}
=-{1\over 2}\lambda^{\underline\mu}\alpha(\tau), \qquad
\delta\lambda^{\underline\mu}
={i\over 2}\dot\theta^{\underline\mu}\alpha(\tau),
\end{equation}
\begin{equation}\label{2.5.16}
\delta x^{\underline a}
={i\over 2}\bar\theta\Gamma^{\underline a}\lambda\alpha(\tau)=
-i\bar\theta\Gamma^{\underline a}\delta\theta,
\end{equation}
\begin{equation}\label{2.5.17}
\delta p^{\underline a} = 0.
\end{equation}
Without losing generality we can always replace the supersymmetry
parameter $\alpha(\tau)$ with the following expression
\begin{equation}\label{2.5.20}
\alpha(\tau)=
-{{4i}\over e}{\bar\lambda}_{\underline\alpha}\kappa^{\underline\alpha},
\end{equation}
where $\kappa_{\underline\alpha}(\tau)$ is an arbitrary Grassmann
spinor.

If we now substitute eq. (\ref{2.5.20}) into
(\ref{2.5.15}) and (\ref{2.5.16}) and take into account
(\ref{2.5.13}) we shall see that the supersymmetry transformations
become
the $\kappa$--symmetry transformations (\ref{2.3.7}) for $x^{\underline
a}$
and $\theta^{\underline\mu}$.

It remains to derive the $\kappa$--variation of the worldvolume field
$e(\tau)$  (\ref{2.3.7}).

To this end we make use of the relation (\ref{2.5.12}), where $e(\tau$)
appears in the formulation under consideration, and
observe that when $\rho^{\underline a}=0$,
the particle momentum is invariant (\ref{2.5.17}) under
the local supersymmetry transformations (\ref{2.2.05}).
For consistency we must require that
the r.h.s. of the Cartan--Penrose representation (\ref{2.5.12})
of $p^{\underline a}$ is also invariant under the supersymmetry
transformations.
To show this we should find the supersymmetry variation of $e(\tau)$
which
cancels the variation of $\lambda$ (\ref{2.5.15}). At this point we
should
also take into account that $\theta$ satisfies the equation of motion
(\ref{2.5.5}). This requirement is justified by the fact that the
worldline
supersymmetry transformations relate the bosonic kinematic equation
(\ref{2.5.10}),
and its solution (\ref{2.5.12}), (\ref{2.5.13}), to the
fermionic dynamical equation (\ref{2.5.5}). In view of (\ref{2.5.10})
and
(\ref{2.5.13}) the general solution of (\ref{2.5.5}) is
\begin{equation}\label{2.5.18}
\dot\theta^{\underline\mu}=\lambda^{\underline\mu}\tilde\psi(\tau),
\end{equation}
where $\tilde\psi(\tau)$ is a Grassmann--odd worldline function.

Then the required worldline supersymmetry variation of $e(\tau)$ is
\begin{equation}\label{2.5.19}
\delta e(\tau)=-ie(\tau)\alpha(\tau)\tilde\psi(\tau).
\end{equation}
Substituting (\ref{2.5.20}) in (\ref{2.5.19}) we get the
$\kappa$--symmetry
variation (\ref{2.3.7}) of $e(\tau)$.

 Therefore, fermionic
$\kappa$--symmetry is nothing but the worldline supersymmetry
transformations of the superparticle dynamical variables when the
auxiliary
field components of the corresponding worldline superfields are
eliminated.
The independent $\kappa$--parameter (or the supersymmetry parameter)
is the (Lorentz--invariant) projection of the spinorial
$\kappa$--symmetry
parameter onto the commuting spinor $\lambda$ (\ref{2.5.20}).

The (irreducible) first class fermionic constraint
which generates $n=1$ worldline supersymmetry (and/or irreducible
$\kappa$--symmetry)
is
\begin{equation}\label{2.5.21}
\lambda^{\underline\alpha}D_{\underline\alpha}=0,\qquad
\{\lambda D,\lambda D\}=
-2ip_{\underline m}\bar\lambda\Gamma^{\underline m}\lambda=0,
\end{equation}
where $D_{\underline\alpha}$ was defined in (\ref{2.3.11}).

This completes the proof of the classical equivalence of the doubly
supersymmetric model based on the action (\ref{2.5.2}) and the standard
superparticle (\ref{2.3.5}), (\ref{2.3.6}) in $N=1$, $D=3$ superspace.

Before going further it is worth mentioning the relation of the
doubly supersymmetric action (\ref{2.5.2}) and (\ref{2.5.3}) with
twistor formulations of (super)particles
\cite{fer,shir}--\cite{sor,gums,bmrs}.

We have already observed similarity between the commuting spinor
variables $\lambda^{\underline\alpha}$ and twistors which manifests
itself through the Cartan--Penrose relation between
$\lambda^{\underline\alpha}$ and $p_{\underline m}$
(\ref{2.5.12}).

If we consider $\lambda^{\underline\alpha}$ and not
$p_{\underline m}$ as independent dynamical variables we can replace
$p_{\underline m}$ in (\ref{2.5.3}) with $\lambda^{\underline\alpha}$.
Then hiding $e(\tau)$ by redefining $\lambda^{\underline\alpha}$
($e^{-{1\over 2}}\lambda^{\underline\alpha} ~
\rightarrow~ \lambda^{\underline\alpha}$)
and dropping out the term with the fermionic
auxiliary fields we get
\begin{equation}\label{2.5.22}
S=\int d\tau \bar\lambda\Gamma_{\underline a}\lambda
\left(\dot x^{\underline a}-
i\dot{\bar\theta}\Gamma^{\underline a}\theta\right).
\end{equation}
This action is one on which the supertwistor formulation
\cite{fer,shir,twbc,twes}
of superparticles in space--times of dimension $D=3,4, 6$ and 10 is
based.
Various versions of the supertwistor formulation have been
developed in a number of papers \cite{shir}--\cite{bmrs,d10p,claus2}.
The doubly supersymmetric
formulation provides us with a natural link between the standard
(\ref{2.3.5}) and the twistor description of superparticles.

\subsubsection{On--shell relationship between superparticles and
spinning particles}
We have seen that in the doubly supersymmetric formulation of
superparticles
the Grassmann superpartners $\chi^{\underline m}$ of the
particle coordinates $x^{\underline m}$ are treated as
auxiliary variables expressed
in terms of $\theta^{\underline\alpha}$ and $\lambda^{\underline\alpha}$
(eq. (\ref{2.5.4})). Note that the equation (\ref{2.5.4})
can be regarded \cite{vz}
as an odd counterpart of the Cartan--Penrose twistor relation
(\ref{2.5.6}),
in fact, they are worldline supersymmetry partners.

On the contrary, in the theory of spinning particles (Section 2.2)
$\chi^{\underline m}$ is a dynamical (spin) variable obeying the
equations
\begin{equation}\label{2.2.09}
\chi^{\underline m}p_{\underline m}=\chi^{\underline m}
(\dot x_{\underline m}-{i\over e}\psi\chi_{\underline m})=0,
\end{equation}
\begin{equation}\label{chi}
\dot\chi^{\underline m}
={1\over e}\psi\dot x^{\underline m}+{{\dot e}\over{2e}}\chi^{\underline
m},
\end{equation}
which are yielded by the spinning particle action (\ref{2.2.8}).

We shall now demonstrate that the equations (\ref{2.2.09}) and
(\ref{chi}) are equivalent to the superparticle equations (\ref{2.5.5}).

Remember that the solution of the fermionic equation in (\ref{2.5.5})
is eq. (\ref{2.5.18}). Substituting  (\ref{2.5.18}) into the expression
for the superparticle momentum (\ref{2.5.14}) and taking into account
(\ref{2.5.4}), i.e. that $\chi^{\underline a}
=\bar\lambda\Gamma^{\underline a}\theta$, we have
\begin{equation}\label{pm}
p^{\underline a}={1\over{e(\tau)}}(\dot x^{\underline a}-
i\dot{\bar\theta}\Gamma^{\underline a}\theta)=
{1\over{e(\tau)}}(\dot x^{\underline a}-i\tilde\psi\chi^{\underline a}).
\end{equation}
We see that in this form the superparticle momentum coincides with
the spinning particle momentum (\ref{2.2.pm}) if we take
$\tilde\psi={1\over{e}}\psi$, both momenta satisfying the same
equation of motion $\dot p^{\underline a}=0$.

Contracting (\ref{pm}) with $\chi^{\underline
a}=\bar\lambda\Gamma^{\underline a}\theta$
and taking into account the Cartan--Penrose relation (\ref{2.5.12}) for
$p_{\underline a}$ we find that the equation (\ref{2.2.09}) is
identically
satisfied.

We now check that the solution
$\chi^{\underline m}=\bar\theta\Gamma^{\underline m}\lambda$ is
compatible
with the equation of motion  (\ref{chi}). For this take the time
derivative
of $\chi^{\underline m}$. Then, in view of (\ref{2.5.18}),
(\ref{2.5.12}),
(\ref{pm}) and that we put
$\tilde\psi={1\over{e}}\psi$, we get
$$
\dot\chi^{\underline m}
={1\over e}\psi\bar\lambda\Gamma^{\underline m}\lambda+
\bar\theta\Gamma^{\underline m}\dot\lambda
$$
\begin{equation}\label{chi1}
={1\over e}\psi (\dot x^{\underline m}-{i\over
e}\psi\chi^{\underline m})+ \bar\theta\Gamma^{\underline
m}\dot\lambda.
\end{equation}
$\dot\lambda$ is found to be
\begin{equation}\label{dotl}
\dot\lambda={{\dot e}\over {2e}}\lambda
\end{equation}
by solving
$$
\dot p^{\underline m}
=\partial_\tau\left({1\over e}\bar\lambda\Gamma^{\underline
m}\lambda\right)
=0.
$$
Hence, the last term in (\ref{chi1}) is equal to
${{\dot e}\over{2e}}\chi^{\underline m}$,
and eq. (\ref{chi1}) coincides with  eq. (\ref{chi}).

We have shown that the dynamical variables of the $N=1$, $D=3$
superparticle and the $n=1$ spinning particle, and their equations of
motion are related to each other by
the Cartan--Penrose (twistor--transform)
expressions (\ref{2.5.4}) and (\ref{2.5.6})
$$
p^{\underline m}={1\over e}\bar\lambda\Gamma^{\underline m}\lambda,
\quad \chi^{\underline m}=\bar\theta\Gamma^{\underline m}\lambda.
$$
Therefore, the doubly supersymmetric action (\ref{2.5.2}) can be
regarded
as a `master' action for both types of the particles. It provides us
with superparticle or spinning particle equations of motion depending on
whether the Grassmann vector $\chi^{\underline m}$ or the  spinor
$\theta^{\underline\mu}$ is chosen as an independent dynamical
variable. The two particles are thus classically equivalent, though
quantum mechanically (as is well known) these are systems with different
spectra of states.

The on--shell relationship
between  $N=1$ superparticles and $n=1$ spinning particles
\cite{vz,stvz},
and between $N=1$ superstrings and $n=1$ spinning strings
\cite{apt,berk}
holds also in space--time dimensions $D=4,6$ and 10.
To demonstrate this relationship
one should gauge fix all but one local worldsheet supersymmetries
(or $\kappa$--symmetries) of the target space supersymmetric objects.
The
remaining local worldsheet supersymmetry is then
identified with $n=1$ worldsheet supersymmetry of the corresponding
spinning objects. The commuting twistor--like spinors relate the
dynamical variables of the two types of models.

It is well known that the equivalence between $D=10$ superstrings and
$D=10$ spinning strings extends to the quantum level so that the
two models describe one and the same quantum string theory, provided
the  NSR spinning strings have been GSO-projected.
The `twistor' transform demonstrated above allows one to utilize
useful properties of both the Green--Schwarz and the NSR formulation
to study the problem of covariant quantization of superstring theory
\cite{berquant,to,apt,tor,dpolyak}.

Another example of quantum equivalent systems (which we have mentioned
in Subsection 2.4) is the equivalence between the free $N=1$, $n=1$
spinning
superparticle and the $N=2$ superparticle in $D=4$ \cite{stvss}.

\subsubsection{The superembedding condition}
We now return to the consideration of the action (\ref{2.5.2})
and repeat its analysis directly
at the worldline superfield level with the purpose of getting the
principal
condition of the superembedding of the particle superworldline into
target superspace and to analyze its properties.

The superembedding condition is derived by varying the action
(\ref{2.5.2})
with respect to the superfield $P_{\underline a}(\tau,\eta)$, and it
reads
that the superparticle moves in target superspace in such a way that
the pullback of the target--space supervielbein vector component
along the Grassmann direction of the superworldline is zero
\begin{equation}\label{2.5.23}
{\cal E}_{\eta}^{\underline a}=DX^{\underline a}-
iD\bar\Theta\Gamma^{\underline a}\Theta=0.
\end{equation}
The consequence of eq. (\ref{2.5.23}) is obtained by
hitting (\ref{2.5.23}) with the derivative $D$
\begin{equation}\label{2.5.24}
\partial_\tau X^{\underline a}-
i\partial_\tau\bar\Theta\Gamma^{\underline a}\Theta
-D\bar\Theta\Gamma^{\underline a}D\Theta=0.
\end{equation}
Eq. (\ref{2.5.24}) implies that the component of
${\cal E}^{\underline a}$ along
the time direction of the worldline is expressed in terms of
the spinor components
${\cal E}^{\underline\alpha}=d\Theta^{\underline\alpha}$ of the target
space
supervielbein (\ref{2.4.3}) along the Grassmann direction of the
worldline
\begin{equation}\label{2.5.25}
{\cal E}^{\underline a}_{\tau}(\tau,\eta)\equiv
\partial_\tau X^{\underline a}-
i\partial_\tau\bar\Theta\Gamma^{\underline a}\Theta
=D\bar\Theta\Gamma^{\underline a}D\Theta.
\end{equation}
This is the superfield version of the Cartan--Penrose relation
(\ref{2.5.6})
which states that ${\cal E}^{\underline a}_{\tau}(\tau,\eta)$ is
light--like
and, hence, the superparticle is massless.

The condition (\ref{2.5.25}) is the only consequence of
eq. (\ref{2.5.23}) which one has in the case under consideration.
We see that the superembedding condition (\ref{2.5.23})
itself does not lead to
the dynamical equations of motion of the superparticle (\ref{2.5.5}).
The superfield form of the dynamical equations is
obtained from the action (\ref{2.5.2}) by varying it with respect to
$X^{\underline a}$ and $\Theta^{\underline\alpha}$
\begin{equation}\label{2.5.26}
DP_{\underline a}=0, \qquad
P_{\underline a}\Gamma^{\underline a}D\Theta=0.
\end{equation}

Therefore, in the case of the $N=1$, $D=3$ superparticle the
superembedding condition (\ref{2.5.5}) `prescribes' the superparticle to
be
massless but does not completely define its
classical dynamics. The classical equations of motion of the
superparticle
should be derived separately, for example, from the action principle as
above.

It turns out that the same situation holds for superparticles in
$N=1$, $D=4,6$ and 10 target superspaces.

\subsubsection{$N=1$, $D=4,6,10$  superparticles}
The $\kappa$--symmetry of the $N=1$, $D=4,6,10$  superparticles has
$n=D-2=2,4,8$ independent parameters, respectively.
Hence, to replace $\kappa$--symmetry  with local worldline supersymmetry
one should consider the embedding into $N=1$, $D=4,6,10$ target
superspace
of a superworldline ${\cal M}_{1,n}$ parametrized by the bosonic time
variable $\tau$ and
$n=D-2$ fermionic variables $\eta^q$ ($q=1,...,D-2$).

Note that, in general, one can consider doubly supersymmetric models in
$D=4,6$ and 10 with the number $n$ of worldvolume supersymmetries being
less than $D-2$. Such models also describe
standard superbranes but with only part of the $\kappa$--symmetries
represented as manifest $n$--extended worldline supersymmetry, the
other $D-2-n$
$\kappa$--symmetries remaining hidden. In the case
of superparticles and superstrings $N=1$ doubly supersymmetric
formulations with $n<D-2$ are used for establishing the classical
relationship between spinning and super objects \cite{vz}--\cite{berk}
as well as for superstring quantization \cite{berquant,tor}.

For instance, the $n=1$ worldline superfield action (\ref{2.5.2})
is also a consistent action for
standard $N=1$, $D=4,6,10$ superparticles, since it does possess
additional fermionic symmetries which complement the $n=1$ worldline
supersymmetry to $n=D-2$ $\kappa$--symmetries.

In the $D=4$ case the additional $\kappa$--symmetry transformations are
\begin{equation}\label{d4}
\delta_\kappa\Theta^{\underline\mu}=
K(\tau,\eta)(\Gamma^5D\Theta)^{\underline\mu}, \quad
\Gamma^5=i\Gamma^0\Gamma^1\Gamma^2\Gamma^3,
\end{equation}
$$
\delta_\kappa X^{\underline m}=-i\bar\Theta\Gamma^{\underline
m}\delta\Theta,
\quad \delta P_{\underline a}=0,
$$
where $K(\tau,\eta)=\kappa(\tau)+\eta b(\tau)$ is an $n=1$ superfield
parameter independent of the $n=1$ supersymmetry parameter
$\Lambda(\tau,\eta)$ (\ref{2.2.4}).

The bosonic component  $b(\tau)$ of $K(\tau,\eta)$
allows one to gauge away one of the four real components of
$\lambda^{\underline\mu}$. The
remaining three components correspond to three independent components
of the $D=4$ lightlike vector $\bar\lambda\Gamma^{\underline m}\lambda$
associated with the superparticle momentum.

The appearance in (\ref{d4}) of the matrix $\Gamma^5$ reflects (in
the language of the Majorana spinors) the presence of the complex
structure inherent to the Weyl representation of the $D=4$ spinors.
(The symmetry with the parameter $b(\tau)$
then becomes local $U(1)$ symmetry acting
on Weyl spinors $\lambda$).
This allows one to promote the second fermionic symmetry (\ref{d4}) to
the second manifest worldsheet supersymmetry by constructing an
$n=2$ superfield action in
$N=1$, $D=4$ superspace as an alternative to (\ref{2.5.2})
(see \cite{stv,ds,ps92}).

In $D=6$ the action (\ref{2.5.2}) should possess 4 fermionic
symmetries one of which is $n=1$ worldline supersymmetry
(\ref{2.2.4}). To show how the other
three look we should introduce the notion of the
$SU(2)$ simplectic Majorana--Weyl spinors (see \cite{kugo,west} for
details).
The spinor index $\underline \alpha=1,...,8$ splits into
the $SU(2)$ index $i=1,2$ and
the index $\alpha=1,...,4$ of a fundamental representation of
$SU^*(4)\sim SO(1,5)$.
By definition the $SU(2)$ simplectic Majorana--Weyl spinor
satisfies a pseudoreality condition
\begin{equation}\label{d6mw}
\overline{\theta^{\alpha i}}:= {\bar\theta}^{\dot\alpha}_{i}
=B^{\dot\alpha}_{~\beta}\theta^{\beta j}\epsilon_{ji},
\end{equation}
where the matrix $B$ is defined by the conditions
$$
B\gamma^{\underline m}B^{-1}=(\gamma^{\underline m})^*, \quad B^*B=-1
$$
and * denotes complex conjugation.

The $4\times 4$ matrices $\gamma^{\underline m}$
replace the $D=6$ Dirac matrices and are analogs of $2\times 2$
$\sigma$--matrices in $D=4$. They are antisymmetric
\begin{equation}\label{d6gamma}
\gamma^{\underline m}_{\alpha\beta}=-\gamma^{\underline
m}_{\beta\alpha}.
\end{equation}
Then {\bf 3} hidden local fermionic symmetries of the action
(\ref{2.5.2}) in $D=6$ are given by
\begin{equation}\label{d6}
\delta\Theta^{\alpha i}=
K_I(\tau,\eta)(\sigma^{I})^i_{~j} D\Theta^{\alpha j},
\end{equation}
where $(\sigma^{I})^i_{~j}$ $(I=1,2,3)$ are the $SU(2)$ Pauli matrices.
$X^{\underline a}$ and $P_{\underline a}$ are transformed as in
(\ref{d4}).

The commuting spinor $\lambda_{\alpha i}$ has now {\bf 8} real
components.
{\bf 5} of these components correspond to five independent components
of a $D=6$ lightlike vector (the particle momentum),
and another three are to be pure gauge
degrees of freedom. Indeed, they are eliminated by local $SU(2)$
transformations
\begin{equation}\label{su2}
\delta \lambda^{\alpha i}=
b_I(\tau)(\sigma^{I})^i_{~j}\lambda^{\alpha j}
\end{equation}
which are part of (\ref{d6}), where $b_I(\tau)=DK_I|_{\eta=0}$.

The $n=4$ superfield generalization of (\ref{2.5.2}), which involves
only one twistor--like spinor $\lambda^{\alpha i}$
and makes the symmetries (\ref{d6})
manifest, was constructed in \cite{ds} by the use of the harmonic
superspace
technique \cite{gikos}.

In the $D=10$ case the situation becomes much more complicated since
to furnish all {\bf 8} fermionic symmetries of the action (\ref{2.5.2})
in an irreducible form
one should deal with the nonassociative octonionic structure
\cite{hsstr1}.
[A corresponding $n=8$ octonionic superfield action which would directly
produce (\ref{2.5.3}) is unknown]. However, it is possible to write down
a reducible set of  $\kappa$--transformations which complement
the $n=1$ local supersymmetry (\ref{2.5.27}) to eight
independent fermionic symmetries of the action (\ref{2.5.2})
in the $D=10$ case \cite{d10p,apt}
\begin{equation}\label{d10}
\delta_\kappa\Theta^{\underline\alpha}=
(D\bar\Theta\Gamma_{\underline a}D\Theta)
(\Gamma^{\underline a}K)^{\underline\alpha}
-2D\Theta^{\underline\alpha}(D\bar\Theta K).
\end{equation}
$$
\delta_\kappa X^{\underline m}
=-i\bar\Theta\Gamma^{\underline m}\delta_\kappa\Theta.
$$
The transformations (\ref{d10}) are reducible since the parameters
$K^{\underline\alpha}(\tau,\eta)$ in the form
$$
K=(\Gamma^{\underline a}K_a)D\Theta,\quad
K_a={\tilde K}(D\bar\Theta\Gamma_{\underline a}D\Theta)
$$
do not contribute to (\ref{d10}). It can be shown that
the number of independent components
in $K^{\underline\alpha}$ is {\bf 7}.

The fermionic superparameter $K^{\underline\alpha}(\tau,\eta)$
contains bosonic
parameters which allow one to eliminate {\bf 7} of the {\bf 16}
components
of $\lambda^{\underline\alpha}$, so that the remaining {\bf 9}
components
again correspond to a lightlike $D=10$ vector.

It is worth mentioning that the appearance
in (\ref{d4})--(\ref{d10})
of extra  bosonic gauge transformations which
reduce the number of independent components of
$\lambda^{\underline\mu}$ is
related to the well known  fact that in $D=3,4,6$ and 10
the  commuting spinors ("twistors") with $n=2(D-2)=2,4,8$ and 16
components
parametrize, respectively, $S^1$, $S^3$, $S^7$ and $S^{15}$ spheres.
(In each case one spinor component can be fixed to be a constant by
scale transformations, which
corresponds to worldline reparametrizations when these spinors
appear in particle models).
These spheres are Hopf fibrations (fiber bundles) which are associated
with the division algebras {\bf R,~C,~H} and {\bf O} of the real,
complex, quaternionic and octonionic numbers.
The bases of these fiber bundles are, respectively,
the spheres $S^1$, $S^2$, $S^4$ and $S^8$, and the fibers are the group
manifolds $Z^2$, $S^1=U(1)$ and $S^3 =SU(2)$,
and a sphere $S^7$ (which is not a group manifold).
The base spheres correspond to and are
parametrized (up to a scaling factor) by
light--like vectors (massless particle momenta) in $D=3,4,6$ and $10$,
respectively,
and the fibers correspond to pure gauge degrees of freedom associated
with additional gauge symmetries of the superparticle models discussed
above.

We have thus shown that the
$n=1$ superfield action (\ref{2.5.2}) is
nonmanifestly $\kappa$--symmetric in $D=3,4,6$ and 10 space--time
dimensions and, hence, describes $N=1$ superparticles. The
generalization
of the action (\ref{2.5.2}) to a $d=2$ worldsheet will describe
$N=1$ $D=3,4,6,10$ tensionless superstrings.

As we have already mentioned there
are different ways of constructing appropriate (classically equivalent)
manifestly $n=D-2$ worldline supersymmetric
actions for $N=1$ superparticles \cite{stv,ds} and $N=1$ superstrings
\cite{hsstr,to,ik,dis}
in $D=4$ and 6 target superspaces which make use
of complex and quaternionic (or harmonic) analyticity structure inherent
to these superspaces.
However, the simplest way (and the
only known one for $N=1$, $D=10$ superparticles \cite{gs} and
superstrings \cite{to,dghs92})
is to write down a straightforward generalization of the action
(\ref{2.5.2}). For this we should define an appropriate  geometry of
an $n=D-2$ worldline superspace ${\cal M}_{1,n}$.

As in the case of $n=1$ worldline superspace (\ref{2.2.4}),
(\ref{2.2.07})
the supergeometry of ${\cal M}_{1,n}$ can always be chosen to be
superconformally flat.
I.e. the worldline supervielbeins have the form
$$
e^\tau=E(\tau,\eta)e^\tau_0=E(\tau,\eta)(d\tau+i\eta^q d\eta^q),
$$
\begin{equation}\label{2.2.027}
e^q=E^{1\over 2}(\tau,\eta)d\eta^q
-ie^\tau_0D^qE^{1\over 2},
\quad q=1,\cdots, n
\end{equation}
(indices $p,q,r,s...$ from the end of the Latin alphabet will be always
reserved for a representation of an internal group transformations
of $\eta^q$).
And one can again work with flat supervielbeins $e^\tau_0$ and
$e^q_0=d\eta^q$ and
flat supercovariant derivatives
\begin{equation}\label{2.5.28}
D_q={\partial\over{\partial\eta^q}}+i\eta_q\partial_\tau, \qquad
\{D_q,D_r\}=2i\delta_{qr}\partial_\tau .
\end{equation}
For the worldline superdiffeomorphisms $z^{\prime M}=z^{\prime M}(z^N)$
of the ${\cal M}_{1,n}$ coordinates $z^M=(\tau,\eta^q)$ to preserve
the conformal structure on ${\cal M}_{1,n}$ they must be restricted
to satisfy the constraint
\begin{equation}\label{sc}
D_q\tau'-i\eta^{\prime r} D_q\eta^{\prime r}=0,
\end{equation}
which implies that the odd supercovariant derivatives transform
homogeneously
under the restricted superdiffeomorphisms
\begin{equation}\label{sc1}
D_q=D_q\eta^{\prime r}D^\prime_r.
\end{equation}

The infinitesimal form of the
superreparametrizations (restricted superdiffeomorphisms) (\ref{sc})
is determined (as in eq. (\ref{2.2.4}))
by a single unconstrained superparameter $\Lambda(\tau,\eta^q)$
\begin{equation}\label{2.5.27}
\begin{array}{rl}
\delta\tau&=\Lambda(\tau,\eta)-{1\over 2}\eta^q D_q\Lambda,
\\
\delta \eta^q &= -{i\over 2}D^q\Lambda,\\
\delta D_q&=-{{1}\over 2}\dot\Lambda D_q+{i\over 4}[D_q,D_r]\Lambda D^r.
\end{array}
\end{equation}

In addition to the bosonic reparametrization parameter
$a({\tau})=\Lambda|_{\eta=0}$ and the worldline supersymmetry parameter
$\alpha_q(\tau)=-iD_q\Lambda|_{\eta=0}$ the superfunction
$\Lambda(\tau,\eta)$
contains parameters of local $SO(n)$ rotations
$b_{qr}(\tau)=[D_q,D_r]\Lambda|_{\eta=0}$. The fermionic coordinates
$\eta^q$
transform under an $n$--dimensional representation of $SO(n)$.
The $SO(n)$ indices are
raised and lowered by the unit matrix and hence there is no distinction
between them.

Let us embed the superworldline ${\cal M}_{1,n}$ into $N=1$, $D=4,6,10$
target
superspace. The image of ${\cal M}_{1,n}$ is described by the worldline
superfields $Z^{\underline M}(z^M)=(X^{\underline m}(\tau,\eta^q),
\Theta^{\underline\mu}(\tau,\eta^q))$ which now have many more
components
than in the $n=1$ case
\begin{equation}\label{2.5.29}
X^{\underline m}(\tau,\eta^q)=x^{\underline m}(\tau)
+i\eta^q\chi^{\underline m}_q(\tau) + ...,
\end{equation}
\begin{equation}\label{2.5.30}
\Theta^{\underline\mu}(\tau,\eta^q)=\theta^{\underline\mu}(\tau)
+\eta^q\lambda^{\underline\mu}_q(\tau) + ...,
\end{equation}
where dots denote terms of higher orders in $\eta^q$.

The pullback onto ${\cal M}_{1,n}$ of the target--space supervielbeins
(\ref{2.3.2}) is
\begin{eqnarray}\label{2.5.31}
{\cal E}^{\underline\alpha}(Z(z^M))&=&
dz^M{\cal E}_{M}^{\underline\alpha}(Z(z))=
dz^M\partial_M\Theta^{\underline\alpha}(z)=
e^\tau_0\partial_\tau\Theta^{\underline\alpha}+
e^q_0D_q\Theta^{\underline\alpha}, \nonumber \\
{\cal E}^{\underline a}(Z(z))&=&
dz^M {\cal E}_{M}^{\underline a}(Z(z)) \nonumber \\
&=&  e^\tau_0\left(\partial_\tau X^{\underline a}-
i\partial_\tau\bar\Theta\Gamma^{\underline a}\Theta \right)+
e^q_0\left(D_q X^{\underline a}-
iD_q\bar\Theta\Gamma^{\underline a}\Theta \right)
\end{eqnarray}

We postulate that the dynamics of massless superparticles in dimensions
$D=4,6,10$ is governed by the superembedding condition similar to
eq. (\ref{2.5.23})
\begin{equation}\label{2.5.32}
{\cal E}_{q}^{\underline a}(Z(z))=D_q X^{\underline a}-
iD_q\bar\Theta\Gamma^{\underline a}\Theta =0.
\end{equation}

Let us analyze the consequences of (\ref{2.5.32}). We shall see that
(\ref{2.5.32}) contains the mass--shell condition of massless
superparticle
dynamics, and that only the leading components
of the superfields
(\ref{2.5.29}) and (\ref{2.5.30}) are independent, while all other
components
are auxiliary and expressed in terms of the leading components and their
derivatives \cite{to,gs2}.

Taking the covariant derivative $D_r$ of (\ref{2.5.32}), symmetrizing
the result with respect to the indices $r$ and $q$ and taking into
account
eq. (\ref{2.5.28}) we get
\begin{equation}\label{2.5.33}
\delta_{qr}(\partial_\tau X^{\underline a}-
i\partial_\tau\bar\Theta\Gamma^{\underline a}\Theta)=
D_q\bar\Theta\Gamma^{\underline a}D_r\Theta,
\end{equation}
where the r.h.s of (\ref{2.5.33}) is automatically symmetric in $q$ and
$r$
because of symmetry properties of the $\Gamma$--matrices in $D=3,4,6$
and 10.

From eq. (\ref{2.5.33}) it follows that
\begin{equation}\label{2.5.34}
{\cal E}^{\underline a}_{\tau}(z^M)=\partial_\tau X^{\underline a}-
i\partial_\tau\bar\Theta\Gamma^{\underline a}\Theta=
{1\over{D-2}}D_q\bar\Theta\Gamma^{\underline a}D_q\Theta,
\end{equation}
and
\begin{equation}\label{2.5.35}
D_q\bar\Theta\Gamma^{\underline a}D_r\Theta=
{1\over{D-2}}\delta_{qr}D_s\bar\Theta\Gamma^{\underline a}D_s\Theta.
\end{equation}
The condition (\ref{2.5.34}) is analogous to (\ref{2.5.25}), and
(\ref{2.5.35}) is an additional one. It, in particular, restricts
the number of independent components of the $n=D-2$ commuting spinors
$\lambda_q^{\underline\mu}=D_q\Theta^{\underline\mu}|_{\eta=0}$ in the
superfield (\ref{2.5.30}) in such a way that ${\cal E}^{\underline
a}_{\tau}$
determined in (\ref{2.5.34}) is lightlike. To check this one should
simply convince oneself that the square of the r.h.s. of (\ref{2.5.34})
is
identically zero due to the properties (\ref{2.5.8}) of the
$\Gamma$--matrices
and by virtue of the relation (\ref{2.5.35}).

Hence, the superparticle
with the superworldline embedding specified by the condition
(\ref{2.5.32}) is massless.

We shall now show that the superembedding condition (\ref{2.5.32})
completely determines the higher components of the superfields
(\ref{2.5.29}) and (\ref{2.5.30}) in terms of their leading components
$x^{\underline m}$ and $\theta^{\underline\mu}$ up to bosonic local
worldline reparametrizations and local $SO(D-2)$ rotations
(\ref{2.5.27}).

To this end one should prove that because of eq. (\ref{2.5.35})
the number of independent components of the bosonic matrix
$D_q\Theta^{\underline\mu}$ is equal (up to the $SO(D-2)$ rotations)
to the number of the independent components of the
light--like vector ${\cal E}^{\underline a}_{\tau}$ (\ref{2.5.34}).

The vector ${\cal E}^{\underline a}_{\tau}$ has
(up to $\tau$--reparametrizations)
$n=D-2$ independent components
and parametrizes an $S^n$--sphere. Indeed, the condition
${\cal E}^{\underline a}_{\tau}
{\cal E}^{\underline b}_{\tau}\eta_{\underline{ab}}=0$
implies that
\begin{equation}\label{2.5.36}
({\cal E}^{\underline {D-1}}_{\tau})^2+
\sum_{i=1}^{i=D-2}({\cal E}^{i}_{\tau})^2=({\cal E}^{\underline
0}_{\tau})^2
\equiv (\partial_\tau X^{\underline 0}-
i\partial_\tau\bar\Theta\Gamma^{\underline 0}\Theta)^2,
\end{equation}
where $i$ stands for $D-2$ transverse spatial directions
and $0$ denotes the time component of the vector.
The r.h.s. of (\ref{2.5.36}) can be put to 1 by gauge fixing,
in an appropriate way, the worldline
bosonic $a(\tau)$--reparametrizations contained in (\ref{2.5.27})
of $X^{\underline 0}$ (\ref{2.2.5}). Thus the remaining
$D-1$ spatial components of the light--like velocity
${\cal E}^{\underline {a}}_{\tau}$, among which
$D-2$ are independent, parametrize an $S^n$--sphere.

The matrix $D_q\Theta^{\underline\mu}$ has $2(D-2)^2$ components, since
the spinor representations which we choose in $D=3,4,6$ and 10 have
dimension $2(D-2)$. ${1\over 2}(D-2)(D-3)+1$ components of this matrix
can be eliminated by the use of $SO(D-2)$ rotations and
$a(\tau)$--reparametrizations (\ref{2.5.27}),
since $D_q$ carries the $SO(D-2)$ index and $D_q\Theta^{\underline\mu}$
transforms homogeneously under finite worldline
reparametrizations and $SO(D-2)$ rotations
\begin{equation}\label{2.5.37}
D_q\Theta^{\underline\mu}(z)=
D_q\eta^{\prime r}D^\prime_r\Theta^{\prime\underline\mu}(z').
\end{equation}
 One can derive eq. (\ref{2.5.37}) from the finite extension (\ref{sc})
of the worldline superreparametrizations (\ref{2.5.27}).

Note that $D_q\eta^{\prime r}$ satisfies the same differential
conditions
as $D_q\Theta^{\underline\mu}$, i.e.
\begin{equation}\label{2.5.38}
\{D_s,D_q\}\eta^{\prime r}-{2i\over{D-2}}\delta_{sq}
{\partial\over{\partial{\tau}}}\eta^{\prime r}=0,
\end{equation}
and algebraic constraints
\begin{equation}\label{2.5.038}
D_q\eta^{\prime r}D_s\eta^{\prime r}+D_s\eta^{\prime r}D_q\eta^{\prime
r}
={1\over{D-2}}\delta_{qs}D_t\eta^{\prime r}D_t\eta^{\prime r},
\end{equation}
which resemble the integrability condition (\ref{2.5.35}). The
conditions
(\ref{2.5.038}) are, in fact, the integrability conditions of the
constraint
(\ref{sc}) for the worldline superreparametrizations to preserve
the conformally flat geometrical structure of the superworldline.

This guarantees that ${1\over 2}(D-2)(D-3)+1$ components of
$D_q\Theta^{\underline\mu}$ can be gauged away by appropriate
$SO(D-2)$--rotations and $a(\tau)$--reparametrizations.

We thus remain with
$2(D-2)^2-{1\over 2}(D-2)(D-3)-1={1\over 2}(D-2)(3D-5)-1$ components.

From (\ref{2.5.35}) it follows that eq. (\ref{2.5.34})  holds for each
value of the index $q$ separately (without the factor $n=D-2$ on the
r.h.s.).
This means that $(D-1)(D-2)$ components of $D_q\Theta^{\underline\mu}$
are expressed in terms of $D-1$ components of the lightlike vector.
Thus the number of components of the matrix
$D_q\Theta^{\underline\mu}$ still remaining undetermined is
${1\over 2}(D-2)(3D-5)-1-(D-1)(D-2)={1\over 2}(D-1)(D-2)-1$.

We now show, following the nice reasoning of refs. \cite{gs92},
that these remaining components are restricted to be zero
by the algebraic equations (\ref{2.5.35}).

Let us consider a particular form of the $\Gamma$--matrices, which
corresponds to a choice of light--cone coordinates in
$D$--dimensional target space. This breaks manifest Lorentz
symmetry $SO(1,D-1)$ of the target space
down to its subgroup $SO(1,1)\times SO(D-2)$. (We have already made
a light--cone splitting of the components of a lightlike vector
when we analyzed the geometrical meaning of the masslessness condition
(\ref{2.5.36})).

Further on,
the case of space--time dimension $D=6$ should be handled separately,
since
there are no Majorana spinors in this dimension and we deal instead with
$SU(2)$ simplectic Majorana--Weyl spinors (\ref{d6mw}).
The choice of an appropriate
realization of the $D=6$ $\Gamma$--matrices differs from that in $D=3,4$
and
10 space--time dimensions. Anyway, the $D=6$ analysis can be performed
along exactly the same lines and gives the same results
as the analysis of the $D=4$ and $D=10$ case
below, and we leave it for the reader as an exercise.

In the dimensions $D=3,4,$ and 10 in the
Majorana representation the
matrices $(C\Gamma^{\underline
a})_{\underline{\alpha}\underline{\beta}}$
are
symmetric and can be chosen as follows
\begin{equation}\label{2.5.39}
C\Gamma^0=\left(
\begin{array}{cc}
1  &  0\\
0  &  1
\end{array}
\right)\,,
\qquad
C\Gamma^i=\left(
\begin{array}{cc}
0  &  \gamma^i\\
(\gamma^i)^T   &  0
\end{array}
\right)\,,
\qquad
C\Gamma^{D-1}=\left(
\begin{array}{cc}
1   &  0\\
0   &   -1
\end{array}
\right)
\, ,
\end{equation}
where $\gamma^i_{qq'}$ are spinor matrices
in a $(D-2)$--dimensional space $(D=3,4,10)$
associated with the space transverse to the particle trajectory.
The indices $i=1,...,D-2$, $q=1,...,D-2$ and $q'=1,...,D-2$
are indices, respectively, of the vector and two (generally
nonequivalent)
spinor representations of
the group $SO(D-2)$ of transformations acting in this space. For
$SO(D-2)$ we use the same indices $q,r,...$ as for the worldline
internal
group $SO(n)$ $(n=D-2)$, since in an appropriate gauge both groups are
identified, as we shall see in a minute.

When $D=4$ the transverse space is two--dimensional.
The group of transverse rotations is $SO(2)$, and it has one
2--dimensional
spinor representation.

In the case of $D=10$ the $16\times 16$ matrices (\ref{2.5.39}) which
act on the 16--component real Majorana--Weyl spinors are analogs
of $D=4$ $\sigma$--matrices rather than the Dirac matrices which
are $32\times 32$ matrices in $D=10$ (see, for example, \cite{bpstv}
for details on $D=10$ and $D=11$ $\Gamma$--matrices).
The transverse space is now 8--dimensional, and
$SO(8)$ has one vector and two different spinor representations, all
three
being related by a famous triality property, which shows up in
properties
of octonions \cite{gg,gt}.

In the basis (\ref{2.5.39}) the matrix ${\cal E}^{\underline\mu}_{q}=
D_q\Theta^{\underline\mu}$
splits into two $(D-2)\times(D-2)$ matrices
\begin{equation}\label{2.5.40}
{\cal E}^{\underline\mu}_{q}=D_q\Theta^{\underline\mu}=
({\cal E}_{q}^{r}, {\cal E}_{q}^{r'}),
\end{equation}
and the eqs. (\ref{2.5.35}) take the form
\begin{equation}\label{2.5.41}
{\cal E}_{q}^{t}{\cal E}_{r}^{t}={1\over{D-2}}
\delta_{qr}{\cal E}_{s}^{t}{\cal E}_{s}^{t}
\end{equation}
\begin{equation}\label{2.5.42}
{\cal E}_{q}^{s}\gamma^{i}_{ss'}{\cal E}_{r}^{s'}+
{\cal E}_{r}^{s}\gamma^{i}_{ss'}{\cal E}_{q}^{s'}
={2\over{D-2}}\delta_{qr}{\cal E}_{s}^{t}\gamma^{i}_{tt'}
{\cal E}_{s}^{t'}
\end{equation}
\begin{equation}\label{2.5.43}
{\cal E}_{q}^{t'}{\cal E}_{r}^{t'}={1\over{D-2}}
\delta_{qr}{\cal E}_{s}^{t'}{\cal E}_{s}^{t'}.
\end{equation}

To proceed with the analysis of eqs. (\ref{2.5.41})--(\ref{2.5.43})
we should first require that the matrix ${\cal E}^{\underline\mu}_{q}$
has the maximum rank $D-2$.
This requirement is of the same nature as one which
we introduced in the case of the $N=1$, $D=3$ superparticle
(subsection 2.5.1) when, to
discard an unphysical ``frozen'' particle solution,
we assumed that all components of the commuting
spinor $\lambda^{\underline\mu}$ cannot be equal to zero simultaneously.

When the rectangular matrix ${\cal E}^{\underline\mu}_{q}$
is split as in (\ref{2.5.40}) the
requirement for it to have the maximum rank is equivalent to the
requirement that the determinant of either ${\cal E}_{q}^{r}$ or
${\cal E}_{q}^{r'}$ is nonzero.

 From (\ref{2.5.41}) and (\ref{2.5.43}) we see that ${\cal E}_{q}^{r}$
and
${\cal E}_{q}^{r'}$ satisfy (up to a normalization) the properties
of the orthogonal matrices $SO(D-2)$. Hence, if, for example,
$\det{{\cal E}_{q}^{r}}\not = 0$, we can use the worldline
transformations
(\ref{2.5.37}) to gauge fix ${\cal E}_{q}^{\alpha}$ to be the unit
matrix
\begin{equation}\label{2.5.44}
{\cal E}_{q}^{r}=D_q\Theta^{r}=\delta^r_q.
\end{equation}
The meaning of this gauge condition is twofold. It identifies the
$SO(D-2)$
group of superworldline transformations with the subgroup $SO(D-2)$ of
the target space Lorentz group $SO(1,D-1)$, and it identifies half of
the target space Grassmann coordinates
$\Theta^{\underline\mu}=(\Theta^{r},\Theta^{r'})$ with the
Grassmann coordinates of the superworldline
\begin{equation}\label{2.5.45}
\Theta^{r}=\eta^q\delta^r_q \quad \Rightarrow
\quad \Theta^r|_{\eta=0}=\theta^{r}(\tau)=0.
\end{equation}
Thus, half of the target space supersymmetries are identified with
worldsurface supersymmetry. It is this half of the supersymmetries of
the target space vacuum which remain unbroken in the presence of
superbranes.

The condition (\ref{2.5.44}), (\ref{2.5.45}) is the superworldsheet
counterpart
of the light--cone condition $(\Gamma^0+\Gamma^{D-1})\theta=0$ often
used
in the standard formulation of superparticles and superstrings to
gauge fix the $\kappa$--symmetry \cite{gsw}.

In the gauge (\ref{2.5.44}) the condition (\ref{2.5.41}) is identically
satisfied and eq. (\ref{2.5.42}) takes the form
\begin{equation}\label{2.5.46}
\gamma^{i}_{qq'}{\cal E}_{r}^{q'}+
\gamma^{i}_{rq'}{\cal E}_{q}^{q'}
={2\over{D-2}}\delta_{q r}\gamma^{i}_{sq'}
{\cal E}_{s}^{q'}.
\end{equation}

It can be easily checked that the general solution of (\ref{2.5.46})
in the dimensions $D-2=2$ and 8 is
\begin{equation}\label{2.5.47}
{\cal E}_{qq'}=\gamma^i_{qq'}V_i(z),
\end{equation}
where $V_i(z)$ is an $SO(D-2)$--vector superfield. Then eq.
(\ref{2.5.43})
is identically satisfied.

Substituting the expressions (\ref{2.5.44}) and (\ref{2.5.47}) for
${\cal E}_{q}^{\underline\alpha}$ into (\ref{2.5.34}) we find that
the vector $V^i$ coincides with the (independent) transverse components
of the light--like particle velocity ${\cal E}^{\underline a}_{\tau}$
(\ref{2.5.36})
\begin{equation}\label{2.5.48}
{\cal E}^{\underline 0}_{\tau}+{\cal E}^{\underline{D-1}}_{\tau}=1,
\quad {\cal E}^{\underline 0}_{\tau}-{\cal
E}^{\underline{D-1}}_{\tau}=V^{i}
V^{i},
\quad {\cal E}^{i}_{\tau}=V^{i}.
\end{equation}

We have thus proved that, up to the worldline superreparametrizations,
the integrability conditions (\ref{2.5.34}) and (\ref{2.5.35})
of the superembedding condition (\ref{2.5.32}) completely express
the components of the spinor matrix ${\cal E}_{q}^{\underline\alpha}
=D_q\Theta^{\underline\alpha}$ in terms of the components of
${\cal E}^{\underline a}_{\tau}=\partial_\tau X^{\underline a}-
i\partial_\tau\bar\Theta\Gamma^{\underline a}\Theta$.
\footnote{We have seen that the independent components of the matrix
${\cal E}_{q}^{\underline\alpha}$ parametrize
the sphere $S^{D-2}$. In Subsection 4.4  we shall demonstrate that this
sphere
can be realized as a compact subspace of a $2(D-2)$--dimensional
coset space ${{SO(1,D-1)}\over{SO(1,1)\times SO(D-2)}}$, and that the
components of ${\cal E}_{q}^{\underline\alpha}$ are associated with
Lorentz--harmonic variables in a spinor representation of $SO(1,D-1)$.}
Then from the form of (\ref{2.5.32}), (\ref{2.5.47}) and (\ref{2.5.48})
it follows that (up to the worldline superreparametrizations)
all higher components of the superfields $X^{\underline a}(z)$ and
$\Theta^{\underline\alpha}(z)$ are expressed through their leading
components
which are dynamical variables in the superparticle model.
This is a remarkable property of the
superembedding condition given the large number of the components which
constitute the worldline superfields when, for instance, $D=10$ and the
number of superworldline Grassmann coordinates is $n=8$.

It implicitly follows from the solution (\ref{2.5.47}) and
(\ref{2.5.48}) of
the integrability conditions (\ref{2.5.34}) and (\ref{2.5.35}) that
the superembedding condition does not contain the
dynamical equations of motion
(\ref{2.5.5}) of the physical particle variables $x^{\underline
m}(\tau)$
and $\theta^{\underline\mu}(\tau)$, i.e.
$\partial_\tau x^{\underline m}$ and
$\partial_\tau\theta^{\underline\mu}$ remain unrestricted.
Hence again, as in the case of the
$N=1$, $D=3$ superparticle, to completely determine the classical
dynamics
of the superparticle one should construct an action from which the
dynamical equations of motion are derived.

The straightforward generalization of the action
(\ref{2.5.2}) which produces the superembedding condition is
\begin{equation}\label{2.5.49}
S=-i\int d\tau d^{D-2}\eta P^q_{\underline a}{\cal E}_{q}^{\underline
a}=
-i\int d\tau d^{D-2}\eta P^q_{\underline a}\left[D_q X^{\underline a}-
iD_q\bar\Theta\Gamma^{\underline a}\Theta\right].
\end{equation}
A difference between eqs. (\ref{2.5.49}) and (\ref{2.5.2}) is
that now $P^q_{\underline a}(\tau,\eta)$ is a Grassmann--odd superfield
for the action (\ref{2.5.49}) to be Grassmann--even.
An appropriate variation of $P^q_{\underline a}$ ensures
the invariance of eq. (\ref{2.5.49}) under the
worldline superreparametrizations.

The  component of $P^q_{\underline a}$
associated with the particle momentum is
\begin{equation}\label{2.5.50}
p^{\underline a}=\epsilon^{q_1...q_{D-2}}D_{q_1}...D_{q_{D-3}}
P^{\underline a}_{q_{D-2}}|_{\eta=0},
\end{equation}
which can be seen from a corresponding term in the component action
obtained
by integrating (\ref{2.5.49}) over $\eta$. In $(\ref{2.5.50})$
$\epsilon^{q_1...q_{D-2}}$ is the totally antisymmetric unit tensor.

The superembedding condition is obtained by varying the action
(\ref{2.5.49}) with respect to $P^q_{\underline a}$. And the variations
of (\ref{2.5.49}) with respect to $X^{\underline a}(z)$ and
$\Theta^{\underline\alpha}(z)$ result in the
superfield equations
\begin{equation}\label{2.5.51}
D_qP^q_{\underline a}=0, \qquad
P^q_{\underline a}\Gamma^{\underline a}D_q\Theta=0,
\end{equation}
which (as one can check) contain
the dynamical equations (\ref{2.5.5}) for the superparticle variables
$x^{\underline m}(\tau)$ and $\theta^{\underline\mu}(\tau)$.

To complete the proof that the action (\ref{2.5.49}) describes standard
massless superparticles
we should convince ourselves that the superfield $P^q_{\underline a}(z)$
does not contain
any extra independent dynamical variables (which would otherwise mean
that
the model has redundant degrees of freedom).

To show this one should notice that
the number of independent components of $P^q_{\underline a}(z)$
substantially
reduces because of an additional local invariance of the action
(\ref{2.5.49})
with respect to the following transformations of $P^q_{\underline a}$
\cite{gs92}
\begin{equation}\label{2.5.52}
\delta P^q_{\underline a}=
D_r(\Lambda^{qrs}\Gamma_{\underline a}D_s\Theta),
\end{equation}
where the tensorial parameter $\Lambda^{qrs}_{\underline\alpha}(z)$ is
totally symmetric and traceless with respect to each pair of
the indices $q,r,s$.

Then, by analyzing the
equations (\ref{2.5.51}),
it can be shown that the components of the superfield
$P^q_{\underline a}(z)$ can either be gauged away by independent local
transformations contained in (\ref{2.5.52}) or are expressed in terms of
components of
$
D_q\Theta^{\underline\alpha}.
$
The reader may find the details of this analysis in the references
\cite{gs92,dghs92}.

In particular, we again get the superparticle equations
of motion (\ref{2.5.5}) and the Cartan--Penrose representation of  the
superparticle momentum (\ref{2.5.50})
\begin{equation}\label{2.5.53}
\delta_{qr}p^{\underline a} =
{1\over{e(\tau)}}D_q\bar\Theta\Gamma^{\underline a}D_r\Theta|_{\eta=0}=
{1\over{e(\tau)}}\bar\lambda_q\Gamma^{\underline a}\lambda_r \quad
\Rightarrow
\end{equation}
$$
\quad \Rightarrow \quad
\bar\lambda_q\Gamma^{\underline a}\lambda_r={1\over{D-2}}\delta_{qr}
\bar\lambda_s\Gamma^{\underline a}\lambda_s.
$$

coordinates

When the auxiliary fields are eliminated the
worldline supersymmetry transformations of the remaining fields take the
form
\begin{equation}\label{2.5.55}
\delta\theta=-\alpha^q(\tau)\lambda_q, \quad
\delta\lambda_q=-i\alpha_q\dot\theta,
\end{equation}
\begin{equation}\label{2.5.56}
\delta x^{\underline a}= -i\bar\theta\Gamma^{\underline a}\delta\theta,
\quad \delta p_{\underline a}=0.
\end{equation}
Taking $\alpha_q(\tau)=-{{2(D-2)i}\over e}\bar\lambda_q\kappa(\tau)$
and using the relation (\ref{2.5.53}) which implies
\begin{equation}\label{2.5.056}
(p^{\underline a}\Gamma_{\underline
a})^{\underline\alpha}_{~~\underline\beta}
={1\over{e(\tau)(D-2)}}\bar\lambda_q\Gamma^{\underline a}\lambda_q
(\Gamma_{\underline a})^{\underline\alpha}_{~~\underline\beta}=
{1\over{e(\tau)}}\lambda_q^{\underline\alpha}\bar\lambda_{q\underline\beta}
\end{equation}
we again recover the
$\kappa$--symmetry transformations (\ref{2.3.7}).

Note that in general the expansion of the symmetric matrix
$\lambda_q^{\underline\alpha}\lambda_q^{\underline\beta}$ in the
basis of $\Gamma$--matrices has additional terms, for instance, in
$D=10$
\begin{equation}\label{2.5.57}
\lambda_q^{\underline\alpha}\lambda_q^{\underline\beta}=
{1\over 16}(\bar\lambda_q\Gamma^{\underline a}\lambda_q)
\tilde\Gamma_{\underline a}^{\underline{\alpha\beta}}
+{1\over {16\cdot 5!}}(\bar\lambda_q\Gamma^{\underline a_1...\underline
a_5}
\lambda_q)
(\tilde\Gamma_{\underline a_1...\underline
a_5})^{\underline{\alpha\beta}},
\end{equation}
where $\tilde\Gamma_{\underline a}^{\underline{\alpha\beta}}$ are the
same
as in eq. (\ref{2.5.39}) but have the vector index down and the spinor
indices
up. (Actually, $\tilde\Gamma_{\underline a}^{\underline{\alpha\beta}}$
are
inverse of $(C\Gamma^{\underline a})_{\underline{\alpha\beta}}$).

In our case the terms with the antisymmetrized product of the
$\Gamma$--matrices vanish
\begin{equation}\label{2.5.58}
\bar\lambda_q\Gamma^{\underline a_1...\underline a_5}\lambda_q=0
\end{equation}
because of the condition (\ref{2.5.53}) on
the commuting spinors, which can be proved using the results of the
analysis of the equations (\ref{2.5.35}),
(\ref{2.5.40})--(\ref{2.5.47}).
The same happens in the dimensions $D=4$ and 6.

This completes the demonstration that the worldline superfield action
(\ref{2.5.49}) based on the superembedding condition (\ref{2.5.32})
describes massless superparticles propagating in flat
$N=1$, $D=3,4,6$ and 10 target superspaces.

\subsection{Coupling to supersymmetric Maxwell fields}
So far we have dealt with free superparticle models. We will now
consider how
the interaction of superparticles with Abelian gauge fields is described
in the superembedding approach \cite{gs92}.

When supersymmetric gauge theory is formulated in superspace the gauge
superpotential
$A_{\underline M}(x,\theta) = (A_{\underline m}, A_{\underline \mu})$
is subject to constraints (see, for instance, \cite{wb,1001,bk,10sym}
and references therein).
This is done to eliminate redundant fields
(especially those of higher spins)
in the component expansion of the superfield $A_{\underline
M}(x,\theta)$.
The constraints are imposed on the field strength of
$A_{\underline M}(x,\theta)$
which is a target space two--form
\begin{equation}\label{fs}
F^{(2)}={1\over 2}dZ^{\underline N}dZ^{\underline M}F_{\underline {MN}}
={1\over 2}dZ^{\underline N}dZ^{\underline M}
(\partial_{\underline M}A_{\underline N}-(-)^{\underline {MN}}
\partial_{\underline N}A_{\underline M}).
\end{equation}
$F_{\underline {MN}}$ is Grassmann--antisymmetric, i.e. it is
antisymmetric when one or both indices are bosonic (of parity 0) and
symmetric when both indices are fermionic (of parity 1).
(In eq. (\ref{fs}) and below the wedge product of external differentials
and forms is implied. The external differential is assumed to act on the
differential
forms from the right, i.e. $d(PQ)=PdQ+(-1)^qdPQ$ for a p--form $P$ and a
q--form $Q$).

We shall also use the expansion of the forms in the supervielbein
basis such as (\ref{2.3.2})
\begin{equation}\label{fs1}
F^{(2)}={1\over 2}{\cal E}^{\underline B}{\cal E}^{\underline A}
F_{\underline {AB}},
\end{equation}
where
\begin{equation}\label{fs2}
F_{\underline {AB}}=(-)^{{\underline M}(\underline B+\underline
N)}
{\cal E}_{\underline A}^{\underline M}
{\cal E}_{\underline B}^{\underline N}F_{\underline {MN}}
\end{equation}
and ${\cal E}_{\underline A}^{\underline M}$ is the supervielbein
matrix inverse to ${\cal E}^{\underline A}=dZ^{\underline M}
{\cal E}^{\underline A}_{\underline M}$.

The constraints are imposed on components of the field strength.
The essential constraint being generic to all super--Yang--Mills
theories
in any space--time dimension is
\begin{equation}\label{symc}
F_{\underline {\alpha\beta}}=0.
\end{equation}
Constraints on other components of $F_{\underline {AB}}$ are obtained
from the consistency of the Bianchi identities $dF^{(2)}=0$ with
the constraint (\ref{symc}). We thus get that
\begin{equation}\label{symc1}
F_{\underline {\alpha a}}
=W_{\underline\beta}
(\Gamma_{\underline {a}})^{\underline\beta}_{~~\underline \alpha}
,
\end{equation}
where $W_{\underline\beta}(Z)$ is a spinorial superfield.

In the case of $N=1$, $D=10$ super--Yang--Mills theory the consistency
of the Bianchi identities in the presence of the constraint (\ref{symc})
also implies the equations of motion for the fields of the SYM
supermultiplet, i.e. the constraint (\ref{symc}) puts $D=10$ SYM on the
mass shell \cite{10sym,twi}.

In \cite{twi} it was observed that $\kappa$--symmetry of the
action of the superparticle coupled to the super--Maxwell field
$A_{\underline M}$ requires that the Maxwell field is
integrable along the lightline trajectories of the superparticle, i.e.
that its field strength is zero along the superparticle trajectories.
This takes place if the gauge field obeys the constraint (\ref{symc}).
Thus $\kappa$--symmetry demands that the gauge field superbackground
is constrained.

In the superembedding formulation light--like integrability means that
the pullback of the field strength $F_{\underline {AB}}$ onto the
superworldline satisfying the superembedding condition (\ref{2.5.32})
vanishes \cite{gs92}.

The pullback is
\begin{equation}\label{pb}
F_{AB}={\cal E}_{A}^{\underline A}{\cal E}_{B}^{\underline B}
F_{\underline {AB}},
\end{equation}
where ${\cal E}_{A}^{\underline A}=({\cal E}_{\tau}^{\underline A},
{\cal E}_{q}^{\underline A})$ are defined in (\ref{2.5.31}).

Taking into account (\ref{2.5.32}) and  (\ref{2.5.33}) we
have
\begin{equation}\label{pb1}
F_{qr}={\cal E}_{q}^{\underline\alpha}{\cal E}_{r}^{\underline\beta}
F_{\underline {\alpha}\underline\beta},
\end{equation}
\begin{equation}\label{pb2}
F_{\tau q}={\cal E}_{\tau}^{\underline\alpha}
{\cal E}_{q}^{\underline\beta}F_{\underline {\alpha}\underline\beta}+
{1\over{D-2}}\bar{\cal E}_{r}\Gamma^{\underline a}{\cal E}_{r}
{\cal E}_{q}^{\underline\beta}F_{\underline a\underline {\beta}}.
\end{equation}
We observe that the components (\ref{pb1}) and (\ref{pb2})
vanish, and, hence,
\begin{equation}\label{pb3}
F_{AB}=0 \quad \Rightarrow \quad
F_{MN}=\partial_MZ^{\underline N}\partial_NZ^{\underline M}
(\partial_{\underline M}A_{\underline N}-(-)^{\underline {MN}}
\partial_{\underline N}A_{\underline M})=0
\end{equation}
if the super--Maxwell field satisfies the constraints (\ref{symc}) and
(\ref{symc1}).

On the other hand, using the properties
of ${\cal E}_{A}^{\underline B}$ discussed in the previous Subsection
it can be shown that $F_{AB}=0$ implies the super--Maxwell constraints.

The condition (\ref{pb3}) can also be regarded as an additional
superembedding condition when the gauge field superbackground
in target superspaces is nontrivial.

We would like to get this condition (\ref{pb3})
from a superparticle action. So we extend the free superparticle action
(\ref{2.5.49}) by an appropriate Maxwell coupling term. The extended
action is
\begin{equation}\label{mca}
S=-i\int d\tau d^n\eta P^q_{\underline a}{\cal E}_{q}^{\underline a}
+
\int d\tau d^n\eta P^M(\partial_MZ^{\underline M}A_{\underline M}
-\partial_M\Phi),
\end{equation}
where $n=D-2$,  $ P^M(Z)$ is a Lagrange multiplier and $\Phi(z)$ is an
auxiliary superworldline field.

The action (\ref{mca}) is invariant under
Abelian gauge transformations $\delta A=d\varphi(Z)$ provided
$\Phi(z)$ transforms under the pullback of $\varphi(Z)$ as a Goldstone
field $\delta\Phi(z)=\varphi(Z(z))$.

The condition (\ref{pb3}) arises as the integrability condition of the
equation of motion
\begin{equation}\label{spm}
{{\delta S}\over{\delta P^M}}=\partial_MZ^{\underline M}A_{\underline M}
-\partial_M\Phi=0.
\end{equation}

The variation of the action (\ref{mca}) with respect to $\Phi$ yields
the equation
\begin{equation}\label{dpm}
\partial_MP^M=0,
\end{equation}
whose general solution is
\begin{equation}\label{dpms}
P^M=\partial_L\tilde\Lambda^{LM}
+{1\over{n!}}\epsilon_{q_1...q_n}\eta^{q_1}....\eta^{q_n}
\delta^M_\tau e,
\end{equation}
where $e$ is
a constant and $\tilde\Lambda^{LM}(z)$ is an arbitrary
Grassmann antisymmetric superfield.

The field $\tilde\Lambda^{LM}(z)$ is a pure gauge if the action
(\ref{mca})
is invariant under the following transformations of $P^M$
\begin{equation}\label{locus}
\delta P^M=\partial_L\Lambda^{LM},
\end{equation}
where $\Lambda^{LM}(z)=-(-)^{LM}\Lambda^{ML}(z)$.

The variation of the action (\ref{mca}) under (\ref{locus}) is
\begin{equation}\label{var}
\delta S= -\int d\tau d^n\eta \Lambda^{AB}F_{AB}=
-\int d\tau d^n\eta \Lambda^{AB}({\cal E}_{A}^{\underline\alpha}
{\cal E}_{B}^{\underline\beta}F_{\underline {\alpha}\underline\beta}+
2{\cal E}_A^{\underline a}
{\cal E}_{B}^{\underline\beta}F_{\underline a\underline {\beta}}+
{\cal E}_A^{\underline a}
{\cal E}_{B}^{\underline b}F_{\underline a\underline b}).
\end{equation}
The last term in (\ref{var}) contains ${\cal E}_q^{\underline a}$ and
hence
can be compensated by an appropriate variation of the Lagrange
multiplier
$P^q_{\underline a}$ in (\ref{mca}), while the first two terms must
vanish
if (\ref{locus}) is a symmetry of the action. This implies the
super--Maxwell
constraints (\ref{symc}) and (\ref{symc1}).

Using the symmetry  (\ref{locus}) one reduces the solution
for $P^M$ to the last term in (\ref{dpms}).
Then substituting this term back into the action (\ref{mca}) we get the
standard minimal coupling Maxwell term
\begin{equation}\label{mini}
S_M=e\int d\tau \dot Z^{\underline M}A_{\underline M},
\end{equation}
where $e$ plays the role of the electric charge. Thus in the
superembedding
formulation the electric charge arises as an integration constant of an
auxiliary superworldline field\footnote{This is analogous to how
gauge coupling constants appear in Kaluza--Klein theories upon
integrating
out extra compact directions.}.
Below we shall see that the string tension appears in the superembedding
approach in a similar way.

\subsection{Superembedding into curved superspaces}
Consider now the propagation of superparticles in curved $N=1$
target superspaces of bosonic dimension $D=3,4,6$ and 10.
Then the super Poincare transformations of $Z^{\underline M}=
(X^{\underline m},\Theta^{\underline\mu})$ are extended to the
target--space
superdiffeomorphisms
\begin{equation}\label{superd}
Z^{\prime\underline M}=Z^{\prime\underline M}(Z).
\end{equation}
The flat supervielbeins ${\cal E}^{\underline A}$  (\ref{2.3.2})
are generalized
to curved supervielbeins
\begin{equation}\label{curve}
E^{\underline A}=dZ^{\underline M}E^{\underline A}_{\underline M}(Z),
\end{equation}
whose leading components correspond to the target space graviton
$e^{\underline a}_{\underline m}(x)=
E^{\underline a}_{\underline m}|_{\theta=0}$ and the gravitino
$\psi^{\underline\alpha}_{\underline m}(x)=
E^{\underline\alpha}_{\underline m}|_{\theta=0}$.

Parallel transport of (spin)--tensors in curved target superspace is
determined by a connection one--form $\Omega^{~~\underline
A}_{\underline B}=
dZ^{\underline M}\Omega^{~~~\underline A}_{\underline M\underline B}$
taking
values in the algebra of the tangent space group which is
the Lorentz group $SO(1,D-1)$.

The geometry of curved superspace is characterized by its torsion
\begin{equation}\label{torsion}
\nabla E^{\underline A}\equiv (dE^{\underline A}+
E^{\underline B}\Omega^{~~\underline A}_{\underline B}) \equiv
T^{\underline A}={1\over 2}E^{\underline C}E^{\underline B}
T^{\underline A}_{\underline {BC}},
\end{equation}
and curvature
\begin{equation}\label{curvature}
R^{~~\underline A}_{\underline B}=d\Omega^{~~\underline A}_{\underline
B}+
\Omega^{~~\underline C}_{\underline B}\Omega^{~~\underline
A}_{\underline C},
\end{equation}
where $\nabla=d+\Omega$ is the covariant external differential in the
curved
target superspace.

The superembedding condition (\ref{2.5.32}) is now imposed
on the superworldline pullback of $E^{\underline A}$
\begin{equation}\label{pull}
E^{\underline A}(Z(z))= e^\tau_0(z)\partial_\tau Z^{\underline
M}E^{\underline A}_{\underline M} +e_0^q(z)D_qZ^{\underline
M}E^{\underline A}_{\underline M}\equiv e_0^BE^{\underline A}_{B}
\end{equation}
and takes the form
\begin{equation}\label{securve}
{E}_{q}^{\underline a}(Z(z))
\equiv D_qZ^{\underline M}E^{\underline a}_{\underline M}=0.
\end{equation}

The worldline superfield action which produces (\ref{securve}) has the
same form as eq. (\ref{2.5.49}) but with ${\cal E}_{q}^{\underline a}$
replaced by ${E}_{q}^{\underline a}$
\begin{equation}\label{acurve}
S=-i\int d\tau d^{D-2}\eta P^q_{\underline a}{E}_{q}^{\underline a}.
\end{equation}
This action is manifestly invariant under the target--space
superdiffeomorphisms and superworldline restricted superdiffeomorphisms
(\ref{2.5.27}). For consistency it must also be invariant under local
variations of $P^q_{\underline a}$ which generalize the transformations
(\ref{2.5.52})
\begin{equation}\label{2.5.52g}
\delta P^q_{\underline b}=
(\delta^{\underline a}_{\underline b}D_r+
\Omega^{~~~\underline a}_{r\underline
b})(\Lambda^{qrs}\Gamma_{\underline a}
E_s),
\end{equation}
where $E_s$ stands for $E_s^{\underline\alpha}
=D_sZ^{\underline M}E_{\underline M}^{\underline\alpha}$, and remember
that
$\Lambda^{qrs}$ is totally symmetric and traceless with respect to the
each pair of indices $q,r,s$.

The variation of (\ref{acurve}) under (\ref{2.5.52g}) is
\begin{equation}\label{varia}
\delta S={i\over 2}\int d\tau d^{D-2}\eta
(\Lambda^{qrs}\Gamma_{\underline b}E_s)\left(E_q^{\underline\alpha}
E_r^{\underline\beta}T^{\underline b}_{\underline\alpha\underline\beta}+
2E_q^{\underline A}E_r^{\underline a}
T^{\underline b}_{\underline a\underline A}\right),
\end{equation}
where $T^{\underline a}_{\underline B\underline A}$ are components of
the
target superspace torsion (\ref{torsion}).

The last term of (\ref{varia}) can be canceled by the following
variation
of $P^q_{\underline a}$
$$
\delta P^q_{\underline a}=(-)^{\underline A}
(\Lambda^{qrs}\Gamma_{\underline b}E_s)
E_r^{\underline A}
T^{\underline b}_{\underline a\underline A},
$$
and the first term of (\ref{varia}) must vanish, which requires the
constraint
on the target--space torsion
\begin{equation}\label{torsionc}
T^{\underline a}_{\underline\alpha\underline\beta}=
-2i(C\Gamma^{\underline a})_{\underline\alpha\underline\beta}.
\end{equation}
This is the basic constraint of the superfield formulation of
supergravity
in any dimension \cite{wb,1001,11sg,10sg,cal}.
Note that the flat superspace torsion (\ref{2.3.3}) is
compatible with this constraint.

To conclude Subsections 2.6 and 2.7 we remark that
in the superembedding formulation the constraints on gauge field and
target space superbackgrounds are not a requirement of local worldsheet
fermionic symmetry (as in the standard approach) but of additional local
symmetries of the action under variations of the Lagrange multipliers
which
ensure the superembedding conditions.

In this regard it is worth mentioning that the case of the $N=1$, $D=3$
superparticle is a special one. The $n=1$ worldline supersymmetry action
in $D=3$ supergravity and super--Maxwell background can be constructed
in such a way that Maxwell coupling is described by the standard minimal
coupling term
\begin{equation}\label{n1d3}
S_{N=1,D=3}=
-i\int d\tau d\eta \left [P_{\underline a}DZ^{\underline M}
{E}_{\underline M}^{\underline a}+eDZ^{\underline M}A_{\underline
M}\right].
\end{equation}
In contrast to the standard $\kappa$--symmetry
formulation (\ref{2.3.5}),
in the form (\ref{n1d3}) the $N=1$, $D=3$ superparticle action does not
have any local symmetries
which would require the target--space and gauge field background
to satisfy the constraints (\ref{symc}) and (\ref{torsionc}).
In this case the constraints should be imposed `by hand' at the
level of equations of motion.

\section{Superstrings}
\setcounter{equation}0
We now turn to the consideration of superstrings in the superembedding
approach, but first let us recall standard forms of the superstring
action.

\subsection{Green--Schwarz formulation}
In this formulation (see \cite{gs,gsw} for details)
superstring dynamics is described by the embedding
into a target superspace
of a bosonic two--dimensional worldsheet parametrized by coordinates
$\xi^m=(\tau,\sigma)$. The requirement of target--space supersymmetry
of the superstring action restricts the possible dimensions of the
target
superspace in which classical superstrings can propagate to $D=2,3,4,6$
and 10
bosonic dimensions and $2N(D-2)$ real fermionic spinor dimensions (where
$N=1,2$ counts the number of spinors). To avoid dealing with boundary
terms, in what follows we restrict ourselves to the discussion of
closed superstrings.

The intrinsic geometry of the worldsheet is described by
the zweibein
\begin{equation}\label{3.1}
e^a(\xi)=d\xi^me^a_m(\xi)
\end{equation}
where $a=0,1$ are $SO(1,1)$ vector indices of the tangent space.

Below we shall also often use the light--cone basis
\begin{equation}\label{3.2}
e^{++}(\xi)=e^0+e^1, \qquad e^{--}(\xi)=e^0-e^1,
\end{equation}
where $(++,--)$ stand for $SO(1,1)$ light--cone vector components. We
reserve a single $(+,-)$ for light--cone components of worldsheet
spinors.

In the light--cone basis the $d=2$ Minkowski metric takes the form
\begin{equation}\label{3.002}
\eta_{++,--}=\eta_{--,++}=-{1\over 2}, \quad
\eta_{++,++}=\eta_{--,--}=0,
\end{equation}
so that
\begin{equation}\label{3.02}
e_{--}=-{1\over 2}e^{++}, \quad e_{++}=-{1\over 2}e^{--}.
\end{equation}

The worldsheet metric is
\begin{equation}\label{3.3}
g_{mn}(\xi)=e^a_m(\xi)e_{an}(\xi)=
-{1\over 2}\left(e^{--}_me^{++}_{n}
+e^{--}_ne^{++}_{m}\right),
\end{equation}
and
\begin{equation}\label{3.03}
\det{g_{mn}}\equiv g=-(\det{e^a_m})^2, \quad \det{e^a_m}=
{1\over 2}\epsilon^{mn}e^{--}_{m}e^{++}_n
\end{equation}
$(\epsilon^{01}=-\epsilon^{10}=1$, $\epsilon^{00}=\epsilon^{11}=0$).

The pullback of the target--space supercovariant forms (\ref{2.3.2})
onto
the worldsheet is{\footnote{For simplicity we again work
in flat target superspace.}
\begin{equation}\label{3.4}
{\cal E}^{\underline a}= d\xi^m{\cal E}_{m}^{\underline a}=
d\xi^m\left(\partial_mx^{\underline m}-
i\partial_m\bar\theta^I\Gamma^{\underline m}\theta^I\right)
\delta_{\underline m}^{\underline a}=
e^a\left(D_ax^{\underline m}-
iD_a\bar\theta^I\Gamma^{\underline m}\theta^I\right)
\delta_{\underline m}^{\underline a},
\end{equation}
\begin{eqnarray}\label{3.5}
{\cal E}^{\underline\alpha I}=d\xi^m{\cal
E}^{\underline{\alpha}I}_{m}&=&
d\xi^m\partial_m\theta^{\underline\mu I}\delta_{\underline\mu}^
{\underline\alpha}
=e^aD_a\theta^{\underline\mu I}
\delta_{\underline\mu}^{\underline\alpha},
\end{eqnarray}
where
\begin{equation}\label{3.6}
D_a\equiv e_a^m(\xi)\partial_m,
\end{equation}
and $e_a^m(\xi)$ is the inverse of $e^a_m(\xi)$
(i.e. $e_a^m(\xi)e^a_n(\xi)=\delta^n_m$). The index $I=1,2$ indicates
that
the number of $\theta$--coordinates can be 1 or 2 depending on whether
we deal with $N=1$ or $N=2$ target superspace.

The $N=2$ Green--Schwarz superstring action is
\begin{equation}\label{3.7}
S=-{T\over 2}\int d^2\xi\sqrt{-g}g^{mn}{\cal E}_{m}^{\underline a}
{\cal E}_{n}^{\underline b}\eta_{\underline{ab}}+ T\int B^{(2)}(\xi),
\end{equation}
Where $T$ is the string tension and $B^{(2)}(\xi)={1\over 2}
d\xi^md\xi^n\partial_mZ^{\underline N}
\partial_nZ^{\underline M}B^{(2)}_{\underline MN}(Z)$
is the worldsheet pullback of the target superspace two--form
\begin{equation}\label{3.8}
B^{(2)}=idx^{\underline a}(d\bar\theta^1\Gamma_{\underline a}\theta^1-
d\bar\theta^2\Gamma_{\underline a}\theta^2)+
d\bar\theta^1\Gamma_{\underline a}\theta^1
d\bar\theta^2\Gamma_{\underline a}\theta^2.
\end{equation}

The physical meaning of $B^{(2)}$ is that, for instance, in $D=10$
it describes a flat limit of the Neveu--Schwarz two--form gauge field
in the superfield formulation of $D=10$ supergravity.
As we see from the action (\ref{3.7}), the strings
couple minimally to $B^{(2)}$, and the string tension is associated with
the
$B^{(2)}$--charge.

The $B^{(2)}$-- term in (\ref{3.7}) is called the Wess--Zumino term
since
it is an integral of an external differential form which does not
contain
the worldsheet metric and is invariant under the target space
supersymmetry
transformations (\ref{2.3.1})
only up to a total derivative (and only in $D=3,4,6$ and $10$ where
the Dirac matrices satisfy eq. (\ref{2.5.8})).

The definite relative coefficient between the two terms of the action
(\ref{3.7}) is required by $\kappa$--symmetry.
The $\kappa$--symmetry transformations of the superstring dynamical
variables are
\begin{eqnarray}\label{3.9}
\delta\theta^I&=&2i\Gamma_m\kappa^{Im}(\xi),
\quad \Gamma_m\equiv {\cal E}_{m}^{\underline a}
\Gamma_{\underline a}\nonumber \\
\delta x^{\underline m}&=&-i\bar\theta^I\Gamma^{\underline
m}\delta\theta^I
\quad \Rightarrow \quad
\delta Z^{\underline M}{\cal E}^{\underline a}_{\underline M}=0,\\
\delta(\sqrt{-g}g^{mn})&
=&-16\sqrt{-g}\left(P^{ml}_{-}\bar\kappa^{1n}\partial_l\theta^1+
P^{ml}_{+}\bar\kappa^{2n}\partial_l\theta^1\right),\nonumber
\end{eqnarray}
where $P^{mn}_{\pm}$ are projectors
\begin{equation}\label{3.10}
P^{mn}_{\pm}={1\over 2}(g^{mn}\pm {1\over\sqrt{-g}}\epsilon^{mn}),
\end{equation}
$$
P^{mn}_{\pm}g_{nl}P^{lm'}_{\pm}=P^{mm'}_{\pm},
\quad P^{mn}_{\pm}g_{nl}P^{lm'}_{\mp}=0,
$$
and the parameters $\kappa^{I\underline\alpha}_m(\xi)$ ($I=1,2$)
are (anti)--self--dual worldsheet vectors
\begin{equation}\label{3.11}
\kappa^{1m}=P^{mn}_{-}\kappa^{1}_n, \qquad
\kappa^{2m}=P^{mn}_{+}\kappa^{2}_n.
\end{equation}
In the light--cone basis (\ref{3.2}) the
components of $\kappa^I$ are
\begin{equation}\label{3.12}
\kappa^1_{--}=e_{--}^m\kappa^1_m, \quad
\kappa^2_{++}=e_{++}^m\kappa^2_m;
\quad
\kappa^1_{++}=0,  \quad
\quad \kappa^2_{--}=0.
\end{equation}

The dynamical equations of motion of the superstring are obtained by
varying (\ref{3.7}) with respect to
$\theta^{\underline\mu}(\xi)$ and  $x^{\underline m}(\xi)$
\begin{eqnarray}\label{3.012}
\Gamma_{m}P^{mn}_-\partial_n\theta^1&=&0, \nonumber \\
\Gamma_{m}P^{mn}_+\partial_n\theta^2&=&0, \nonumber \\
\partial_m\left [\sqrt{-g}\left(g^{mn}\partial_nx^{\underline m}+
2iP^{mn}_-\partial_n\bar\theta^1\Gamma^{\underline m}\theta^1
+2iP^{mn}_+\partial_n\bar\theta^2\Gamma^{\underline
m}\theta^2\right)\right]
&=&0.
\end{eqnarray}

Varying the action (\ref{3.7}) with respect to the worldsheet
metric we get
\begin{equation}\label{3.13}
{1\over 2}
g_{mn}\left( g^{lp}{\cal E}_{l}^{\underline a}
{\cal E}_{p}^{\underline b}\eta_{\underline{ab}}\right)=
{\cal E}_{m}^{\underline a}{\cal E}_{n}^{\underline
b}\eta_{\underline{ab}}
\equiv G_{mn}.
\end{equation}
Rewritten in the light--cone basis eqs. (\ref{3.13}) take the form of
the
Virasoro constraints
\begin{equation}\label{3.14}
{\cal E}_{--}^{\underline a}{\cal E}_{--\underline a}=0,
\quad
{\cal E}_{++}^{\underline a}{\cal E}_{++\underline a}=0.
\end{equation}

From eqs. (\ref{3.13}) it also follows that the intrinsic worldsheet
metric
$g_{mn}$ can be identified with the worldsheet metric
$G_{mn}={\cal E}_{m}^{\underline a}{\cal E}_{n}^{\underline b}
\eta_{\underline{ab}}$
induced by embedding into the target superspace
\begin{equation}\label{3.15}
{1\over\sqrt{-g}}g_{mn}={1\over\sqrt{-G}}G_{mn}.
\end{equation}
The two metrics can be made equal by gauge fixing the
invariance of the action (\ref{3.7}) under the Weyl rescaling of
$g_{mn}$
($g_{mn} \rightarrow \phi(\xi)g_{mn}$).

Using (\ref{3.15}) we can eliminate the metric $g_{mn}$ from the
superstring
action and rewrite it in the Nambu--Goto form
\begin{equation}\label{3.16}
S=-T\int d^2\xi\sqrt{-det G_{mn}}+T\int B^{(2)}(\xi).
\end{equation}

It is instructive to note that in this form of the superstring action
it is more natural to replace the (anti)--self--dual worldsheet
vector parameters $\kappa^{I\underline\alpha}_m(\xi)$ of the
$\kappa$--symmetry transformations (\ref{3.9}) with
worldsheet scalar parameters
\begin{equation}\label{3.17}
\kappa^{I\underline\alpha}(\xi)\equiv
2i\left(\Gamma_{m}\kappa^{Im}\right)^{\underline\alpha},
\end{equation}
which satisfy the condition
\begin{equation}\label{3.18}
\kappa^{I\underline\alpha}=
{1\over 2}(1+\bar\Gamma)^{I\underline\alpha}_{~~J\underline\beta}
\kappa^{J\underline\beta},
\end{equation}
where ${1\over 2}(1+\bar\Gamma)$ is the spinor projection matrix
$$
{1\over 2}(1+\bar\Gamma){1\over 2}(1+\bar\Gamma)={1\over
2}(1+\bar\Gamma),
$$
\begin{equation}\label{3.19}
\bar\Gamma={1\over{2\sqrt{-\det{G_{mn}}}}}
\epsilon^{m_1m_2}\Gamma_{m_1}\Gamma_{m_2}{\cal K}, \quad
(\bar\Gamma)^2=1,
\quad Tr\bar\Gamma=0,
\end{equation}
$$
{\cal K}^{IJ}=\left(
\begin{array}{cc}
1   &  0\\
0   &   -1
\end{array}
\right),
$$
and, as in (\ref{3.9}),
$
\Gamma_m={\cal E}^{\underline a}_{m}\Gamma_{\underline a}
$
defines the worldsheet pullback of the target space Dirac matrices.

Eq. (\ref{3.18}) implies that, again as in the case of the
superparticles,
the number of independent $\kappa$--transformations is half the
number of components of $\theta$ and is equal to $N(D-2)$.

The property (\ref{3.18})
of the parameters (\ref{3.17}) can be checked using the
(anti)--self--duality
of $\kappa^I_m$ (\ref{3.11}) where in the projectors (\ref{3.10})
the metric $g_{mn}$ is replaced by $G_{mn}$ \footnote{
Note that the $\kappa$--parameters in the form (\ref{3.17}),
(\ref{3.18}) can
also be used directly in (\ref{3.9}). This would lead to a more
complicated form of a redefined variation of the intrinsic metric
$g_{mn}$. See \cite{bst1} for details on this form of
$\kappa$--transformations
in the case of a supermembrane.}.

Following our strategy we would like to replace the $\kappa$--symmetry
transformations with local supersymmetry transformations.
In the case of the $N=2$ superstrings the
worldsheet supersymmetry transformations should be parame\-trized
by $n=D-2$ worldsheet chiral and $n=D-2$ antichiral Majorana--Weyl
fermionic
parameters, which can be denoted as $(n,n)$--supersymmetry.

In the case of the $N=1$ superstrings (when one of the
$\theta$--coordinates
in (\ref{3.7}) is put to zero)
the worldsheet supersymmetry
should be of a heterotic type $(n,0)$ and have $n=D-2$ chiral or
antichiral parameters.
The reason why in the $N=1$ case the worldsheet supersymmetry associated
with $\kappa$--symmetry should have the heterotic (chiral) structure is
that in this case there is only one $\kappa$--symmetry parameter which
is either the anti--self--dual (antichiral) or self--dual (chiral)
worldsheet vector (see eq. (\ref{3.10})--(\ref{3.12})).

To realize local supersymmetry on the worldsheet we should extend the
latter
to a supersurface ${\cal M}_{2;n,n}$ parametrized by two bosonic
coordinates $\xi^m=(\tau,\sigma)$ and $N(D-2$) real fermionic
coordinates
$\eta^{\alpha q}=(\eta^{-q},
\eta^{+q'})$ (where $\alpha=-,+$ are the spinor (chirality) indices in
the
light--cone basis and $q,q'=1,...,n$ are indices of internal $SO(D-2)$
group
transformations which can be independent for $\eta^-$ and $\eta^+$).
Thus real $\eta^{\pm q}$ satisfy the $d=2$
Weyl condition
\begin{equation}\label{3.20}
\eta^{\pm q}=\pm\gamma^0\gamma^1\eta^{\pm q}=\pm\gamma^3\eta^{\pm q},
\end{equation}
where $\gamma^a$ $(a=0,1)$ and $\gamma^3$ are $d=2$ Dirac matrices which
can be chosen to have the form of (\ref{2.5.9}).

In the next subsection we consider the superembedding
of a superworldsheet which describes $N=1$ superstrings.

\subsection{Doubly supersymmetric $N=1$ superstrings}

To describe the embedding of a supersurface ${\cal M}_{2,n}$ with
d=2 bosonic coordinates $\xi^m=(\tau,\sigma)$ and $n=D-2$ real
fermionic coordinates $\eta^{-q}$ into $N=1$, $D=3,4,6,10$ target
superspace
we should first specify the geometrical properties of ${\cal M}_{2,n}$
appropriate for the description of the superstrings.

As in the case of the superparticles,
we would like to make life as simple as possible and to deal with an
${\cal M}_{2,n}$
geometry which would be almost flat and, at the same
time, preserved by {\it restricted} worldsheet superdiffeomorphisms
containing
{\it unrestricted} $d=2$ local reparametrizations
$\delta\xi^m=a^m(\tau,\sigma)$ and local
supersymmetry transformations
$\delta\eta^{-q}=\alpha^{-q}(\tau,\sigma)$. In $d=2$ superspaces
this can always be achieved starting from the most general form of the
supervielbeins, imposing suitable (supergravity) constraints on the
superworldsheet torsion and restricting the number of independent
supervielbein components by partially gauge fixing the general
worldsheet
superdiffeomorphisms \cite{supercon}.

A suitable choice of ${\cal M}_{2,n}$ geometry is as follows.

The constraints on the torsion $T^A={1\over 2}
e^Ce^BT^A_{BC}\equiv{\cal D}e^A(z)$ (where the covariant external
differential
${\cal D}=\delta^A_Bd+\omega^{~A}_{B}(z)$ contains a tangent space
connection)
are
\begin{equation}\label{3.2.00}
T^{--}_{-q,-r}=-i\delta_{qr}, \qquad T^{++}=0.
\end{equation}

The worldsheet supervielbeins $e^A(z)$ are
\begin{eqnarray}\label{3.2.1}
e^a(z^M)&=&\left(d\xi^m-id\eta^{-q}e_{-q}^m(z)\right)e^a_m(z),
\nonumber \\
e^{-q}(z^M)&=&d\eta^{-q},
\end{eqnarray}
and the supercovariant derivatives $D_A$ (which form the tangent basis
dual to (\ref{3.2.1})) are
\begin{equation}\label{3.2.2}
D_{++}=e_{++}^m(z)\partial_m, \quad
D_{--}={1\over n}D_{-q}e_{-q}^m(z)\partial_m,
\end{equation}
\begin{equation}\label{3.2.3}
D_{-q}=\partial_{-q}+ie_{-q}^m(z)\partial_m,
\end{equation}
where $e_{++}^m(z)$ and ${1\over n}D_{-q}e_{-q}^m(z)$ form the
vector--vector
components of the inverse supervielbein in the light--cone basis
\begin{equation}\label{3.2.0}
\left(e_{--}^m,e_{++}^m\right)=
\left({1\over n}D_{-q}e_{-q}^m,e_{++}^m\right).
\end{equation}

So the supervielbein matrix is
\begin{equation}\label{3.2.01}
e^{~A}_M=
\left(
\begin{array}{cc}
e^a_m   &  0\\
-ie^a_me_{-r}^m\delta^{-r}_\mu   &   \delta^{-q}_\mu
\end{array}
\right)
\end{equation}
and its inverse is
\begin{equation}\label{3.2.02}
e^{~M}_A=
\left(
\begin{array}{cc}
e^m_a   & 0 \\
 ie_{-q}^m  &   \delta_{-q}^\mu
\end{array}
\right),
\end{equation}

Note that $e_{++}^m$ can be transformed to the component
$\delta_{++}^m$
of the unit matrix by appropriate tangent space
transformations $e_{++}^{\prime m}=L_{++}^{~~a}(z)e_{a}^m$.
Therefore,  $e_{++}^m$, and hence
$e^{++}_m$, are completely auxiliary. This allows one to consider
an even simpler form \cite{dghs92} of supervielbeins and covariant
derivatives
(\ref{3.2.1})--(\ref{3.2.02}). However, we prefer to keep $e_{++}^m$
arbitrary since together with $e_{--}^m$ (at $\eta^{-q}=0$)
they are identified with the components of the
worldsheet zweibein (\ref{3.1})--(\ref{3.3}).

In accordance with the torsion constraints (\ref{3.2.00})
the supercovariant derivatives (\ref{3.2.2}),
(\ref{3.2.3}) are required to
satisfy the `flat' superalgebra
\begin{equation}\label{3.2.4}
\{D_{-q},D_{-r}\}={2i}\delta_{qr}D_{--},\quad  \{D_{-q},D_{--}\}=0.
\end{equation}
This imposes the constraint on the form of the supervielbein components
$e_{-q}^m(z)$
\begin{equation}\label{3.2.5}
D_{-q}e_{-r}^m+D_{-r}e_{-q}^m={2\over n}\delta_{qr}D_{-s}e_{-s}^m.
\end{equation}

The form of (\ref{3.2.2}) and (\ref{3.2.3}) is preserved (up to local
tangent space transformations $D'_a=L_a^{~~A}D_A,
~D'_{-q}=L_{-q}^{~-p}D_{-p}$
which include Weyl rescaling)
by restricted superdiffeomorphisms whose infinitesimal form is
\begin{eqnarray}\label{3.2.6}
\delta\xi^m&=&\Lambda^m-{1\over 2}e_{-q}^mD_{-q}\Lambda^{--}, \quad
\delta\eta^{-}_q=-{i\over 2}D_{-q}\Lambda^{--},\\
\delta D_{-q}&=&{i\over 2}(D_{-q}D_{-r}\Lambda^{--}) D_{-r},\nonumber\\
\delta e_{-q}^m&=&
-iD_{-q}\Lambda^m+{i\over 2}(D_{-r}\Lambda^{--})D_{-q}e_{-r}^m,
\end{eqnarray}
where $\Lambda^m(\xi,\eta)$ and $\Lambda^{--}(\xi,\eta)$ are independent
worldsheet superfunction parameters. $\Lambda^{--}(\xi,\eta)$ is a
self--dual vector in the same sense
as the ones in eqs. (\ref{3.12}).

In the Wess--Zumino gauge $D_{-q}|_{\eta=0}=\partial_{-q}$ one can use
local transformations with parameters
$$
D_{[-q_1}D_{-q_2]}\Lambda^m|_{\eta=0},~...,~D_{[-q_1}\cdots
D_{-q_{n-1}]}
\Lambda^m|_{\eta=0}
$$
to put all components of $e_{-q}^m(z)$ to zero except for
\begin{equation}\label{3.2.7}
{1\over n}D_{-q}e_{-q}^m|_{\eta=0}=e_{--}^m(\xi),
\end{equation}
which is a light--cone component of the inverse zweibein
(\ref{3.2}). In this gauge the constraint (\ref{3.2.5}) is
identically satisfied.
If we then impose on the worldsheet zweibein
the conformal gauge $e^m_a(\tau,\sigma)=\delta^m_a$, the covariant
derivatives
(\ref{3.2.2}) and (\ref{3.2.3}) become flat, and the
superdiffeomorphisms
(\ref{3.2.6}) reduce to the chiral (heterotic) superconformal $(n,0)$
transformations
\begin{equation}\label{3.2.06}
\delta\xi^{--}\equiv
\delta(\tau-\sigma)=\Lambda^{--}(\xi^{--},\eta^{-q})
-{1\over 2}\eta^{-q}D_{-q}\Lambda^{--}, \quad
\delta\eta^{-}_q=-{i\over 2}D_{-q}\Lambda^{--};
\end{equation}
$$
\delta\xi^{++}\equiv \delta(\tau+\sigma)=\Lambda^{++}(\xi^{++}).
$$

We are now in a position to write down the superembedding condition for
the $N=1$, $D=3,4,6,10$ superstring. It again prescribes that
in the basis (\ref{3.2.1}) the Grassmann
component of the superworldsheet pullback of the target space
supervielbein
${\cal E}^{\underline a}(Z)$ (\ref{2.3.2})  is zero
\begin{equation}\label{3.2.8}
{\cal E}_{-q}^{\underline a}(Z(z))=D_{-q} X^{\underline a}-
iD_{-q}\bar\Theta\Gamma^{\underline a}\Theta =0.
\end{equation}

The only difference between eq. (\ref{3.2.8}) and the superembedding
condition
for the $N=1$ superparticles is that now $D_{-q}$ contains the zweibein
components $e_{-q}^m(z)$.

The analysis of the superembedding condition (\ref{3.2.8}) is carried
out
in exactly the same way as the superparticle case (Subsection 2.5.3) and
results in the same conclusions \cite{to,dghs92}:

i) All higher components in the expansion of the superfields
$X^{\underline m}(\xi,\eta)$ and $\Theta^{\underline\mu}(\xi,\eta)$
are expressed in terms of the leading components $x^{\underline m}(\xi)$
and
$\theta^{\underline\mu}(\xi)$
which are dynamical variables of the superstrings.

ii) The dynamical equations of motion (\ref{3.012}) of
$x^{\underline m}(\xi)$ and
$\theta^{\underline\mu}(\xi)$ do not follow from the superembedding
condition.

iii) One of the Virasoro constraints (\ref{3.14}) of the $N=1$
superstrings
appears as a consequence of (\ref{3.2.8}) which produces the
Cartan--Penrose
relation
\begin{equation}\label{3.2.9}
\delta_{qr}{\cal E}_{--}^{\underline a}\equiv \delta_{qr}\left
(D_{--} X^{\underline a}-
iD_{--}\bar\Theta\Gamma^{\underline a}\Theta\right)=
D_{-q}\bar\Theta\Gamma^{\underline a}D_{-r}\Theta
\end{equation}
\begin{equation}\label{3.2.09}
\Rightarrow \quad {\cal E}^{\underline a}_{--}{\cal E}_{--{\underline
a}}=0.
\end{equation}

iv) Because of the relation (\ref{3.2.9}), the $\kappa$--symmetry
(\ref{3.9})
of the $N=1$ Green--Schwarz formulation ($\theta^2=0$)
is identified with local worldsheet supersymmetry.

v) The superstring dynamical equations and the second Virasoro condition
should be derived from a worldsheet superfield action in $N=1$,
$D=3,4,6,10$
target superspace.

We start to construct this action by writing down the term which
produces the superembedding condition (\ref{3.2.8})
\begin{equation}\label{3.2.10}
S_0=-i\int d^2\xi d^{n}\eta P^{-q}_{\underline a}
{\cal E}_{-q}^{\underline a}=
-i\int d^2\xi d^{n}\eta P^{-q}_{\underline a}\left[D_{-q} X^{\underline
a}-
iD_{-q}\bar\Theta\Gamma^{\underline a}\Theta\right],
\end{equation}
where $n=D-2$.

To this action one should, in principle,
add a Lagrange multiplier term which takes into
account the superworldsheet geometry constraint (\ref{3.2.5}), but since
this constraint is identically satisfied in the Wess--Zumino gauge
(\ref{3.2.7}) such a term is purely auxiliary and does not contribute
to the component superstring action \cite{dghs92}.
We therefore skip this term.

As in the superparticle case the action (\ref{3.2.10}) is invariant
under
the local transformations of the Lagrange multiplier
\begin{equation}\label{3.2.11}
\delta P^{-q}_{\underline a}(z)=
(D_{-r}+i\partial_me_{-r}^m)(\Lambda^{qrs}_{++}\Gamma_{\underline a}
D_{-s}\Theta),
\end{equation}
where the parameter $\Lambda^{qrs}_{++\underline\alpha}(z)$ is
totally symmetric and traceless with respect to the each pair of the
$SO(n)$
indices $q,r,s$.

Again, as in the superparticle case, this symmetry allows one to reduce
the number of independent components of the superfield
$P^{-q}_{\underline a}(z)$, and to express the remaining ones in terms
of
components of
$\Theta^{\underline\mu}(z)$ by solving the equations of motion
\begin{equation}\label{3.2.12}
(D_{-q}+i\partial_me_{-q}^m)P^{-q}_{\underline a}=0, \qquad
P^{-q}_{\underline a}\Gamma^{\underline a}D_{-q}\Theta=0.
\end{equation}

The action (\ref{3.2.10}), however, does not describe the fully fledged
$N=1$, $D=3,4,6,10$ superstrings \cite{bstv}.
As we shall see in the next subsection,
it describes so called null (or tensionless) superstrings, extended
objects
which are characterized by having zero tension and a degenerate
worldsheet metric \cite{shi}. The reason is that we have not yet
incorporated a Wess--Zumino--like term into the worldsheet superfield
action,
which will generate the string tension dynamically.

\subsubsection{Null superstrings}.

Null strings \cite{shi} (and  in general null (super)branes or
tensionless branes)
have attracted certain attention from various
points of view  \cite{zh1,zh2,noc,zb0,bn,l,t,lu}.
As far as their properties are concerned the null strings are closer to
the massless particles than to the strings.
Actually, they describe a continuum
of massless particles moving along a degenerate light--like
surface  \cite{zh1,l}.
Moreover the null strings do
not need any critical dimension of space--time to live in
\cite{noc,zb0} and
the null superstrings  \cite{zh1,bn,l}, in contrast to the ordinary
superstrings, do not require a Wess--Zumino term to be
$\kappa$--symmetric.

It has been assumed  \cite{zh2,l,lu}
that the null strings and branes
may be regarded as a high energy limit of the ordinary
string  thus providing a way for describing strings
beyond the Planck's scale.

We now show that the action (\ref{3.2.10}) describes null superstrings
in $N=1$, $D=3,4,6$ and 10 superspaces \cite{bstv}.

For simplicity consider the case when the superworldsheet
in (\ref{3.2.10}) has only one fermionic direction (i.e. $n=1$).
This case directly corresponds to an $N=1$, $D=3$ superstring
but the action (\ref{3.2.10}) with $n=1$
describes $N=1$, $D=4,6$ and 10 superstrings as well, though in these
higher dimensions only one of $D-2$ $\kappa$--symmetries is manifest.
The form of the hidden
$\kappa$ symmetries is exactly the same as in the case of
superparticles (\ref{d4})--(\ref{d10}).

Let us integrate
(\ref{3.2.10}) over the worldsheet fermionic variable and solve for
algebraic
equations of motion of auxiliary variables analogous to eqs.
(\ref{2.5.4}).
Then the action reduces to
\begin{equation}\label{3.3.1}
S_0=\int d^2\xi p^{--}_{\underline a}
\left[e_{--}^m(\xi)(\partial_m x^{\underline a}-
i{\partial_m\bar\theta}\Gamma^{\underline a}\theta)-\bar\lambda_-
\Gamma^{\underline a}\lambda_-\right],
\end{equation}
where
$$
p^{--}_{\underline a}(\xi)=P^{--}_{\underline a}(z)|_{\eta=0},
\quad \lambda_-^{\underline\mu}=D_-\Theta^{\underline\mu}(z)|_{\eta=0},
\quad e_{--}^m(\xi)=D_-e_{-}^m(z)|_{\eta=0}.
$$

As in the superparticle case,
from the action (\ref{3.3.1}) we derive the Cartan--Penrose relations
$$
p^{--}_{\underline a}={e^{-4}(\xi)}\bar\lambda_-
\Gamma_{\underline a}\lambda_-
$$
\begin{equation}\label{3.3.2}
{\cal E}^{\underline a}_{--}(\xi)=e_{--}^m(\xi)(\partial_m x^{\underline
a}-
i{\partial_m\bar\theta}\Gamma^{\underline a}\theta)=\bar\lambda_-
\Gamma^{\underline a}\lambda_-={1\over{e^{-4}}}p^{--\underline a},
\end{equation}
where $e^{-4}(\xi)$ is a proportionality coefficient with four `upper
minus' $SO(1,1)$ indices.

Eq. (\ref{3.3.2}) implies that ${\cal E}^{\underline a}_{--}$ satisfies
the
Virasoro constraint (\ref{3.2.09}).

In addition the variation of (\ref{3.3.1}) with respect to $e_{--}^m$
gives
\begin{equation}\label{3.3.3}
p^{--}_{\underline a}(\partial_m x^{\underline a}-
i{\partial_m\bar\theta}\Gamma^{\underline a}\theta)=0,
\end{equation}
which in view of (\ref{3.3.2}) implies
\begin{equation}\label{3.3.4}
e_{--}^n{\cal E}^{\underline a}_{n}{\cal E}_{m\underline a}=0.
\end{equation}
 From eq. (\ref{3.3.4}) it follows that (if $e_{--}^n$
is nonzero, which we assume) the induced worldsheet metric
\begin{equation}\label{3.3.5}
G_{mn}={\cal E}^{\underline a}_{m}{\cal E}_{n\underline a}
\end{equation}
is degenerate
$$
\det{G_{mn}}=0.
$$
This is a characteristic feature of the null strings and branes.

Making use of the relations (\ref{3.3.2}) and introducing instead of
$e_{--}^m$ the worldvolume vector
$$
V^m(\xi)=\sqrt{e^{-4}}e_{--}^m,
$$
we can rewrite the action (\ref{3.3.1}) so that it takes the form of the
null
superstring action considered, for example, in \cite{l}
\begin{equation}\label{3.3.6}
S_0=\int d^2\xi V^m V^n\left(\partial_m x
^{\underline m}-i\partial_m\bar\theta\Gamma^{\underline m}\theta\right)
\left(\partial_n x
_{\underline m}-i\partial_n\bar\theta\Gamma_{\underline m}\theta\right).
\end{equation}
One gets exactly the same form (\ref{3.3.6}) of the null superstring
action  from the action (\ref{3.2.10}) with $n=D-2$,
though, the procedure becomes a bit more
complicated because of the much greater number of auxiliary fields
entering the
initial superfield action.

Thus the doubly supersymmetric action (\ref{3.2.10}) describes $N=1$,
$D=3,4,6,10$ null superstrings. Note that in contrast to the
Green--Schwarz
action the null superstring action (\ref{3.3.6})
does not contain the Wess--Zumino term. However it is still
$\kappa$--symmetric
(which is the relic of the linearly realized worldsheet supersymmetry
of the action (\ref{3.2.10})).
The $\kappa$--symmetry transformations are
$$
\delta_\kappa\theta^\alpha=iV^m\left(\partial_m x
^{\underline m}-i\partial_m\bar\theta\Gamma^{\underline m}\theta\right)
\gamma_{{\underline m}\beta}^\alpha\kappa^\beta,\quad \delta_\kappa
x^m=-i\bar\theta\Gamma^m\delta_\kappa\theta,
$$
$$
\delta_\kappa V^m=2V^m V^n\partial_n\bar\theta\kappa.
$$

To get a doubly supersymmetric action for the fully fledged $N=1$
superstrings we should add to the action (\ref{3.2.10}) a term which
would
generate the string tension and make the worldsheet metric
nondegenerate.
Such a term should reproduce the Wess--Zumino term of the Green--Schwarz
formulation.

The treatment of the string tension as a dynamical variable has been
discussed in various aspects \cite{zh2,t,t1}. In particular, it allows
one
to construct superbrane actions which are manifestly target--space
supersymmetric \cite{t,t2,t3,t1}, i.e. do not contain the
standard Wess--Zumino term.
The replacement of tension with the worldvolume p-form field
is also useful when one deals with branes ending on other branes
\cite{t2}.

\subsubsection{String tension generating mechanism}
Before considering the doubly supersymmetric case let us first show how
one
can extend the null superstring component action (\ref{3.3.1})
to describe the standard $N=1$ superstring. This construction is similar
to
that of references \cite{t} and is analogous to the super--Maxwell
coupling
of the superparticles discussed in Subsection 2.6.

One introduces an `electromagnetic' field
$A_m(\xi)$ on the superstring worldsheet and constructs a worldsheet
two--form which is invariant under the target space supersymmetry
transformations (\ref{2.3.1}). The appropriate two--form is
\begin{equation}\label{3.4.1}
{\cal F}^{(2)}=e^{--}e^{++}{\cal E}^{\underline a}_{--}
{\cal E}^{\underline b}_{++}
\eta_{\underline{ab}}+B^{(2)}+dA.
\end{equation}
It includes the worldsheet pullback of the Wess--Zumino form (\ref{3.8})
(where $\theta^2=0$ because we deal with $N=1$), i.e.
\begin{equation}\label{3.4.01}
B^{(2)}=idx^{\underline a}d\bar\theta\Gamma_{\underline a}\theta,
\end{equation}
and
$e^{++}=d\xi^me_m^{++}(\xi)$ and $e^{--}=d\xi^me_m^{--}(\xi)$ are the
light--cone components of the zweibein (\ref{3.1}), (\ref{3.2}).

Since under the supersymmetry transformations $B^{(2)}$ transforms as
a total derivative
$$
\delta B^{(2)}=id\left[(dx^{\underline a}
-{i\over 3}d\bar\theta\Gamma^{\underline a}\theta)
\bar\theta\Gamma_{\underline a}\epsilon\right]
$$
the vector field $A_m(\xi)$ must also vary as
$$
\delta A=-i(dx^{\underline a}
-{i\over 3}d\bar\theta\Gamma^{\underline a}\theta)
\bar\theta\Gamma_{\underline a}\epsilon
$$
to cancel the variation of $B^{(2)}$ in (\ref{3.4.1}).

The vector field $A_m(\xi)$ should not be a new
propagating worldsheet field since our aim is to
describe the ordinary superstrings which do not
carry such fields. Hence we must require that on the mass shell
$A_m(\xi)$ is expressed in terms of superstring dynamical variables.
This is achieved by assuming that the two--form (\ref{3.4.1}) vanishes
on the mass shell, i.e.
\begin{equation}\label{3.4.2}
dA=-e^{--}e^{++}{\cal E}^{\underline a}_{--}{\cal E}^{\underline b}_{++}
\eta_{\underline{ab}}-B^{(2)}.
\end{equation}
Eq. (\ref{3.4.2}) implies that the field strength of $A_m(\xi)$
is not independent and hence does not describe new redundant
physical degrees of
freedom. From eq. (\ref{3.4.2}) it also follows that its r.h.s. is
a closed form, which is true for any two--form on a $d=2$ bosonic
manifold.

To get eq. (\ref{3.4.2}) as an equation of motion we add to the action
(\ref{3.3.1}) the term
\begin{equation}\label{3.4.3}
S_T=\int d^2\xi\phi(\xi)\epsilon^{mn}\left[
e_m^{--}e_n^{++}
{\cal E}^{\underline a}_{--}{\cal E}^{\underline b}_{++}
\eta_{\underline{ab}}+B_{mn}^{(2)}+\partial_mA_n
\right],
\end{equation}
where $\phi(\xi)$ is a Lagrange multiplier whose variation
gives rise to eq. (\ref{3.4.2}).

The extended action (\ref{3.3.1})+(\ref{3.4.3})
\begin{equation}\label{3.4.003}
S=S_0+S_T
\end{equation}
is invariant under new local transformations found in \cite{to,dghs92}
\begin{equation}\label{3.4.03}
\delta p^{--\underline a}
=\Lambda^{--}_{++}(\xi)\epsilon^{mn}e_m^{++}e_n^{--}
({\cal E}^{\underline a}_{--} + \bar\lambda_-\Gamma^{\underline
a}\lambda_-)
\phi(\xi)
\end{equation}
$$
\delta e^n_{++}= \Lambda^{--}_{++}(\xi)e^n_{--} ,\qquad \delta
e^n_{--}=0.
$$
If $\phi(\xi)$ is nonzero this symmetry allows one to
eliminate $p^{--}_{\underline a}$
on the mass shell when the eqs. (\ref{3.3.2}) hold.

Let us now see how the string tension appears, and
the Green--Schwarz action is recovered.
Varying (\ref{3.4.3}) with respect to $A_m$ we get
\begin{equation}\label{3.4.4}
\partial_m\phi=0 \quad \Rightarrow \quad \phi=T,
\end{equation}
where $T$ is an integration constant.

If $T$ is nonzero, and we substitute (\ref{3.4.4}) into the action
(\ref{3.4.3}), the term $\epsilon^{mn}\partial_{m}A_n$ becomes a
total derivative and can be skipped. As a result the action
$S_T$ (\ref{3.4.3})
reduces to the $N=1$ Green--Schwarz action (\ref{3.7}) (with
$\theta^2=0$).
The first term of the Green--Schwarz action is nothing but the first
term
in (\ref{3.4.3}), which can be checked using the expressions
(\ref{3.2})--(\ref{3.03}). We see that $T$ is the string tension.

Moreover, when $\phi$ (and hence $T$) is
nonzero the local transformations (\ref{3.4.03}) allow one to
put $p^{--}_{\underline a}$ satisfying (\ref{3.3.2}) to zero. Thus the
term $S_0$  (\ref{3.3.1}) drops out of the extended action
(\ref{3.4.003})
and the latter coincides with the $N=1$ Green--Schwarz action
\begin{equation}\label{n1gs}
S_{N=1}=T\int d^2\xi\epsilon^{mn}\left[
e_m^{--}e_n^{++}
{\cal E}^{\underline a}_{--}{\cal E}^{\underline b}_{++}
\eta_{\underline{ab}}+B_{mn}^{(2)}
\right].
\end{equation}

On the contrary, when $T=0$ the second term $S_T$ (\ref{3.4.3})
disappears from
(\ref{3.4.003}) and it reduces to the null superstring action
(\ref{3.3.1}).
Therefore, eq. (\ref{3.4.003}) is the generic action which
is nonsingular in the
tensionless limit and describes superstrings with all values of the
tension.

Our next step is to lift the action $S_T$ on to the worldsheet
superspace.

So now we consider the two--form (\ref{3.4.1}) given on the
superworldsheet ${\cal M}_{2,n}$, i.e. all its ingredients depend on
$z^M=(\xi^m,\eta^{-q})$.

\subsubsection{Weil triviality}
An interesting observation concerning properties of the two--superform
(\ref{3.4.1}) on ${\cal M}_{2,n}$ was made in \cite{to,tor}.

The two--superform ${\cal F}^{(2)}$ is closed on ${\cal M}_{2,n}$ when
the
superembedding condition (\ref{3.2.8}) is satisfied. The closure of
a two--form on ${\cal M}_{2,n}$ is a nontrivial property
because of the presence of $n$ Grassmann directions.

Let us check this property, but first rewrite (\ref{3.4.1}) in
a more appropriate form
\begin{equation}\label{3.5.1}
{\cal F}^{(2)}={\cal E}^{\underline a}\wedge e^{++}{\cal
E}_{++\underline a}
+B^{(2)}+dA.
\end{equation}
By definition of the pullback
\begin{equation}\label{3.5.01}
{\cal E}^{\underline a}= e^{--}{\cal E}^{\underline a}_{--}+
e^{++}{\cal E}_{++}^{\underline a}+e^{-q}{\cal E}^{\underline a}_{-q},
\end{equation}
we see that if the superembedding condition ${\cal E}^{\underline
a}_{-q}=0$
(\ref{3.2.8}) holds, the form (\ref{3.5.1}) coincides with
(\ref{3.4.1}).

Then, in view of the definition of $B^{(2)}$ (\ref{3.4.01}),
$$
d{\cal F}^{(2)}=d\left({\cal E}^{\underline a}e^{++}
{\cal E}_{++\underline a}\right)
+dB^{(2)}=
$$
\begin{equation}\label{3.5.2}
=-(d{\cal E}^{\underline a})e^{++}{\cal E}_{++\underline a}+
{\cal E}^{\underline a}d(e^{++}{\cal E}_{++\underline a})+
i{\cal E}^{\underline a}d\bar\Theta\Gamma_{\underline a}d\Theta.
\end{equation}
Because of the target--space torsion constraint (\ref{2.3.3}) and
the expansion (\ref{3.5.01}), the first term
of (\ref{3.5.2}) can be rewritten as
\begin{equation}\label{3.5.3}
-id\bar\Theta\Gamma_{\underline a}d\Theta e^{++}{\cal E}_{++\underline
a}=
-id\bar\Theta\Gamma_{\underline a}d\Theta{\cal E}^{\underline a}+
id\bar\Theta\Gamma_{\underline a}d\Theta
\left(e^{--}{\cal E}^{\underline a}_{--}+e^{-q}{\cal E}^{\underline
a}_{-q}
\right).
\end{equation}
If the superembedding condition (\ref{3.2.8}) and its consequences
(\ref{3.2.9}) and (\ref{3.2.09}) are satisfied, the last two terms
of (\ref{3.5.3}) vanish, and the first term cancels the last term in
(\ref{3.5.2}).

By virtue of the worldsheet torsion constraints (\ref{3.2.00})
and the superembedding conditions (\ref{3.2.8})--(\ref{3.2.09})
the second term in (\ref{3.5.2}) also vanishes
$$
{\cal E}^{\underline a}d(e^{++}{\cal E}_{++\underline a})=
e^{--}{\cal E}^{\underline a}_{--}e^{++}e^{-q}
\nabla_{-q}{\cal E}_{++\underline a} \sim
$$
\begin{equation}\label{3.5.03}
\sim e^{--}{\cal E}^{\underline a}_{--}e^{++}e^{-q}D_{++}\bar\Theta
\Gamma_{\underline a}D_{-q}\Theta=0.
\end{equation}
The transition to the expression after the similarity sign $\sim$ has
been
made by getting the expression for $e^{++}e^{-q}
\nabla_{-q}{\cal E}_{++\underline a}$ as a component of the tangent
space torsion constraint (\ref{2.3.3}), $d{\cal E}^{\underline a}=
-id\Theta\Gamma^{\underline a}d\Theta$, in the worldsheet supervielbein
basis and taking into account (\ref{3.2.00}) and
(\ref{3.2.8})--(\ref{3.2.09}).

We thus get
\begin{equation}\label{3.5.4}
d{\cal F}^{(2)}=\Phi^{-q}_{\underline a}{\cal E}^{\underline a}_{-q}+
\Phi^{--\{qr\}}_{\underline a}D_{-r}{\cal E}^{\underline a}_{-q}=0,
\end{equation}
where $\Phi$ are some three--superforms.

The variation of ${\cal F}^{(2)}$ with respect to the ${\cal M}_{2,n}$
superdiffeomorphisms $z^{\prime M}=z^M+\delta z^M(\xi,\eta)$ is
\begin{equation}\label{3.5.5}
\delta {\cal F}^{(2)}=d(i_\delta {\cal F}^{(2)}) +i_\delta d {\cal
F}^{(2)}.
\end{equation}

Now take the integral of ${\cal F}^{(2)}$ over the two--dimensional
slice
of ${\cal M}_{2,n}$ such that $\eta^{-r}=0$, i.e. take the integral
over the ordinary worldsheet
\begin{equation}\label{3.5.6}
S_T=T\int_{{\cal M}_{2}} {\cal F}^{(2)}.
\end{equation}
Eq. (\ref{3.5.6}) is nothing but the action (\ref{3.4.3}) where
$\phi(\xi)$ is replaced with $T$
(the term with $A_m(\xi)$ is a total derivative and can be skipped).

The variation of (\ref{3.5.6}) with respect to the worldsheet
superdiffeomorphisms (\ref{3.5.5}) is (up to a total derivative)
\begin{equation}\label{3.5.7}
\delta S_T= T\int_{{\cal M}_{2}}i_\delta d {\cal F}^{(2)}.
\end{equation}
Since  $d {\cal F}^{(2)}$ (\ref{3.5.4})
is `proportional' to the superembedding condition, the variation of
$S_T$ can be canceled by an appropriate variation of the Lagrange
multiplier
of the action (\ref{3.2.10}) if we take the sum of the two actions.

Thus, the action
\begin{equation}\label{3.5.8}
S=-i\int d^2\xi d^{n}\eta P^{-q}_{\underline a}
{\cal E}_{-q}^{\underline a}+T\int_{{\cal M}_{2}} {\cal F}^{(2)}
\end{equation}
possesses local $n=D-2$ worldsheet supersymmetry though the second term
is
not an integral over the whole superworldsheet. As we have learned
in the previous subsection the component version of this action
describes
$N=1$, $D=3,4,6,10$ superstrings.

The doubly supersymmetric action in the form (\ref{3.5.8}) was proposed
in \cite{to}. The property of ${\cal F}^{(2)}$ to be a closed superform
up to
the superembedding condition and the corresponding nonmanifest
superdiffeomorphism invariance of the action (\ref{3.5.8}) was called
Weil triviality, a property which had been found to be useful for
studying
chiral anomalies in super Yang--Mills theories \cite{bpt}. Actually,
the action (\ref{3.5.6}) is a particular example of the generalized
actions of the group manifold (rheonomic) approach to the description
of supersymmetric field theories \cite{gma} and superbranes \cite{bst}.

\subsubsection{Worldsheet superfield action}
To make the action $S_T$ (\ref{3.5.6})
manifestly supersymmetric on the superworldsheet
one constructs a Lagrange multiplier term \cite{dghs92}
which produces the on--shell condition
\begin{equation}\label{3.5.9}
 {\cal F}^{(2)}=0.
\end{equation}

The superstring action which includes such a  term is
\begin{equation}\label{3.5.10}
S=S_0+S_T=-i\int d^2\xi d^{n}\eta P^{-q}_{\underline a}
{\cal E}_{-q}^{\underline a}
+\int d^2\xi d^n\eta P^{MN}{\cal F}^{(2)}_{MN},
\end{equation}
where now
\begin{equation}\label{3.5.010}
S_T=\int d^2\xi d^n\eta P^{MN}{\cal F}^{(2)}_{MN}=\int d^2\xi d^n\eta
P^{MN}
\left[{\cal E}^{\underline a}_{[M} e^{++}_{N\}}{\cal E}_{++\underline a}
+B^{(2)}_{MN}+\partial_{[M}A_{N\}}\right].
\end{equation}
$[,\}$ denotes graded  antisymmetrization of the superworldsheet indices
(i.e. if one or both of the indices $M,N$ are bosonic they are
antisymmetrized, and if both of them are fermionic they are
symmetrized).

In addition to all symmetries discussed above the action (\ref{3.5.10})
is invariant under the local transformations of the Lagrange multiplier
$P^{MN}$
\begin{equation}\label{3.5.11}
\delta P^{[MN\}}=\partial_L\Lambda^{[LMN\}}(z).
\end{equation}

Indeed, the variation of the $S_T$  term (\ref{3.5.010})
with respect to (\ref{3.5.11}) is
\begin{equation}\label{3.5.12}
\delta S_T=\int d^2\xi d^n\eta \Lambda^{[LMN\}}(d{\cal F}^{(2)})_{LMN},
\end{equation}
where $d{\cal F}^{(2)}$ has the form of eq. (\ref{3.5.4}), and, hence,
the variation
(\ref{3.5.12}) is canceled by an appropriate variation of the
Lagrange multiplier $P^{-q}_{\underline a}$ in (\ref{3.5.10}).

Varying the action (\ref{3.5.010}) with respect to $A_M$ we get
\begin{equation}\label{3.5.13}
 \partial_MP^{[MN\}}=0,
\end{equation}
whose general solution is
\begin{equation}\label{3.5.14}
P^{[MN\}}=\partial_L\tilde\Lambda^{[LMN\}}(z)
+{1\over{n!}}\epsilon_{q_1...q_n}\eta^{-q_1}....\eta^{-q_n}
\delta^{[M}_{--}\delta^{N\}}_{++}T,
\end{equation}
where $T$ is a constant.

The first term in (\ref{3.5.14}) can be put to zero by gauge fixing the
local transformations (\ref{3.5.11}). Then, substituting (\ref{3.5.14})
(with $\tilde\Lambda=0$) back into the action (\ref{3.5.10})
we see that upon the $\eta$--integration it reduces to (\ref{3.5.8}).

We have thus demonstrated that the worldsheet superfield action
(\ref{3.5.10})
describes the $N=1$, $D=3,4,6,10$ superstrings.

Note that in the case when the superworldsheet has only one
fermionic direction, i.e. when $n=1$, there exists another form of the
superfield action $S_T$ \cite{to},
which involves only one Grassmann component
$B^{(2)}_{-,++}$ of the ${\cal M}_{2,1}$ pullback of the two--form
$B^{(2)}$
(\ref{3.4.01}). In this form the full action is
\begin{equation}\label{3.5.15}
S=S_0+S_T=-i\int d^2\xi d\eta_- P^{--}_{\underline a}
{\cal E}_{-}^{\underline a}
+T\int d^2\xi d\eta_- sdet{(e)}B^{(2)}_{-,++},
\end{equation}
where $sdet{(e)}$ is the superdeterminant of the supervielbein matrix
$e^A_M$
(\ref{3.2.01}).

In Subsection 3.2 (eqs. (\ref{3.2.8})--(\ref{3.2.09}))
we demonstrated that the superembedding condition gives rise only to one
of the superstring Virasoro constraints. The second constraint, namely
${\cal E}^{\underline a}_{++}{\cal E}_{++\underline a}=0$,
arises as a result of the variation of the
full action (in any of its forms, (\ref{3.5.8}), (\ref{3.5.10}) or
(\ref{3.5.15})) with respect to the supervielbein component $e^m_{--}$.
A completely `twistorized' version of the doubly supersymmetric action
for the $N=1$ superstrings, where both Virasoro constraints have a
twistor origin, was constructed in \cite{bcsv}.

\subsection{Coupling to supergravity background}
The generalization of the superstring action (\ref{3.5.10}) to describe
a superstring propagating in curved target superspace introduced in
Subsection 2.7 is almost straightforward \cite{to,dghs92}.
In (\ref{3.5.10}) and (\ref{3.5.010}) one should replace the flat
supervielbeins with curved ones ${\cal E}_{-q}^{\underline a}\rightarrow
E_{-q}^{\underline a}$, consider $B^{(2)}(Z)$ as a two--form gauge
superfield, whose leading component $B_{\underline m\underline n}(X)$
is the Neveu--Schwarz gauge potential entering the supergravity
multiplet,
and introduce a dilaton superfield
$\Phi(Z)$ coupling by redefining ${\cal F}^{(2)}(Z)$ as follows
\begin{equation}\label{hatb}
{\cal F}^{(2)}_{MN}=e^\Phi
{E}^{\underline a}_{[M} e^{++}_{N\}}{E}_{++\underline a}
+B^{(2)}_{MN}+\partial_{[M}A_{N\}}.
\end{equation}
The superstring action takes the form
\begin{equation}\label{3.5.0100}
S=S_0+S_T=-i\int d^2\xi d^{n}\eta P^{-q}_{\underline a}
{E}_{-q}^{\underline a}
+\int d^2\xi d^n\eta P^{MN}{\cal F}^{(2)}_{MN}.
\end{equation}
By construction (\ref{3.5.0100})
is superdiffeomorphism invariant on the superworldsheet
and in the target superspace, and it must also be invariant under
the variation (\ref{3.5.11}) of the Lagrange multiplier $P^{MN}$.
This means that the variation (\ref{3.5.12}) of the action
(\ref{3.5.0100}) must be compensated by an appropriate variation
of the Lagrange multiplier $P^{-q}_{\underline a}$. For this,
$d{\cal F}^{(2)}$ must have the form of the l.h.s. of eq. (\ref{3.5.4}),
i.e.
\begin{equation}\label{3.5.4b}
d{\cal F}^{(2)}=\Phi^{-q}_{\underline a}{E}^{\underline a}_{-q}+
\Phi^{--\{qr\}}_{\underline a}D_{-r}{E}^{\underline a}_{-q}.
\end{equation}

Performing the direct computation of the external differential of
(\ref{hatb}) one finds that it has the form of eq. (\ref{3.5.4b}) if the
superbackground satisfies the supergravity torsion constraint
(\ref{torsionc})
\begin{equation}\label{torsionc1}
T^{\underline a}_{\underline\alpha\underline\beta}=
-2i(C\Gamma^{\underline a})_{\underline\alpha\underline\beta},
\end{equation}
and the components of the field strength $H^{(3)}=dB^{(2)}$ of the
two--from
gauge superfield are constrained as follows
\begin{equation}\label{b2constr}
H_{\underline\alpha\underline\beta\underline a}
=2ie^{\Phi(Z)}(C\Gamma_{\underline
a})_{\underline\alpha\underline\beta},
\qquad
H_{\underline\alpha\underline\beta\underline\gamma}=0
\end{equation}
(where $\Phi(Z)$ is the dilaton field).

Thus, as in the case of the Green--Schwarz formulation,
the consistency of coupling the doubly supersymmetric strings to
the supergravity background demands that the superbackground obeys the
supergravity constraints (\ref{torsionc1}) and (\ref{b2constr}). In
$D=10$
these constraints put $N=1$ supergravity on the mass shell
\cite{10sg,shat}.

\subsection{Heterotic fermions}
So far we have dealt with classical $N=1$ superstrings in space--time
of dimensions $D=3,4,6$ and 10. Quantum consistency of superstring
theory
singles out 10--dimensional space--time. In addition, to be anomaly
free a closed $N=1$, $D=10$ superstring should be heterotic
\cite{heter},
i.e. it should contain extra (heterotic) matter on its worldvolume
which upon quantization produces an $N=1$, $D=10$ super--Yang--Mills
field
$A_{\underline M}(Z)$ taking its values in the adjoint representation of
the
gauge group $SO(32)$ or $E_8\times E_8$. Such worldsheet matter fields
can be 32 chiral fermions $\psi^{-{\cal A}}(\xi)$, where `$-$' is the
chiral
spinor index of the $d=2$ Lorentz group $SO(1,1)$ and ${\cal
A}=1,\cdots,32$
is the index of the 32--dimensional vector representation of the
target--space gauge group $SO(32)$. Therefore,
the chiral fermions can be minimally coupled to
the $SO(32)$ gauge superfield $A^{[{\cal A}{\cal B}]}_{\underline
M}(Z)$.
Note that $\psi^{-{\cal A}}(\xi)$ cannot be minimally
coupled to a gauge field of the  $E_8\times E_8$ group whose
lowest--dimensional representation is $496=2\times 248$.
The $E_8\times E_8$ gauge fields appear in heterotic string theory
in a much more subtle way (see, for instance, \cite{gsw})
and we shall not describe them here.

Our aim now is to introduce
the heterotic fermions into the superembedding approach.

In the Green--Schwarz formulation the heterotic fermion term, which is
added to the $N=1$, $D=10$ superstring action (\ref{n1gs}), is
\begin{equation}\label{3.6.1}
S_{\psi}={i\over 2}T\int d^2\xi\psi^-e^m_{--}(\xi)
(\partial_m-\partial_mZ^{\underline M}A_{\underline M})\psi^{-}.
\end{equation}

Varying (\ref{3.6.1}) with respect to $\psi^-$ we get the equation of
motion
which has the form of a chirality condition on $\psi^-$
\begin{equation}\label{3.6.2}
e^m_{--}(\partial_m-\omega_{m}^{~-+}-\partial_mZ^{\underline
M}A_{\underline M})\psi^{-} =0,
\end{equation}
where $\omega_{m}^{~-+}(\xi)$ is a worldsheet spin connection.

 In the conformal gauge $e^m_{--}(\xi)=\delta^m_{--}$ and
when
$A_{\underline M}=0$ eq. (\ref{3.6.2}) reduces to
\begin{equation}\label{3.6.3}
\partial_{--}\psi^{-{\cal A}}=0
\end{equation}
which implies that the heterotic fermions are worldsheet left--movers
\begin{equation}\label{3.6.4}
\psi^{-{\cal A}}=\psi^{-{\cal A}}(\xi^{++})=\psi^{-{\cal
A}}(\tau+\sigma).
\end{equation}
For comparison let us note that in the conformal
gauge the equations of motion of $\theta(\xi)$
deduced from the $N=1$ Green--Schwarz action (\ref{n1gs}) imply that
$\theta(\xi)$ are right--movers, i.e.
$$
\theta^{\underline\alpha}=\theta^{\underline\alpha}(\xi^{--})
=\theta^{\underline\alpha}(\tau-\sigma).
$$

Under the $\kappa$--symmetry transformations (\ref{3.9})--(\ref{3.12})
the heterotic fermions transform as follows
\begin{equation}\label{3.6.5}
\delta_\kappa\psi^-=\bar A_{\underline \alpha}
({\cal E}^{\underline a}_{--}\Gamma_{\underline a})^{\underline \alpha}
_{~\underline \beta}\kappa_{++}^{\underline \beta}\psi^-.
\end{equation}
Therefore, they are inert with respect to the $\kappa$--transformations
when
the SYM background is switched off
(note that when $N=1$, from (\ref{3.9})--(\ref{3.12}) it follows that
$\delta_\kappa e^m_{--}=0$).

We would like to extend the heterotic fermion action (\ref{3.6.1})
and the chirality condition (\ref{3.6.2}) to the superworldsheet
formulation
of the heterotic string whose $N=1$, $D=10$ target--superspace part is
described by the action (\ref{3.5.10}).

The generalization of the chirality condition (\ref{3.6.2}) is
straightforward.
We consider the heterotic fermions as worldsheet superfields
$\Psi^{-{\cal A}}(\xi^m,\eta^{-q})$ and assume that on the mass shell
they
satisfy the equation of motion
\begin{equation}\label{3.6.6}
\left[D_{-q}-D_{-q}Z^{\underline M}A_{\underline M}\right]\Psi^{-}=0,
\end{equation}
where the covariant derivative $D_{-q}$ was determined in eq.
(\ref{3.2.3}).

It can be easily checked that in the Wess--Zumino gauge (\ref{3.2.7}),
eq. (\ref{3.6.6}) reduces to eq. (\ref{3.6.2}), and hence,
in the conformal gauge and in the absence of the SYM background, the
heterotic fermions $\Psi^{-}$ are chiral (\ref{3.6.4}) and do not
transform
under the `left--moving' conformal supersymmetry transformations
(\ref{3.2.06}),
which is in agreement with their $\kappa$--symmetry properties
(\ref{3.6.5}).

Note that, as usual, $\kappa$--symmetry, or (in the worldsheet
superfield
formulation) the integrability of (\ref{3.6.6}) requires the SYM
constraints
(\ref{symc}) and (\ref{symc1})
on the background gauge field $A_{\underline M}$.

Now the problem is to get the equation (\ref{3.6.6}) from a worldsheet
$n$-extended superspace action.

When $n=1$, the action (\ref{3.6.1}) admits the straightforward
generalization
\begin{equation}\label{3.6.7}
S_\psi=-{T\over 2}\int d^2\xi d\eta_-\Psi^-(D_--D_-Z^{\underline M}
A_{\underline M})\Psi^-.
\end{equation}

When $n=2$ there is a generalization of (\ref{3.6.7}) in terms of
complex chiral $n=2$ superfields \cite{to}.

For the worldsheet
supersymmetry formulation with  $n>2$ different forms
of the superfield action of chiral fermions were proposed
\cite{st,howe,is}.
The most
elegant formulation, which we shall sketch for $n=8$,
was proposed by P. Howe \cite{howe}.

Let us split the 32--dimensional index ${\cal A}$
of $\psi^{-\cal A}(z)$ into an index $i=1,...,8$
and an index ${\cal A}^\prime=1,2,3,4$, and identify the index $i$
with the index of the vector representation of the group of internal
isomorphisms $SO(8)$ of the $n=8$ worldsheet supersymmetry (remember
that
$\eta^{-q}$ transform under a spinor representation of $SO(8)$).
The index ${\cal A}^\prime$ labels the vector representation of a
target superspace gauge group $SO(4)$ and
$A^{[{\cal A}^\prime{\cal B}^\prime]}_{\underline M}(Z)$ are now
$SO(4)$--gauge
fields which the heterotic fermions can couple to.
Note that the possible minimal coupling of the heterotic fermions is
now reduced from $SO(32)$ gauge coupling to $SO(4)$ coupling.
This is a shortcoming of the present formulation which, however, seems
to be
akin to the general problem of directly coupling $E_8\times E_8$ gauge
fields
mentioned at the beginning of this subsection.

Assume that {\it off the mass shell} the superfield $\psi^{-i{\cal
A}'}(z)$
satisfies the constraint
\begin{equation}\label{3.6.8}
\hat\nabla_{-q}\psi^{-i}\equiv \left(\delta^{i}_j D_{-q}
-\omega^{~~~~-i}_{-q,-j}-\delta^{i}_jD_{-q}Z^{\underline M}
A_{\underline M}\right)\psi^{-j}=\gamma^{i}_{qq'} P^{q'},
\end{equation}
where $\omega^{~~~~-i}_{-q,-j}$ is an $SO(1,1)\times SO(8)$
superworldsheet
connection,
the $SO(8)$ gamma--matrices have already been introduced in
(\ref{2.5.39}), and  $P^{q'{\cal A}'}(z)$ is a superfield in the
spinor representation of $SO(8)$ `dual' to that of $\eta^{-q}$.

On the mass shell $P^{q'{\cal A}'}(z)$ should vanish
and then
eq. (\ref{3.6.8}) reduces to the chirality condition analogous to
(\ref{3.6.6})
\begin{equation}\label{3.6.08}
P^{q'}=\hat\nabla_{-q}\psi^{-i}=0,
\end{equation}
thus, providing us with the equations of motion of the
heterotic fermions.

It seems instructive to note that the constraint (\ref{3.6.8}) resembles
the superembedding condition (\ref{3.2.8}),
and, indeed, (\ref{3.2.8}) reduces
to the exactly the same form when the local worldsheet supersymmetry
is gauge fixed as in eqs. (\ref{2.5.44}) and (\ref{2.5.45}).
To see this, take the transverse part of (\ref{3.2.8}) which, in view of
eqs. (\ref{2.5.39})--(\ref{2.5.45}), is
\begin{equation}\label{3.6.100}
D_{-q}X^i-i\eta^{-p}\gamma^i_{pq'}D_{-q}\Theta^{q'}_-
-i\gamma^i_{qq'}\Theta^{q'}_-=0,
\end{equation}
$$
i=1,...,8; \quad q=1,...,8; \quad q'=1,...,8\, .
$$
The second term of (\ref{3.6.100}) can be `absorbed' by redefining the
transverse coordinate
$X^i=\hat X^i-i\eta^{-p}\gamma^i_{pq'}\Theta^{q'}_-$, and
we finally get
\begin{equation}\label{3.6.101}
D_{-q}\hat X^i=2i\gamma^i_{qq'}\Theta^{q'}_-,
\end{equation}
whose form is the same as that of the constraint (\ref{3.6.8}).

The constraint (\ref{3.6.8}) can be explicitly solved \cite{howe}, the
solution
being in terms of an unconstrained superfield
$V^{-8{\cal A}'}_{q'}(z)$ (with
eight $SO(1,1)$ minuses)
\begin{equation}\label{3.6.9}
\psi^{-i}=\gamma^{i}_{qq'}\Delta^7_{-q}V^{-8}_{q'},
\end{equation}
where $\Delta^7_{-q}$ is a differential operator of a 7-th power in
$\hat\nabla_{-q}$ whose explicit form the reader can find in ref.
\cite{howe}.

The action which gives rise to eq. (\ref{3.6.08}) is
\begin{equation}\label{3.6.10}
S_\psi=\int d^2\xi d^8\eta_-~ sdet({e})V^{-8}_{q'}P^{q'},
\end{equation}
where under $P^{q'}$ one should imply eq. (\ref{3.6.8}) with $\psi^{-i}$
being replaced by its `prepotential' (\ref{3.6.9}).

This completes the construction of the manifest
doubly supersymmetric formulation of the $N=1$, $D=10$ heterotic string,
and we now briefly discuss the case of

\subsection{Type II superstrings}
As we have already mentioned at the end of Subsection 3.1,
to replace the $\kappa$--symmetry of the $N=2$
superstring action in the Green--Schwarz formulation (\ref{3.7}) with
local
worldsheet supersymmetry one should deal with an $(n,n)$
worldsheet superspace
parametrized by bosonic coordinates $\xi^m$ $(m=0,1)$ and $2n=2(D-2$)
worldsheet
spinor coordinates $\eta^{\alpha q}$ ($\alpha=-,+;$ $q=1,...,D-2$).
The basic condition specifying the  superembedding of such a
supersurface
into an $N=2$ target
superspace parametrized by supercoordinates
$Z^{\underline M}=(X^{\underline m},\Theta^{I\underline\mu})$ $(I=1,2)$
is
$$
E^{\underline a}_{\alpha q}=
D_{\alpha q}Z^{\underline M}E^{\underline a}_{\underline M}(Z(z))=0,
$$
where (as above) $D_{\alpha q}$ is a
spinorial covariant derivative on the worldsheet
supersurface and $E^{\underline a}_{\underline M}$ are components of the
target--space supervielbein.

The content and consequences of this superembedding condition were
analyzed
in the references \cite{gs2,bpstv}.
It was shown that in the case of an $N=2$, $D=3$
superstring the superembedding condition does not contain dynamical
equations
of motion of the superstring coordinates and, hence, a doubly
supersymmetric
action for the $N=2$, $D=3$ superstring can be constructed in the form
analogous to those of the $N=1$ superstrings \cite{to,dghs92}. In the
case
of IIA and IIB superstrings in ten--dimensional space--time the
superembedding
condition contains the dynamical equations, which was demonstrated in
detail in refs. \cite{gs2,bpstv}. This is similar to the appearance of
the
equations of motion in the constraints of higher dimensional
super--Yang--Mills
and supergravity theories, as we have mentioned above. In such cases
standard
superfield actions cannot be constructed, since, for example, if one
tried
to introduce the superembedding condition into an action with a Lagrange
multiplier, Lagrange multiplier components would contain propagating
(redundant) degrees of freedom which would be provided with kinetic
terms by
the dynamical equations contained in the superembedding condition.

Though the standard
superfield actions cannot be constructed in the cases when the
superembedding condition puts a superbrane theory on the mass shell, in
these
cases the superembedding condition contains exhaustive information about
the classical dynamics of the superbranes. This is very useful
when one deals with new brane objects for which a detailed worldvolume
theory has not yet been developed, as it was, for instance with
D--branes,
and especially with the M--theory five--brane \cite{hs1,hs2}.

We now leave the superstrings and, upon the discussion of general
geometrical
grounds of (super)embedding\footnote{Mathematically profound
presentation of surface theory
the reader may find in \cite{kn}, and its extension to superembeddings
in
\cite{bpstv,aspects}.},
turn to the consideration of M--theory branes,
the $D=11$ supermembrane and the super--5--brane, whose classical
properties
are completely determined by the superembedding condition.

\section{Basic elements of (super)surface theory}
\setcounter{equation}0

\subsection{Bosonic embedding}
Consider a $(p+1)$--dimensional bosonic surface ${\cal M}_{p+1}$
embedded
into a $D$--dimensional curved bosonic target space $\underline{\cal
M}_D$.

A local orthogonal basis in the cotangent space of ${\cal M}_{p+1}$ is
given by a vielbein one--form
\begin{equation}\label{4.1}
e^a(\xi^m)=d\xi^me^a_m(\xi),
\end{equation}
which transforms under the vector representation of $SO(1,p)$. The
matrix
$e^a_m(\xi)$ and its inverse $e_a^m(\xi)$ relate the local orthogonal
basis on ${\cal M}_{p+1}$ to the local coordinate basis $d\xi^m$. They
have
the following properties
$$
e^a_m(\xi)e^{mb}(\xi)=\eta^{ab}, \quad \eta^{ab}= (-,+,...,+),
$$
\begin{equation}\label{4.2}
e^a_me^b_n\eta_{ab}=g_{mn}(\xi),
\end{equation}
where $g_{mn}(\xi)$ is a metric on ${\cal M}_{p+1}$.

We assume that the bosonic surface has zero torsion, i.e. its connection
$\omega^{~~a}_b(\xi)=d\xi^m\omega_{mb}^{~~~a}$
satisfies the condition
\begin{equation}\label{4.02}
T^a=de^a+e^b\omega^{~a}_b=0.
\end{equation}

Similarly, in the target space $\underline{\cal M}_D$ we introduce
a local orthogonal cotangent basis given by a vielbein one--form
\begin{equation}\label{4.3}
E^{\underline a}(x^{\underline m})
=dx^{\underline m}E^{\underline a}_{\underline m}(x),
\end{equation}
which transforms under the vector representation of $SO(1,D-1)$ and
whose components satisfy the conditions
$$
E^{\underline a}_{\underline m}E^{\underline m\underline b}
=\eta^{\underline{ab}}, \quad \eta^{\underline{ab}}= (-,+,...,+),
$$
\begin{equation}\label{4.03}
E^{\underline a}_{\underline m}E^{\underline b}_{\underline n}
\eta^{\underline{ab}}=g_{\underline{mn}}(x),
\end{equation}
where $g_{\underline{mn}}(x)$ is a metric on
$\underline{\cal M}_D$ which is assumed to be torsion free as well
\begin{equation}\label{4.4}
T^{\underline a}=dE^{\underline a}
+E^{\underline b}\Omega^{~~\underline a}_{\underline b}=0.
\end{equation}

To embed the surface ${\cal M}_{p+1}$ into the target space
$\underline{\cal M}_D$ means to relate the intrinsic geometry of
${\cal M}_{p+1}$ to the geometry of $\underline{\cal M}_D$.
To this end we consider how the local $\underline{\cal M}_D$ frame
$E^{\underline a}$ behaves on the surface, i.e consider
the pullback of $E^{\underline a}$ on to
${\cal M}_{p+1}$
\begin{equation}\label{4.5}
E^{\underline a}(x(\xi))=d\xi^m\partial_mx^{\underline m}
E^{\underline a}_{\underline m}=e^aE_a^{\underline a},
\end{equation}
where
\begin{equation}\label{4.6}
E_a^{\underline a}(x(\xi))\equiv e_a^m\partial_mx^{\underline m}
E^{\underline a}_{\underline m}.
\end{equation}
Since the orthogonal frame $E^{\underline a}$ is determined up to the
local
$SO(1,D-1)$ transformations,
rotating it by an appropriate $SO(1,D-1)$ matrix
$u^{~\underline a}_{\underline b}(\xi)$,
it is always possible to adapt its pullback on ${\cal M}_{p+1}$
in such a way that (up to a rescaling factor)
\begin{equation}\label{4.7}
E^a\equiv E^{\underline b}u^{~a}_{\underline b}=e^a,
\end{equation}
\begin{equation}\label{4.8}
\quad E^i\equiv E^{\underline b}u^{~i}_{\underline b}
=e^bE_b^{\underline b}u^{~i}_{\underline b}=0,
\end{equation}
where we have split the upper index of the  $SO(1,D-1)$ matrix
$u^{~\underline a}_{\underline b}$ into the index $a=0,1,...,p$
(of the $\underline{\cal M}_D$ directions parallel
to ${\cal M}_{p+1}$) and the index
$i=1,...,D-p-1$ (of the $\underline{\cal M}_D$ directions transverse to
${\cal M}_{p+1}$). These indices are, respectively, the indices
of the subgroups $SO(1,p)$ and $SO(D-p-1)$ of $SO(1,D-1)$
\begin{equation}\label{4.9}
u^{~\underline a}_{\underline b}=\left(u^{~a}_{\underline b},
~u^{~i}_{\underline b}\right).
\end{equation}
Note that, by definition, the $SO(1,D-1)$ matrix
$u^{~\underline a}_{\underline b}$
satisfies the orthogonality conditions
\begin{equation}\label{4.10}
u^{~\underline a}_{\underline c}u^{~\underline b}_{\underline d}
\eta^{\underline{cd}}=\eta^{\underline{ab}}, \quad
u_{\underline a}^{~\underline c}u_{\underline b}^{~\underline d}
\eta_{\underline{cd}}=\eta_{\underline{ab}},
\end{equation}
which are preserved by the independent left and right $SO(1,D-1)_L\times
SO(1,D-1)_R$
transformations of $u^{~\underline a}_{\underline b}$
\begin{equation}\label{4.11}
{\hat u}^{~\underline a}_{\underline b}=
(O_{L})^{~\underline d}_{\underline b}~u_{\underline d}^{~\underline c}~
(O^{-1}_{R})^{~\underline a}_{\underline c}.
\end{equation}

The meaning of eqs. (\ref{4.7}) and (\ref{4.8}) is the following.
When the surface ${\cal M}_{p+1}$ is embedded into $\underline{\cal
M}_D$
the vielbein frame $E^{\underline a}$ of $\underline{\cal M}_D$
can always be chosen in such a way that  the  pullback onto ${\cal
M}_{p+1}$
of $p+1$ of its components coincides with a given local frame $e^a(\xi)$
on ${\cal M}_{p+1}$, and the other $D-p-1$ components of $E^{\underline
a}$
are orthogonal to the surface. With such a choice one identifies
a subgroup $SO(1,p)$ of the target space group $SO(1,D-1)_R$  acting on
$u^{~\underline a}_{\underline b}$ on the right (see eq. (\ref{4.11}))
with
the group $SO(1,p)$ of rotations in the (co)tangent space of
${\cal M}_{p+1}$.

Eqs. (\ref{4.7}) and (\ref{4.8}) are invariant under
the transformations of $SO(1,p) \times SO(D-p-1)$ of the
group $SO(1,D-1)_R$ (\ref{4.11}).
Thus, $SO(1,D-1)_R$ is broken down to $SO(1,p) \times SO(D-p-1)$,
where $SO(D-p-1)$ acts in the subspace of $\underline{\cal M}_D$
transverse
to the surface ${\cal M}_{p+1}$. Therefore, the $SO(1,D-1)$ matrix
(\ref{4.9}) which brings the pullback of
$E^{\underline a}$ to the form (\ref{4.7}) and (\ref{4.8}) is determined
up to the $SO(1,p) \times SO(D-p-1)$ rotations. This means that its
elements,
which are called (Lorentz--vector) harmonics \cite{gikos,sok,bh},
parametrize a coset manifold ${{SO(1,D-1)}\over{SO(1,p) \times
SO(D-p-1)}}$.

We should stress that the left group $SO(1,D-1)_L$ of (\ref{4.11}),
which is
associated with the local Lorentz transformations of the target--space
vielbein (\ref{4.3}), remains
unbroken. This ensures the Lorentz--covariant description of the
embedding.

The equations (\ref{4.7}) and (\ref{4.8})
can be regarded as embedding
conditions. Their structure reflects the fact that the embedded surface
${\cal M}_{p+1}$ breaks the local Lorentz symmetry $SO(1,D-1)_R$
of the target space $\underline{\cal M}_D$ down to
$SO(1,p) \times SO(D-p-1)$ acting in the directions
parallel and transversal to ${\cal M}_{p+1}$. Eq. (\ref{4.7}) defines
the vielbein $e^a$ on ${\cal M}_{p+1}$ as the vielbein induced by
embedding.
Using (\ref{4.03}), (\ref{4.7}) and (\ref{4.8})
one can easily check that $E_a^{\underline a}$ defined in (\ref{4.6})
are orthonormal\footnote{ Notice that in the case of strings
the eq. (\ref{4.12}) is equivalent to the Virasoro constraints.}
\begin{equation}\label{4.12}
E_a^{\underline a}E_{b\underline a}=\eta_{ab},
\end{equation}
and, hence,
the
${\cal M}_{p+1}$ metric (\ref{4.2}) is the induced metric
\begin{equation}\label{4.13}
g_{mn}(\xi)=\partial_mx^{\underline m}\partial_nx^{\underline n}
g_{\underline{mn}}(x(\xi)).
\end{equation}

Because of the orthonormality property (\ref{4.12}) the components of
$E_a^{\underline a}$ can be identified with the $u_a^{~\underline a}$
components of the $SO(1,D-1)$ harmonic matrix inverse to (\ref{4.9})
(as usual, the
Lorentz indices are raised and lowered by the Minkowski metrics)
\begin{equation}\label{4.14}
E_a^{\underline a}(x(\xi))=u_a^{~\underline a}(\xi)\equiv
\eta^{\underline{ab}}u^{~b}_{\underline b}\eta_{ba}.
\end{equation}
Remember that by virtue of (\ref{4.10}) the matrix inverse to
the orthogonal matrix $u^{~\underline a}_{\underline b}$ is
\begin{equation}\label{4.0014}
(u^{-1})^{~\underline a}_{\underline b}=\eta^{\underline{ac}}
u^{~\underline d}_{\underline c}\eta_{\underline{db}}
\equiv (u_a^{~\underline a},u_i^{~\underline a}).
\end{equation}
To simplify notation we shall always skip the superscript ``-1'' of the
components of the inverse $SO(1,D-1)$ matrices when their lower index is
decomposed into the $SO(1,p)\times SO(D-p-1)$ indices. This should not
lead to a confusion if we keep in mind that only
the upper index of the `direct' $SO(1,D-1)$ matrices can be subject to
$SO(1,p)\times SO(D-p-1)$ splitting, as in eq. (\ref{4.9}).

We shall now show that the spin connection $\omega^{~a}_b$ of the
surface
${\cal M}_{p+1}$ is related to the pullback of the spin connection
$\Omega^{~\underline a}_{\underline b}$ of $\underline{\cal M}_D$
by local $SO(1,D-1)$ transformations of the latter.

Consider the integrability condition of eq. (\ref{4.5}) by
taking the ${\underline{\cal M}}_D$
external covariant differential $\nabla=d+\Omega$ of its
left-- and  right--hand side. Then, in view
of (\ref{4.4}), (\ref{4.02}) and (\ref{4.14}), we get
\begin{equation}\label{4.015}
e^b\omega^{~a}_bu_{a}^{~\underline b}=
e^b{\nabla} E_b^{~\underline b}\equiv
e^b (du_b^{~\underline b}
+u_{b}^{~\underline a}\Omega^{~\underline b}_{\underline a}).
\end{equation}
Multiplying eq. (\ref{4.015}) by $u^{~a}_{\underline b}$
and taking into account (\ref{4.10}) and (\ref{4.14}) we arrive at
the expression\footnote{Remember that the spin connections are
antisymmetric
with respect to the Lorentz--group indices.
Note also that with such a choice of $\omega_b^{~a}$ the covariant
differential
$\left({\hat\nabla} u_b^{~\underline b}\right)u^{~a}_{\underline b}=
(du_b^{~\underline b}
+u_b^{~\underline a}\Omega^{~\underline b}_{\underline a}
-\omega_b^{~c}u_c^{~\underline b})
u^{~a}_{\underline b}$, which contains both connections, is identically
zero $\left({\hat\nabla} u_b^{~\underline b}\right)u^{~a}_{\underline b}
\equiv0$.}
\begin{equation}\label{4.15}
\omega^{~a}_b=\left({\nabla} u_b^{~\underline
b}\right)u^{~a}_{\underline b}
\equiv
du_b^{~\underline b}u^{~a}_{\underline b}
+u_b^{~\underline a}\Omega^{~\underline b}_{\underline a}
u^{~a}_{\underline b},
\end{equation}
where $du_b^{~\underline b}u^{~a}_{\underline b}$ are $SO(1,p)$-valued
components of the $SO(1,D-1)$ Cartan form
\begin{equation}\label{4.16}
\Omega_{0\underline a}^{~~\underline b} \equiv
d(u^{-1})_{\underline a}^{~\underline c}u^{~\underline b}_{\underline
c}=
-(u^{-1})_{\underline a}^{~\underline c}du^{~\underline b}_{\underline
c}
\end{equation}
($(u^{-1})_{\underline a}^{~\underline c}$
was defined in (\ref{4.0014})).
$\Omega_{0\underline a}^{~~\underline b}$ identically
satisfies the Maurer--Cartan (zero--curvature) equations
\begin{equation}\label{maurer}
d\Omega_{0\underline a}^{~~\underline b}
+\Omega_{0\underline a}^{~~\underline c}
\Omega_{0\underline c}^{~~\underline b}=0.
\end{equation}
The integrability of the Maurer--Cartan equations is the basis for
relating
(super)branes to integrable systems
\cite{lr,om,barnes,zhelt,hoppe}--\cite{wz}.

As a generalization of (\ref{4.16}) we also introduce a `covariantized
Cartan'
form
(which is, of course, nothing but the Lorentz--transformed
spin connection $\Omega$ on $\underline{\cal M}_D$)
\begin{equation}\label{hatom}
\hat\Omega_{\underline a}^{~\underline b}\equiv \nabla
(u^{-1})_{\underline a}^{~\underline c}u^{~\underline b}_{\underline c}=
d(u^{-1})_{\underline a}^{~\underline c}u^{~\underline b}_{\underline c}
+(u^{-1})_{\underline a}^{~\underline c}\Omega^{~\underline
d}_{\underline c}
u_{\underline d}^{~\underline b}.
\end{equation}
Written in terms of the connection $\hat\Omega$,
the curvature two--form on $\underline{\cal M}_D$
\begin{equation}\label{curva}
R_{\underline a}^{~\underline b}\equiv
d\Omega_{\underline a}^{~\underline b}
+\Omega_{\underline a}^{~\underline c}
\Omega_{\underline c}^{~\underline b}
\end{equation}
takes the form
\begin{equation}\label{maurer1}
(u^{-1})_{\underline a}^{~\underline c}
R_{\underline c}^{~\underline d}u_{\underline d}^{~\underline b}=
d\hat\Omega_{\underline a}^{~\underline b}
+\hat\Omega_{\underline a}^{~\underline c}
\hat\Omega_{\underline c}^{~\underline b}.
\end{equation}

From  (\ref{4.15}), (\ref{hatom}) and (\ref{maurer1}) it follows that
the curvature $R_a^{~b}$ of the surface ${\cal M}_{p+1}$ is
\begin{equation}\label{curva1}
R_a^{~b}\equiv d\omega^{~b}_a+\omega^{~c}_a\omega^{~b}_c=
 u_{a}^{~\underline c}
R_{\underline c}^{~\underline d}u_{\underline d}^{~b}-\hat\Omega_a^{~i}
\hat\Omega_i^{~b},
\end{equation}
where $i=1,...,D-p-1$ are the transverse indices (see eq. (\ref{4.9}))
and
\begin{equation}\label{hatomi}
\hat\Omega_a^{~i}=\left(\nabla u_a^{~\underline b}\right)u_{\underline
b}^{~i}
\equiv d u_a^{~\underline b}u_{\underline b}^{~i}
+u_{a}^{~\underline c}\Omega^{~\underline d}_{\underline c}
u_{\underline d}^{~i}, \quad
\hat\Omega_i^{~a}=-\delta_{ij}\eta^{ab}\hat\Omega_b^{~j}.
\end{equation}
Note that though only the target--space covariant derivative enters the
expression (\ref{hatomi}) for $\hat\Omega_a^{~i}$, it is also covariant
with respect to the local $SO(1,p)$ transformations on the surface,
since
the term with the surface spin connection is identically zero because
of the orthogonality properties of the harmonics
$$
\omega_a^{~b}u_b^{~\underline a}u_{\underline a}^{~i}\equiv 0.
$$

In surface theory the equation (\ref{curva1}), which expresses the
surface curvature in terms of the target--space connection and
curvature,
is called the Gauss equation. The other two determining equations
of surface theory contained in (\ref{maurer1}) are the Codazzi and Ricci
equations (see \cite{kn,barnes} for details).
We shall present the explicit form
of these equations for superembeddings in the next subsection.

Note that when the target space is flat
$R^{~\underline b}_{\underline a}=0$ and
$\Omega_{\underline a}^{~\underline b}$ can be gauge fixed to zero,
the induced spin connection
$\omega^{~b}_a$ on ${\cal M}_{p+1}$ coincides with the $SO(1,p)$
components
of the Cartan form $\Omega_{0}$
\begin{equation}\label{4.17}
\omega^{~~b}_a=du_a^{~\underline b}u^{~b}_{\underline b}
\end{equation}
and
$$
R_{ab}=\Omega_{0a}^{~~i}\Omega_{0b}^{~~i} \quad {\rm in~~flat}
\qquad\underline{\cal M}_D.
$$

Let us now derive the restriction on components of $\hat\Omega_a^{~~i}$
(\ref{hatomi}) which follows from the integrability condition
(\ref{4.015}).
To this end we multiply (\ref{4.015}) by $u^{~i}_{\underline b}$, then
the left hand side of (\ref{4.015}) becomes zero and we get
\begin{equation}\label{4.18}
e^a\hat\Omega_a^{~~i}=e^ae^b\hat\Omega_{b,a}^{~~i}=0 \quad \Rightarrow
\quad \hat\Omega_{b,a}^{~~i}=\hat\Omega_{a,b}^{~~i},
\end{equation}
which implies that $\hat\Omega_{a,b}^{~~i}$  is symmetric in the
$SO(1,p)$
indices $a$ and $b$. This symmetric matrix is called the second
fundamental
form of the surface, and its trace $h^i=\eta^{ab}\hat\Omega_{a,b}^{~~i}$
characterizes average extrinsic curvatures of the surface in the
target space \cite{kn,barnes}.

To summarize, the conditions (\ref{4.7}) and (\ref{4.8}) whose
integrability
leads to eqs. (\ref{4.15}), (\ref{maurer1}) (containing
(\ref{curva1})), and (\ref{4.18}) completely determine the
geometrical properties of embedding a general (pseudo) Riemann surface
${\cal M}_{p+1}$  into
a (pseudo) Riemann manifold $\underline{\cal M}_D$ in terms of the
vielbein, connection and curvature forms of $\underline{\cal M}_D$
pulled back
on to ${\cal M}_{p+1}$. In other words eqs. (\ref{4.7}), (\ref{4.8}),
(\ref{4.15}) and (\ref{4.18}) determine the induced geometry on the
embedded surface.

From the physical point of view the induced geometry of the surface
means
that, though the surface is a curved manifold and hence contains a
$(p+1)$--dimensional gravity,
this gravity is not `fully fledged' in the sense that
the graviton on ${\cal M}_{p+1}$ is described by the induced metric
(\ref{4.13}) and, hence, is a composite field built from the
worldsurface scalars $x^{\underline m}(\xi)$ \footnote{An interpretation
of General Relativity as induced gravity in the worldvolume of a
brane has been discussed in \cite{induced}.}.

We would like the embedded surface ${\cal M}_{p+1}$ to be associated
with
the worldvolume of a p--brane. The worldvolume which describes the
classical
motion of the p--brane in the target space is of a minimal volume. This
is an additional condition of embedding ${\cal M}_{p+1}$ which is
equivalent to the p--brane equations of motion.

In the geometrical language the minimal embedding of
${\cal M}_{p+1}$ is specified by the requirement that the trace of
the second fundamental form
$\hat\Omega_{a,b}^{~~i}$ determined in (\ref{hatomi}) and
(\ref{4.18}) is zero
\begin{equation}\label{4.19}
\eta^{ab}\hat\Omega_{a,b}^{~~i}=0,
\end{equation}
which implies that the average extrinsic curvatures of ${\cal M}_{p+1}$
in
$\underline{\cal M}_D$ are zero.

Let us now show that eq. (\ref{4.19}) is equivalent to the p--brane
equations
of motion which one gets, for example, from the Nambu--Goto action
(\ref{2}).
Using the definition (\ref{hatomi}) of $\hat\Omega_{a,b}^{~~i}$
and the relations
(\ref{4.14}), (\ref{4.2}), (\ref{4.03}) and  (\ref{4.13})
we can rewrite eq. (\ref{4.19}) as follows
\begin{eqnarray}\label{4.20}
\eta^{ab}\hat\Omega_{a,b}^{~~i}&=&
\eta^{ab}\left(\nabla_aE_b^{\underline b}\right)u^{~i}_{\underline b}
=\eta^{ab}\nabla_a\left(e^m_b\partial_mx^{\underline m}
E_{\underline m}^{\underline a}\right)E_{\underline a\underline n}
E^{\underline n \underline b}u^{~i}_{\underline b}\\ \nonumber
&=&\eta^{ab}e_a^n\partial_n\left(e^m_b\partial_mx^{\underline m}
g_{\underline{mn}}(x)
\right)E^{\underline n \underline b}u^{~i}_{\underline b}+
\eta^{ab}e^m_a\partial_mx^{\underline m}g_{\underline{mn}}
\left(\nabla_bE^{\underline n \underline b}\right)u^{~i}_{\underline b}
\\ \nonumber
&=& \left[{1\over{\sqrt{-g}}}\partial_m
\left(\sqrt{-g}g^{mn}\partial_nx^{\underline m}
g_{\underline{mn}}(x)\right)
-g^{mn}\partial_mx^{\underline m}\partial_nx^{\underline l}g_{\underline
m
\underline p}
\Gamma_{\underline l\underline n}^{\underline p}\right]
E^{\underline n \underline b}u^i_{\underline b}=0,
\end{eqnarray}
where $\nabla_a=e_a^m(\partial_m
+\partial_mx^{\underline m}\Omega_{\underline m})$,
and
$\Gamma_{\underline l\underline n}^{\underline p}(x)$ is the
D-dimensional
Christoffel symbol
\begin{equation}\label{4.21}
\Gamma_{\underline l\underline n}^{\underline p}(x)\equiv
(\nabla_{\underline l}E^{\underline p}_{\underline b})
E^{\underline b}_{\underline n}
=\left(\partial_{\underline l}E^{\underline p}_{\underline b}
-\Omega_{\underline l \underline b}^{~~\underline c}
E^{\underline p}_{\underline c}\right)
E^{\underline b}_{\underline n}.
\end{equation}

In the square brackets of the last line of (\ref{4.20}) one can
recognize
the standard equations of motion of a bosonic p--brane propagating
in a D--dimensional target space. In (\ref{4.20}) they are projected on
to
the $D-p-1$ directions transverse to the p--brane. This projection
reflects the
fact that, because of the (p+1)--dimensional worldvolume
reparametrization invariance,
among the $D$ equations obtained from the Nambu--Goto action (\ref{2})
only $D-p-1$ transverse equations
are independent, and these are explicitly singled out
by the geometrical embedding condition (\ref{4.19}), (\ref{4.20}),
while the other $p+1$ equations (`parallel' to the brane) are
identically
satisfied. To see this let us in the last line
of (\ref{4.20}) substitute
$u^{~i}_{\underline b}$
with $u^{~c}_{\underline b}$, which corresponds to projecting the
equations
of motion of $x^{\underline m}(\xi)$ along the brane worldvolume. Then
`moving' in the direction opposite to that which we followed developing
eq. (\ref{4.20}) we get
$$
\left[{1\over e}\partial_m\left(eg^{mn}\partial_nx^{\underline m}
g_{\underline{mn}}(x)\right)
-g^{mn}\partial_mx^{\underline m}\partial_nx^{\underline l}g_{\underline
m
\underline p}
\Gamma_{\underline l\underline n}^{\underline p}\right]
E^{\underline n \underline b}u^{~c}_{\underline b}=
$$
$$
=\eta^{ab}\left[\left(\nabla_a
E_{b}^{\underline a}\right)u^{~c}_{\underline
a}-e_a^m\omega_{mb}^{~~c}\right]
=\eta^{ab}\left[\left(\nabla_a
u_{b}^{~\underline a}\right)u^{~c}_{\underline a}
-e_a^m\omega_{mb}^{~~c}\right]
\equiv 0,
$$
which is identically zero by virtue of the
embedding condition (\ref{4.15}) for the ${\cal M}_{p+1}$ spin
connection.

Thus, in the embedding approach the bosonic p--brane equations of motion
are described by the minimal embedding condition (\ref{4.19}) on
components
of the ${\cal M}_{p+1}$ pullback (\ref{hatom})
of the target space connection form
$\hat\Omega_{\underline a}^{~\underline b}$.

We shall now generalize the results of this subsection to the
description
of superembeddings \cite{bpstv,hs1}.

\subsection{Superembedding}
Consider a supermanifold $\underline{\cal M}_{D,2n}$ locally
parametrized
by supercoordinates $Z^{\underline M}
=(X^{\underline m},\Theta^{\underline\mu})$ $({\underline m}=0,...,D-1;~
{\underline\mu}=1,...,2n)$.
When $\underline{\cal M}_{D,2n}$ is
flat the index ${\underline\mu}$ is associated
with an irreducible spinor representation of $SO(1,D-1)$ and, in
addition,
can also include
an internal index corresponding to $N$-extended supersymmetry,
as in the case of the $N=2$, $D=10$ superstrings (Section 3).

We assume that the supergeometry on $\underline{\cal M}_{D,2n}$ is such
that
it describes a $D$--dimensional supergravity (see Subsections 2.7 and
3.3).
This implies that
the structure group of the tangent superspace of $\underline{\cal
M}_{D,2n}$
is $SO(1,D-1)$ (or more precisely $Spin(1,D-1)$) and the supervielbein
components
\begin{equation}\label{sviel}
E^{\underline A}(Z)=dZ^{\underline M}E^{\underline A}_{\underline M}=
(E^{\underline a}, E^{\underline\alpha})
\end{equation}
satisfy the torsion constraint (\ref{torsionc1}).

We are interested in embedding into the target superspace
$\underline{\cal M}_{D,2n}$
a supersurface ${\cal M}_{p+1,n}$ locally parametrized by
supercoordinates $z^M=(\xi^m,\eta^{\mu})$.
With such an embedding we would like to describe the dynamics of
superbranes whose
$n$-dimensional $\kappa$--symmetry is replaced by more fundamental
worldvolume supersymmetry. So,
we take a supersurface with
the number of fermionic directions
$\eta^{\mu}$ which is half the number of the
target superspace fermionic directions.\footnote{One can study the
embedding
of a more general class of supersurfaces with the dimension $d_f$ of the
odd
(fermionic) subspace being less than $n$. As we have seen with the
examples
of superparticles (Subsection 2.5.4) and superstrings (Subsection
3.2.1),
such embeddings may also describe superbranes which preserve half
of the target--space supersymmetries, but there may exist
superembeddings which describe BPS states preserving a lower number
of target space supersymmetries. Let us also mention that the embedding
of supersurfaces with $d_f>n$, in general, seems to be too
restrictive to describe physically interesting models
(see, however, \cite{1/4}).}.

We should stress that the consideration of general properties of
superembeddings carried out in this subsection is schematic, as far as
spinors and $\Gamma$--matrices are concerned, since it is pretty hard
to describe with one and the same set of exact formulae different
types of spinor representations (Majorana, Majorana--Weyl,
simplectic--Majorana, etc.) which are used in each given case of
superembedding depending on $D$ and $p$.

A local cotangent frame on ${\cal M}_{p+1,n}$ is given by the
supervielbein
\begin{equation}\label{4.022}
e^A(z)=dz^Me_M^A=(e^a,~e^{\alpha q})
\end{equation}
whose components are transformed
under vector and spinor representations of the
structure group $SO(1,p)\times SO(D-p-1)$ of ${\cal M}_{p+1,n}$
(remember that $D$ is the bosonic dimension of the target superspace).
The index $a=0,1,...,p$ is the index of the vector representation of
$SO(1,p)$, while  $\alpha$ and $q$ are the indices of
irreducible spinor representations
of $SO(1,p)$ and $SO(D-p-1)$, respectively, such that
$\dim{Spin(1,p)}\times\dim{Spin(D-p-1)}=n$. These requirements on
the structure group of ${\cal M}_{p+1,n}$ and its spinor representations
put restrictions on the dimensions of supersurfaces which can be
embedded
into a given target superspace. The analysis of possible superembeddings
of this kind shows that they correspond to all known superbranes
(super--p--branes, D--branes and M--branes) and their dimensional
reductions
(possibly accompanied by a dualization).
This results in a brane scan \cite{hs1} similar to the standard one
\cite{scan}.

At the moment we do not make any assumption concerning
the supergeometry of
${\cal M}_{p+1,n}$, since, as we shall see, it is determined by the
superembedding condition.

Consider the pullback on to ${\cal M}_{p+1,n}$ of the
$\underline{\cal M}_{D,2n}$
supervielbein $E^{\underline A}$
\begin{equation}\label{4.22}
E^{\underline a}=
e^Ae^{~M}_A\partial_MZ^{\underline M}E_{\underline M}^{\underline
a}\equiv
e^a E_a^{\underline a}+e^{\alpha q} E_{\alpha q}^{\underline a},
\end{equation}
\begin{equation}\label{4.23}
E^{\underline\alpha}=
e^Ae^{~M}_A\partial_MZ^{\underline M}
E_{\underline M}^{\underline \alpha}\equiv
e^a E_a^{\underline \alpha}+e^{\alpha q} E_{\alpha
q}^{\underline\alpha},
\end{equation}
where $e^{~M}_A(z)$ is the supervielbein matrix inverse to
$e_{M}^{~A}(z)$.

As in the bosonic case we postulate that the superembedding is
such that with an appropriate choice of the local frame $e^a$ on
${\cal M}_{p+1,n}$ and by the use
of local $SO(1,D-1)$ rotations (\ref{4.9})--(\ref{4.11}) one can
direct the vector component pullback (\ref{4.22}) of
$E^{\underline a}$ on the
supersurface such that
\begin{equation}\label{4.07}
E^a\equiv E^{\underline b}u^{~a}_{\underline b}(z)=e^a(z),
\end{equation}
\begin{equation}\label{4.08}
\quad E^i\equiv E^{\underline b}u^{~i}_{\underline b}(z)
=e^bE_b^{\underline b}u^{~i}_{\underline b}=0.
\end{equation}
Eqs. (\ref{4.22}), (\ref{4.07}) and (\ref{4.08}) imply
that
$e^a(z)$ is the induced supervielbein on
${\cal M}_{p+1,n}$
since it coincides with the pullback of $E^{\underline a}$.
We also see that
\begin{equation}\label{4.014}
E_a^{\underline a}(Z(z))=u_a^{~\underline a}(z).
\end{equation}

But what is most important, in view of (\ref{4.22})--(\ref{4.08}),
we get the basic superembedding condition
\begin{equation}\label{4.24}
E_{\alpha q}^{\underline a}\equiv e_{\alpha q}^M\partial_MZ^{\underline
M}
E_{\underline M}^{\underline a}=0.
\end{equation}
Thus, the superembedding condition (\ref{4.24}) is a natural consequence
of the embedding carried out in such a way that
the bosonic {\sl cotangent} subspace of  ${\cal M}_{p+1,n}$ is a
subspace of
the bosonic {\sl cotangent} subspace of $\underline{\cal M}_{D,2n}$.
Alternatively,
eq. (\ref{4.24}) implies that the fermionic {\sl tangent} subspace of
${\cal M}_{p+1,n}$ is a subspace of the fermionic {\sl tangent} subspace
of
$\underline{\cal M}_{D,2n}$.

The embedding conditions (\ref{4.07}),  (\ref{4.08}) and  (\ref{4.24})
imply that
(as in the bosonic case (\ref{4.13})) the supersurface metric
\begin{equation}\label{4.024}
g_{mn}(z)\equiv e^{~a}_me^{~b}_n\eta_{ab}
=\partial_mZ^{\underline M}E_{\underline M}^{\underline a}
\partial_nZ^{\underline N}
E_{\underline N}^{\underline b}\eta_{\underline{ab}}
=E_{m}^{\underline a}E_{n\underline a}
\end{equation}
is the induced metric.

In addition to the vector components
the supervielbein (\ref{sviel}) contains the spinor components
(\ref{4.23}).
When the vector components are transformed by the  $SO(1,D-1)$ matrix
$u_{\underline b}^{~\underline a}$, the
spinor components
must be {\sl simultaneously} transformed by a matrix
$v_{\underline\beta}^{~\underline\alpha}$ of the spinor representation
of $SO(1,D-1)$. The matrices $u_{\underline b}^{~\underline a}$ and
$v_{\underline\beta}^{~\underline\alpha}$ are related to each other by
the standard $D$--dimensional $\Gamma$--matrix defining conditions
\begin{equation}\label{4.25}
\Gamma^{\underline{b}}_{\underline{\gamma}\underline{\delta}}
u^{~\underline{a}}_{\underline{b}}
\equiv
v^{~\underline{\alpha}}_{\underline{\gamma}}
\Gamma^{{\underline a}}_{\underline{\alpha}\underline{\beta}}
v^{~\underline{\beta}}_{\underline{\delta}}.
\end{equation}
Eq. (\ref{4.25}) implies that $u_{\underline b}^{~\underline a}$ are
composed
of $v_{\underline\beta}^{~\underline\alpha}$, which have $D(D-1)/2$
independent components corresponding to $D(D-1)/2$ independent
generators
of $Spin(1,D-1)$.

The choice of the $\Gamma$--matrices depends on the dimensions
$D$ and $p$, but
for the embedding under consideration they can always be
{\sl schematically} represented
by the following $SO(1,p)\times SO(D-p-1)$ invariant set of
matrices (compare with eq. (\ref{2.5.39})).
In the directions parallel to the supersurface
\begin{equation}\label{4.26}
\Gamma^a_{\underline\alpha\underline\beta}=\left(
\begin{array}{cc}
\gamma^a_{\alpha\beta}\delta_{q}^{r}  &  0\\
0  &  \gamma^{a\alpha\beta}\delta_{q'}^{r'}
\end{array}
\right)\,, \quad a=0,1,...,p\, ,
\end{equation}
and in the directions normal to the supersurface
\begin{equation}\label{4.27}
\Gamma^i_{\underline\alpha\underline\beta}=\left(
\begin{array}{cc}
0  &  \delta^\alpha_\beta\gamma^i_{qq'}\\
\delta_\alpha^\beta(\gamma^i)^{q'q}   &  0
\end{array}
\right)\,, \quad i=1,...,D-p-1\,,
\end{equation}
where $\gamma^a_{\alpha\beta}$ are $d=p+1$ $\gamma$--matrices, and
the matrices $\gamma^i_{qq'}$ obey the $SO(D-p-1)$ Clifford algebra,
the indices $q$ and $q'$ corresponding to (in general non--equivalent)
spinor representations of $SO(D-p-1)$. Note that in (\ref{4.26}) and
(\ref{4.27}) the $SO(1,D-1)$ spinor indices $\underline\alpha$,
$\underline\beta$ split into the $SO(1,p)\times SO(D-p-1)$ spinor
indices
$(\alpha q)$ of the ${\cal M}_{p+1,n}$ odd subspace and the
$SO(1,p)\times SO(D-p-1)$ spinor indices
$\alpha':=(\alpha q')$ `transverse' to the supersurface.

The spinor components of the supervielbein pullback (\ref{4.23})
transformed
with
\begin{equation}\label{4.028}
v_{\underline\beta}^{~\underline\alpha}(z)
\equiv
\left(v_{\underline\beta}^{~\alpha q},
~v_{\underline\beta}^{~\alpha'}\right)
\end{equation}
take the form
\begin{equation}\label{4.28}
E^{\underline\beta}v_{\underline\beta}^{~\underline\alpha}=
\left(E^{\underline\beta}v_{\underline\beta}^{~\alpha q},
E^{\underline\beta}v_{\underline\beta}^{~\alpha'} \right),
\quad {\rm where} \quad v_{\underline\beta}^{~\alpha'}\equiv
v_{\underline\beta,\alpha q'}.
\end{equation}
In the same way as we have done with the vector basis (\ref{4.07}),
we can always make a choice of a spinor basis  $e^{\alpha q}$
on ${\cal M}_{p+1,n}$ such that
\begin{equation}\label{4.29}
E^{\underline\beta}v_{\underline\beta}^{\alpha q}=e^{\alpha q},
\end{equation}
which, in view of (\ref{4.23}), implies
\begin{equation}\label{4.30}
E_{a}^{\underline\beta}v_{\underline\beta}^{~\alpha q}=0,
\end{equation}
\begin{equation}\label{4.31}
E_{\beta r}^{\underline\beta}v_{\underline\beta}^{~\alpha q}
=\delta^{~\alpha}_\beta\delta^{~q}_r,
\end{equation}
and, hence,
\begin{equation}\label{4.32}
E^{\underline\beta}v_{\underline\beta}^{~\alpha'}\equiv
e^{\alpha q}h_{\alpha q}^{~~\alpha'}+
e^aE_a^{\underline\beta}v_{\underline\beta}^{~\alpha'},
\end{equation}
where
\begin{equation}\label{4.33}
h_{\alpha q}^{~~\alpha'}(z)\equiv h_{\alpha q,\beta q'}(z),
\end{equation}
is an $SO(1,p)\times SO(D-p-1)$--valued matrix on ${\cal M}_{p+1,n}$.
From eqs. (\ref{4.30})--(\ref{4.33}) we derive that
\begin{equation}\label{4.34}
E_{\alpha q}^{\underline\beta}=
v_{\alpha q}^{~~\underline\beta}
+h_{\alpha q}^{~~\alpha'}~v^{~~\underline\beta}_{\alpha'},
\end{equation}
where by $v_{\alpha q}^{~~\underline\beta}$ and
$v^{~\underline\alpha}_{\alpha'}$ we have defined the
components of the matrix
\begin{equation}\label{4.35}
(v^{-1})_{\underline\beta}^{~\underline\alpha}\equiv
\left(v^{~~\underline\alpha}_{\alpha q},
~v^{~\underline\alpha}_{\alpha'}\right)
\end{equation}
inverse to
$v_{\underline\alpha}^{~\underline\beta}\equiv
\left(v_{\underline\alpha}^{~\alpha q},
~v_{\underline\alpha}^{~\alpha'}\right)$.

The inverse matrices (\ref{4.0014}) and (\ref{4.35}) are related by the
equation similar to (\ref{4.25})
\begin{equation}\label{4.36}
\Gamma^{\underline{b}}_{\underline{\gamma}\underline{\delta}}
(u^{-1})^{~\underline{a}}_{\underline{b}}
\equiv
(v^{-1})^{~\underline{\alpha}}_{\underline{\gamma}}
\Gamma^{{\underline a}}_{\underline{\alpha}\underline{\beta}}
(v^{-1})^{~\underline{\beta}}_{\underline{\delta}}.
\end{equation}

Note that the embedding conditions (\ref{4.07}), (\ref{4.08}),
(\ref{4.29})
and (\ref{4.32}) are defined up to the local $SO(1,p)\times SO(D-p-1)$
rotations which reduce the number of independent components of
$u^{~\underline{a}}_{\underline{b}}(z)$ and
$v_{\underline\alpha}^{~\underline\beta}(z)$ from $D(D-1)/2$ down to
$D(D-1)/2-(p+1)p/2-(D-p-1)(D-p-2)/2=(p+1)(D-p-1)$.
Hence, as in the bosonic case, the matrices
$u^{~\underline{a}}_{\underline{b}}$ and
$v_{\underline\alpha}^{~\underline\beta}(z)$ are, respectively,
Lorentz--vector
and Lorentz--spinor harmonics which parametrize the coset space
 ${{SO(1,D-1)}\over{SO(1,p) \times SO(D-p-1)}}$ \cite{ghs,gds,bpstv}.

As we shall see in Section 5 the physical meaning of the
supersurface field
$h_{\alpha q}^{~~\alpha'}(z)$ is that it appears in (\ref{4.32})
due to the presence of gauge fields
on the worldvolume of superbranes, such as Dirac--Born--Infeld vector
fields
of the D--branes and the self--dual antisymmetric gauge field
of the M--5--brane. $h_{\alpha q}^{~~\alpha'}(z)$ is expressed in terms
of the field strengths of these fields \cite{hs1,hs2,bst}.

For the ordinary super--p--branes (where there is no worldvolume matter
other than scalar and spinor fields
associated with the transverse oscillations of the brane
in the target superspace) $h_{\alpha q}^{~\alpha'}(z)$ is either
auxiliary (as in the case of an $N=1$, $D=4$ supermembrane \cite{hrs})
or zero
\cite{bpstv}.
We shall demonstrate this for the
$N=1$ superparticles and superstrings in the next subsection,
and for the $D=11$ supermembrane in Subsection 5.1.

We now consider the consequences of the superembedding conditions
(\ref{4.07})--(\ref{4.24}), (\ref{4.29})--(\ref{4.34}) for the
properties
of the induced supergeometry of ${\cal M}_{p+1,n}$.
As in the bosonic case (eq. (\ref{4.015})),  we take the
$\underline{\cal M}_{D,2n}$ external differential $\nabla=d+\Omega$ of
the vector supervielbein pullback (\ref{4.22}).
Then, in view of (\ref{4.014}),
(\ref{4.24}) and the target space
supergravity torsion constraint
\begin{equation}\label{torsionc2}
T^{\underline a}\equiv \nabla E^{\underline a}=dE^{\underline a}
+E^{\underline b}\Omega_{\underline b}^{~\underline a}
=-iE^{\underline\alpha}\Gamma^{\underline a}_
{\underline\alpha\underline\beta}E^{\underline\beta},
\end{equation}
where all other components of the two--form $T^{\underline a}$ are
(conventionally) put to zero \footnote{Provided that
the basic torsion constraint
$T^{\underline a}_{\underline\alpha\underline\beta}
=-2i\Gamma^{\underline a}_
{\underline\alpha\underline\beta}$ is imposed, all other torsion
constraints
are derived by solving for the torsion Bianchi identities. In addition,
by redefining the supervielbeins and superconnection one can always put
all components of $T^{\underline a}$ except for
$T^{\underline a}_{\underline\alpha\underline\beta}$ to zero
(see \cite{sohnius}--\cite{bandzim}, \cite{10sg,11sg,cal}
for a detailed analysis of supergravity constraints in various
dimensions).},
we get
\begin{equation}\label{4.37}
-iE^{\underline\alpha}\Gamma^{\underline b}_
{\underline\alpha\underline\beta}E^{\underline\beta}=
T^au_{a}^{~\underline b}-e^b\left(\omega^{~a}_bu_{a}^{~\underline
b} -{\nabla} u_b^{~\underline b}\right).
\end{equation}
Remember that $T^a={1\over 2}e^Be^AT^a_{AB}=de^a+e^b\omega_b^{~a}$ is
the supersurface torsion and $E_a^{~\underline b}=u_{a}^{~\underline b}$
(\ref{4.014}).

 From (\ref{4.37}) it follows that
(up to an appropriate redefinition
of $T^a$)
the ${\cal M}_{p+1,n}$ spin connection $\omega^{~a}_b$ can always be
identified with ${\nabla} u_b^{~\underline b}u_{\underline b}^{~a}$,
as in the bosonic case (\ref{4.15})
\begin{equation}\label{4.038}
\omega^{~a}_b={\nabla} u_b^{~\underline b}u_{\underline b}^{~a}
\equiv (du_b^{~\underline b}+u_b^{~\underline a}
\Omega_{\underline a}^{~\underline b})u_{\underline b}^{~a}\equiv
\hat\Omega_b^{~a}.
\end{equation}
Then (\ref{4.37}) reduces to
\begin{equation}\label{4.38}
T^a\equiv{1\over 2}e^Be^AT^a_{AB}
=-iE^{\underline\alpha}\Gamma^{\underline a}_
{\underline\alpha\underline\beta}E^{\underline\beta}u_{\underline
a}^{~a}
\equiv T^{\underline a}u_{\underline a}^{~a}
\end{equation}
and
\begin{equation}\label{4.39}
e^a{\nabla} E_a^{~\underline b}u_{\underline b}^{~i}=
e^a{\nabla} u_a^{~\underline b}u_{\underline b}^{~i}
\equiv e^a\hat\Omega_a^{~i}= -iE^{\underline\alpha}\Gamma^{\underline
a}_
{\underline\alpha\underline\beta}E^{\underline\beta}u_{\underline
a}^{~i}.
\end{equation}


The equations (\ref{4.38}) and (\ref{4.39}) are the integrability
conditions
of the superembedding conditions (\ref{4.08}), (\ref{4.24}),
(\ref{4.29})
and (\ref{4.038}).

In particular, eq. (\ref{4.38}) implies that the superembedding
conditions
(\ref{4.24}), (\ref{4.29}) and (\ref{4.038}) require
the supersurface torsion to satisfy constraints, whose form depends
on the concrete superembedding under consideration \cite{bpstv}
(see Section 5).

Thus, if target space supergeometry is constrained to be of the
supergravity type, the superembedding induces supergeometry
on the supersurface such that it corresponds to a worldvolume (induced)
supergravity. The supermultiplet of the induced supergravity includes
the graviton and gravitino which are composed from
components of a worldvolume matter supermultiplet. The matter
supermultiplet
consists of
the $D-p-1$ scalar modes $x^i(\xi)$ and the $n\over 2$
spinor modes $\theta^{\alpha'}$ of the brane transverse fluctuations,
and it may also contain worldvolume gauge fields (as in the case of the
D--branes and the M5--brane).

Let us now briefly discuss the consequences of the integrability
equation
(\ref{4.39}).
In each given case of superembedding, eq. (\ref{4.39}) allows one
to find out whether or not the superembedding condition (\ref{4.24})
contains the superbrane equations of motion.

 From eq. (\ref{4.39}), in view of
(\ref{4.25}), (\ref{4.27}), (\ref{4.29}) and (\ref{4.32}), it
follows that, as in the bosonic case (\ref{4.18}), the vector component
of the superform $\hat\Omega^{~i}_a=e^A\hat\Omega^{~~i}_{A,a}$ is
symmetric
with respect to the $SO(1,p)$ vector indices
\begin{equation}\label{4.41}
e^ae^b\hat\Omega^{~~i}_{b,a}=0, \quad \Rightarrow \quad
\hat\Omega^{~~i}_{b,a}=\hat\Omega^{~~i}_{a,b}.
\end{equation}
By analogy with the bosonic embedding we can identify
$\hat\Omega^{~~i}_{a,b}\equiv e_a^{~M}\partial_MZ^{\underline M}
\nabla_{\underline M}E_b^{\underline a}u_{\underline a}^{~i}$ with
components of the second fundamental form of the supersurface
${\cal M}_{p+1,n}$ defined as
\begin{equation}\label{4.42}
K_{AB}^{~~C'}=e^{~M}_A\left(\partial_MZ^{\underline
M}\nabla_{\underline M} E_B^{\underline
A}+\omega_{MB}^{~~~C}E_C^{\underline A}
\right)E_{\underline A}^{~C'}
\end{equation}
where
$$
E_{\underline A}^{~C'}
\equiv (u_{\underline a}^{~i},~v_{\underline\alpha}^{~\alpha'}
-v_{\underline\alpha}^{~\alpha q}h_{\alpha q}^{~~\alpha'}), \quad
E_B^{\underline A}E_{\underline A}^{~C'}\equiv 0, \quad
\alpha'=\left({}_{\beta r'}\right)
$$

In many cases of embedding ${\cal M}_{p+1,n}$ into
$\underline{\cal M}_{D,2n}$
the integrability conditions of (\ref{4.24}) require additional
restrictions on the components of $\hat\Omega_{A,a}^{~~i}$,
such as a  (possibly inhomogeneous) analog of the
tracelessness of its vector part $\eta^{ab}\hat\Omega_{b,a}^{~~i}=0$,
which is equivalent to the bosonic field equations (\ref{4.20}),
and the condition on the spinor part of the following type
\cite{bzm,bpstv}
\begin{equation}\label{4.042}
(\gamma^a)^{\beta\alpha}\hat\Omega_{\alpha q,a}^{~~~i}=0.
\end{equation}
As we shall see in the next section, eq. (\ref{4.042}) amounts
to the fermionic field equations, and, hence, is a superembedding
counterpart
of the bosonic minimal embedding condition (\ref{4.19}).

Therefore, when integrability requires (\ref{4.042}) or its
inhomogeneous generalization due to the structure of the supergravity
background,
the superembedding condition  (\ref{4.24}) (or equivalently
(\ref{4.07})~)
contains the equations of motion of the
corresponding superbrane.
This is in contrast to the bosonic embedding
where the minimal embedding condition (\ref{4.19}), being equivalent
to the dynamical equations for the bosonic brane, is always imposed as
an additional condition.

Note that eq. (\ref{4.042}) is a generic minimal superembedding
condition, and
it can be also used (as an addition condition)
to find the superbrane equations of motion in the
cases where the basic superembedding condition (\ref{4.24}) is
off--shell.

To find out whether the superembedding condition (\ref{4.24}) puts
a given superbrane theory on the mass shell one should examine each case
separately. In the next section we shall make such an analysis for the
M--theory branes.

To conclude the description of the induced supergeometry of
${\cal M}_{p+1,n}$ we should identify the $SO(D-p-1)$ connection
$A_q^{~r}=dz^MA_{Mq}^{~~~r}(z)={1\over 4}A^{ij}(\gamma_{ij})_q^{~r}$
acting on the spinor components $e^{\alpha q}$
of the ${\cal M}_{p+1,n}$ supervielbein (\ref{4.022}). Note that
the $Spin(1,p)$ connection $\omega_\alpha^{~\beta}(z)$ is
given by the standard relation
\begin{equation}\label{4.43}
\omega_\alpha^{~\beta}(z)={1\over 4}\omega^{ab}
(\gamma_{ab})_\alpha^{~\beta}={1\over 4}\hat\Omega^{ab}
(\gamma_{ab})_\alpha^{~\beta}.
\end{equation}

To find the form of $A_q^{~r}$ let us take the
$\underline{\cal M}_{D,2n}$ external differential $\nabla$ of
eq. (\ref{4.23}) and multiply the result by
$v_{\underline\alpha}^{~\beta r}$. In view of eqs. (\ref{4.30})--
(\ref{4.34}) we get
\begin{eqnarray}\label{4.44}
T^{\underline\alpha}v_{\underline\alpha}^{~\beta r}&=&
T^{\beta r}
+e^a(\nabla E_a^{\underline\alpha})v_{\underline\alpha}^{~\beta r}+
e^{\alpha q}h_{\alpha q}^{~~\alpha'}(\nabla
v_{\alpha'}^{\underline\alpha})
v_{\underline\alpha}^{~\beta r} \\
&+&e^{\alpha q}
\left[(\nabla v_{\alpha q}^{\underline\alpha})
v_{\underline\alpha}^{~\beta r}-\delta_{q}^{~r}\omega_\alpha^{~\beta}-
\delta^{~\beta}_{\alpha}A_q^{~r}\right], \nonumber
\end{eqnarray}
where $T^{\beta r}\equiv de^{\beta r}+e^{\alpha
r}\omega_\alpha^{~\beta}+
e^{\beta q}A_q^{~r}$ is the spinor component of the ${\cal M}_{p+1,n}$
torsion form.

Using the harmonic relations (\ref{4.25}) and (\ref{4.36})
one can show that
\begin{equation}\label{4.45}
(\nabla v_{\alpha q}^{\underline\alpha})
v_{\underline\alpha}^{~\beta r}={1\over 4}\delta_q^r
\hat\Omega^{ab}(\gamma_{ab})_{\alpha}^{~\beta}
+{1\over 4}\delta_\alpha^{~\beta}
\hat\Omega^{ij}(\gamma_{ij})_q^{~r},
\end{equation}
where $\hat\Omega^{ab}$ and $\hat\Omega^{ij}$ are components of
the target--space connection form defined in (\ref{hatom})
and (\ref{4.038}).

In eqs. (\ref{4.038}) and (\ref{4.43})
we have identified $\omega^{ab}$ with $\hat\Omega^{ab}$.
Then from eqs. (\ref{4.44}) and (\ref{4.45})
it follows that we can identify the
$SO(D-p-1)$ connection form on ${\cal M}_{p+1,n}$ with $\hat\Omega^{ij}$
\footnote{Our choice of the supersurface connections (\ref{4.43}) and
(\ref{4.46}) is such that
$[(d+\Omega+\omega+A)v_{\alpha q}^{~\underline\beta}]
v^{~\beta r}_{\underline\beta}\equiv 0$
(see also the footnote to eq. (\ref{4.15})). This
differs from the choice of refs. \cite{hs2,hsw1}
where the form of the connections has been fixed in such a
way that the supersurface torsion components
$T^c_{Ab}$ and $T^{\gamma s}_{\alpha q,\beta r}$ vanish.},
i.e.
\begin{equation}\label{4.46}
A^{ij}=\hat\Omega^{ij}, \quad
A_q^{~r}={1\over 4}\hat\Omega^{ij}(\gamma_{ij})_q^{~r},
\end{equation}
and eq. (\ref{4.44}) reduces to the expression defining $T^{\beta r}$
as the induced torsion of ${\cal M}_{p+1,n}$
\begin{equation}\label{4.044}
T^{\beta r}=
T^{\underline\alpha}v_{\underline\alpha}^{~\beta r}
-e^a(\nabla E_a^{\underline\alpha})v_{\underline\alpha}^{~\beta r}-
e^{\alpha q}h_{\alpha q}^{~~\alpha'}(\nabla
v_{\alpha'}^{~\underline\alpha})
v_{\underline\alpha}^{~\beta r}.
\end{equation}

The curvature forms associated with the induced connections
$\omega_{a}^{~b}$ and $A_i^{~j}$
are obtained
by taking the corresponding components of the pullback (\ref{maurer1})
of the target space curvature. $R_a^{~b}$ has the
form similar to the Gauss equation (\ref{curva1})
\begin{equation}\label{curva2}
R_a^{~b}\equiv d\omega^{~b}_a+\omega^{~c}_a\omega^{~b}_c=
 u_{a}^{~\underline c}
R_{\underline c}^{~\underline d}u_{\underline d}^{~b}-\hat\Omega_a^{~i}
\hat\Omega_i^{~b},
\end{equation}
and $R_i^{~j}$ is
determined by the so--called Ricci equation
\begin{equation}\label{4.47}
R_i^{~j}\equiv dA_i^{~j}+A_i^{~k}A_k^{~j}= u_{i}^{~\underline c}
R_{\underline c}^{~\underline d}u_{\underline d}^{~j}-
\hat\Omega_i^{~a}\hat\Omega_a^{~j}.
\end{equation}
The last equation contained in (\ref{maurer1}) is the Codazzi equation
which
specifies the form of the supersurface covariant differential of
$\hat\Omega_a^{~i}$
\begin{equation}\label{4.48}
{\cal D}\hat\Omega_a^{~i}\equiv
d\hat\Omega_a^{~i}+\omega_a^{~b}\hat\Omega_b^{~i}+\hat\Omega_a^{~j}A_j^{~i}=
 u_{a}^{~\underline c}
R_{\underline c}^{~\underline d}u_{\underline d}^{~i}.
\end{equation}
From eq. (\ref{4.48}) we see that in the flat target (super)space
($R=0$)
the (super)embedding is such that the external covariant differential of
$\hat\Omega_a^{~i}$ on the (super)surface vanishes.

To summarize, the equations (\ref{4.07}), (\ref{4.08}), (\ref{4.29}),
(\ref{4.038}) and (\ref{4.46})--(\ref{4.48}) completely determine the
induced
geometrical properties of the embedded supersurface in terms of the
pullbacks
of the target--superspace supervielbein and superconnection adapted to
the supersurface by the use of the Lorentz harmonics.

Note that, analyzing the general properties of the superembedding, we
have
from the beginning imposed the constraints (\ref{torsionc2})
on target--superspace geometry, which we find more instructive.
But instead of introducing the target--superspace constraints, we might
from the beginning specify the supergeometry on the supersurface. Then
the
constraints on the geometry of target superspace would arise as
consequences
of the integrability of the superembedding condition.

These two initial options of choosing the superspace constraints for
studying the
superembeddings are equivalent, and depending on the problem considered
it is
convenient to use one option or another.

This equivalence follows from the general property of the basic
superembedding condition
(\ref{4.24}). If we only impose this condition and do not assume any
constraints on
the supersurface and target--superspace torsion, the spinor components
of the
torsion $T^a$ and the pullback of $T^{\underline a}$ are related by the
following
identity
\begin{equation}\label{ti}
T_{\alpha q,\beta r}^aE_a^{\underline a}=E^{~\underline\alpha}_{\alpha
q}
E^{~\underline\alpha}_{\beta
r}T_{\underline\alpha\underline\beta}^{\underline a},
\end{equation}
which can be easily derived from (\ref{4.22}) by taking the covariant
external
differential of both of its sides and taking into account (\ref{4.24}).

We shall now demonstrate how in the generic case
to extract $\kappa$--symmetry transformations
from general local superdiffeomorphisms of the embedded supersurface
\cite{stv,bpstv,bsv,hs2,hsw1,hrs}.

\subsection{$\kappa$--symmetry
from the point of view of superembedding}

As we have already mentioned (see eqs. (\ref{4}), (\ref{2.3.7}) and
(\ref{3.9}) as examples) for all known superbranes
in the Green--Schwarz--type formulation the $\kappa$--symmetry
transformations of the superbrane coordinates $Z^{\underline M}(\xi)$
in target superspace
have the following generic form
\begin{equation}\label{4.49}
\delta_\kappa Z^{\underline M}E_{\underline M}^{\underline \alpha}=
{1\over 2}(1+\bar\Gamma)^{\underline \alpha}_{~\underline\beta}
\kappa^{\underline\beta},
\end{equation}
\begin{equation}\label{4.049}
\delta_\kappa Z^{\underline M}E_{\underline M}^{\underline a}=0,
\end{equation}
where ${1\over 2}(1+\bar\Gamma)^{\underline \alpha}_{~\underline\beta}$
is a spinor projection matrix specific to each type of
superbranes.

At the same time, in the superembedding approach $Z^{\underline M}(z^M)$
($z^M=(\xi^m,\eta^\mu)$)
are worldvolume superfields transformed as scalars under the
${\cal M}_{p+1,n}$
superdiffeomorphisms $z'{}^M=z'{}^M(z)$, their infinitesimal variations
being
\begin{equation}\label{4.50}
\delta Z^{\underline M}=\delta z^M\partial_MZ^{\underline M}=
\delta z^Me_M^A(e_A^N\partial_NZ^{\underline M}).
\end{equation}
Multiplying (\ref{4.50}) by $E_{\underline M}^{\underline A}$ and
splitting the index $A$ on the vector  and spinor
 indices ($\underline a$, $\underline\alpha$) we get
\begin{equation}\label{4.51}
\delta Z^{\underline M}E_{\underline M}^{\underline\alpha}=
\delta z^Me_M^A(e_A^N\partial_NZ^{\underline M}
E_{\underline M}^{\underline\alpha} )
\equiv \delta z^Me_M^AE_A^{\underline\alpha},
\end{equation}
\begin{equation}\label{4.52}
\delta Z^{\underline M}E_{\underline M}^{\underline a}=
\delta z^Me_M^A(e_A^N\partial_NZ^{\underline M}
E_{\underline M}^{\underline a} )
\equiv \delta z^Me_M^AE_A^{\underline a}.
\end{equation}

Because of the superembedding condition (\ref{4.24}),
eq. (\ref{4.52}) reduces to
\begin{equation}\label{4.53}
\delta Z^{\underline M}E_{\underline M}^{\underline a}=
\delta z^Me_M^aE_a^{\underline a}.
\end{equation}

Comparing (\ref{4.53}) with (\ref{4.049}) we see that to
get $\kappa$--symmetry transformations from the
worldvolume superdiffeomorphisms we should consider such variations of
$z^M$ for which the Grassmann--even superdiffeomorphisms are zero
\begin{equation}\label{4.54}
\delta z^Me_M^a=0=\delta Z^{\underline M}E_{\underline M}^{\underline
a}.
\end{equation}
Then only the Grassmann--odd part $\delta z^Me_M^{\alpha q}$ of
the  ${\cal M}_{p+1,n}$
superdiffeomorphisms contributes to the variation (\ref{4.51}), and
we get
\begin{equation}\label{4.55}
\delta Z^{\underline M}E_{\underline M}^{\underline\alpha}=
\delta z^Me_M^{\alpha q}E_{\alpha q}^{\underline\alpha}.
\end{equation}

Since $\delta z^Me_M^{\alpha q}$ are arbitrary variations we can
(without
losing generality) replace them with
\begin{equation}\label{4.56}
\delta z^Me_M^{\alpha q}\equiv
\kappa^{\underline\beta}(z)v_{\underline\beta}^{~\alpha q}.
\end{equation}
Eq. (\ref{4.55}) takes the form
\begin{equation}\label{4.57}
\delta Z^{\underline M}E_{\underline M}^{\underline\alpha}=
\kappa^{\underline\beta}(z)
\left(v_{\underline\beta}^{~\alpha q}E_{\alpha
q}^{~\underline\alpha}\right).
\end{equation}
If we now recall the superembedding conditions (\ref{4.31}),
(\ref{4.34})
($E_{\alpha q}^{\underline\alpha}=v_{\alpha q}^{~\underline\alpha}
+h_{\alpha q}^{~~\alpha'}~v^{~~\underline\alpha}_{\alpha'}$)
we shall
notice that
$v_{\underline\beta}^{~~\alpha q}E_{\alpha q}^{\underline\alpha}=
P_{\underline\beta}^{~\underline\alpha}(z)$ is a projector
($P^{~\underline \gamma}_{\underline\beta}
P^{~\underline \alpha}_{\underline\gamma}
=P^{~\underline \alpha}_{\underline\beta}$). This projector
can be identified (possibly, up to linear
transformations) with ${1\over 2}(1+\bar\Gamma)$ from (\ref{4.49}), i.e.
\begin{equation}\label{4.057}
{1\over 2}(1+\bar\Gamma)^{\underline\alpha}_{~\underline\beta}=
v_{\underline\beta}^{~~\alpha q}E_{\alpha q}^{\underline\alpha}.
\end{equation}

We conclude that the leading ($\eta^\mu=0$) components
of the infinitesimal
superdiffeomorphisms (\ref{4.57}) and (\ref{4.54})
yield the $\kappa$--symmetry transformations (\ref{4.49}) and
(\ref{4.50}).

Here it is also the place to note that in the superembedding formulation
of the superbranes we have not introduced Wess--Zumino terms as
independent
objects. The reason is that in superbrane actions the Wess--Zumino term
is introduced to
ensure the nonmanifest $\kappa$--symmetry of the worldvolume action,
while
in this formulation the construction is manifestly worldvolume
superdiffeomorphism invariant.
In the superembedding approach the Wess--Zumino term  shows up
in the structure of the bosonic equations of motion, which we will see
with
the examples of the M--theory branes.

\subsection{$N=1$, $D=10$ superparticles and superstrings from
the general perspective of superembedding}

Let us now connect the general properties of the superembeddings
considered
in the previous subsection with the doubly supersymmetric description
of $N=1$, $D=10$ superparticles and superstrings given in
Sections 2 and 3. Superembeddings which describe superparticles
have been discussed in \cite{banur}, and the $N=1,2$, $D=10$
superstrings
have been studied in \cite{bpstv}.

In these cases the target superspace is ${\underline{\cal M}}_{10,16}$
(for simplicity, we take it to be flat),
and the embedded supersurface is ${\cal M}_{d,8}$ (where $d=1$
for the superparticles and $d=2$ for the superstrings).

The target--space structure group $SO(1,9)$ is broken by the presence of
the superparticle or the superstring down to $SO(1,1)\times SO(8)$.
The explicit
realization of the $\Gamma$--matrices (\ref{4.26}) and  (\ref{4.27}) is
given
by eq. (\ref{2.5.39}), with the matrices $C\Gamma^0$ and $C\Gamma^9$,
or $C\Gamma^{\pm\pm}=C(\Gamma^0\pm \Gamma^9)$
corresponding to the $SO(1,1)$ subgroup of $SO(1,9)$. The indices
$a=(--,++)$ and $\alpha=(-,+)$ are, the lightcone indices
of the vector and the spinor representation of $SO(1,1)$, respectively.
In the case of the superparticle the worldline coordinate
$\tau$ is identified with the coordinate
$\xi^{--}$ of the superworldsheet $\xi^m=(\xi^{--},\xi^{++})$.
And recall that
the odd coordinates $\eta^{-q}$ of ${\cal M}_{d,8}$ are one--component
worldline or worldsheet (chiral) spinors.

The $SO(1,1)\times SO(8)$ splitting of the
Lorentz--harmonic matrices (\ref{4.9}), (\ref{4.028}) is
\begin{equation}\label{4.58}
u^{~\underline a}_{\underline b}=\left(u^{--}_{\underline b},
u^{++}_{\underline b},
~u^{~i}_{\underline b}\right)
\end{equation}
\begin{equation}\label{4.59}
v_{\underline\beta}^{~\underline\alpha}(z)
=
\left(v_{\underline\beta}^{~-q},~v_{\underline\beta}^{~+q'}\right).
\end{equation}

The harmonics parametrize the coset space
${{SO(1,9)}\over{SO(1,1)\times SO(8)}}$.

In the flat target superspace the projection (\ref{4.28}) of the
spinor supervielbein ${\cal
E}^{\underline\alpha}=d\Theta^{\underline\alpha}$
is
\begin{equation}\label{4.60}
d\Theta^{\underline\beta}v_{\underline\beta}^{~\underline\alpha}=
\left(d\Theta^{\underline\beta}v_{\underline\beta}^{~-q},
d\Theta^{\underline\beta}v_{\underline\beta}^{~+q'}\right).
\end{equation}
In (\ref{4.60})
we can recognize the Lorentz--covariant version of the splitting
of $D_q\Theta^{\underline\mu}$ discussed in eqs.
(\ref{2.5.40})--(\ref{2.5.44}), when we made the analysis of
the superembedding condition for the $N=1$ superparticle.

To get (\ref{2.5.40}) from (\ref{4.60}) we should simply take
$v_{\underline\beta}^{~\underline\alpha}
=\delta_{\underline\beta}^{~\underline\alpha}$
which implies $u_{\underline b}^{~\underline a}
=\delta_{\underline b}^{~\underline a}$. This, of course, breaks both
the right and the left $SO(1,9)$ group associated with the Lorentz
harmonics (see eq. (\ref{4.11}) and below) and, hence, breaks the
manifest
target--superspace covariant description of the superembedding.
Then, for instance,  eq. (\ref{4.31}) reduces to eq. (\ref{2.5.44}).

Furthermore, comparing (\ref{4.32}) with (\ref{2.5.40}),
(\ref{2.5.47}) and (\ref{2.5.48}) (where
$\tau\equiv --$) we find that
\begin{equation}\label{4.61}
h_{-q,-q'}=\gamma^i_{qq'}{\cal E}^i_{--},
\end{equation}
where ${\cal E}^i_{--}$ are  `transverse' components of the vector
supervielbein pullback ${\cal E}^{\underline a}_{--}=
\partial_{--}X^{\underline a}
-i\partial_{--}\bar\Theta\Gamma^{\underline a}\Theta$ (\ref{2.5.48}).

Now notice that by taking $u_{\underline b}^{~\underline a}
=\delta_{\underline b}^{~\underline a}$ (and
$v_{\underline\beta}^{~\underline\alpha}
=\delta_{\underline\beta}^{~\underline\alpha}$) we do not (in general)
have the embedding condition (\ref{4.08}) (i.e. ${\cal E}^i_{--}\not
=0$)
for an adapted local frame such that
\begin{equation}\label{4.62}
{\cal E}_{--}^{\underline b}u_{\underline b}^{~i}=0.
\end{equation}
To get (\ref{4.62}) we must perform an appropriate $SO(1,9)$
transformation
of ${\cal E}^{\underline a}_{--}$, as has been explained in the previous
subsections. Once this has been done, the tensor $h_{-q,-q'}$
(\ref{4.61})
vanishes.

Thus, in the case of the $N=1$ superparticles and superstrings,
when the pullback of the target--space supervielbein $E^{\underline A}$
is
adapted (by the Lorentz harmonics)
to the superworldline or the superworldsheet such that eqs.
(\ref{4.07}), (\ref{4.08}) and (\ref{4.29}) are satisfied, the
$h$--component
of $E^{\underline\beta}_{-q}$ is zero and (in the flat target
superspace)
we have
\begin{equation}\label{4.63}
{\cal E}^{\underline\beta}_{-q}\equiv D_{-q}\Theta^{\underline\beta}
=v^{~\underline\beta}_{-q}.
\end{equation}
From (\ref{4.63}) it follows that its leading ($\eta^{-q}=0$) component,
which
is the commuting spinor variable $\lambda^{~\underline\beta}_{-q}$,
coincides with components of the inverse Lorentz--spinor harmonics
 \begin{equation}\label{4.64}
\lambda^{\underline\beta}_{-q}(\xi)=v^{~\underline\beta}_{-q}|_{\eta=0}.
\end{equation}
And from (\ref{4.014}) it follows that
\begin{equation}\label{4.65}
{\cal E}^{\underline a}_{--}\equiv
\partial_{--}X^{\underline a}
-i\partial_{--}\bar\Theta\Gamma^{\underline a}\Theta=u_{--}^{~\underline
a}
\end{equation}
coincides with a light--like component of the
inverse Lorentz--vector harmonics.

Using the defining harmonic relations (\ref{4.36}) one can convince
oneself
that ${\cal E}^{\underline a}_{--}$ and ${\cal
E}^{\underline\beta}_{-q}$
are related by the Cartan--Penrose formula (\ref{2.5.33})
$$
\delta_{qr}u_{--}^{~\underline a}=v_{-q}C\Gamma^{\underline a}v_{-r}.
$$

In the Subsection 2.5.4
we have demonstrated that in $D=10$ the independent components
of the matrix ${\cal E}^{\underline\beta}_{-q}$ parametrize the $S^8$
sphere,
while from eq. (\ref{4.63}) we see that ${\cal
E}^{\underline\beta}_{-q}$
coincides with components of
Lorentz harmonics $v^{~\underline\beta}_{-q}$
(which are new auxiliary variables introduced in the superembedding
formulation). As we have mentioned,
the harmonics
$(v^{-1})^{~\underline\beta}_{\underline\alpha}=(v^{~\underline\beta}_{-q},
v^{~\underline\beta}_{+q'})$ parametrize larger (16--dimensional) coset
space
${{SO(1,9)}\over{SO(1,1)\times SO(8)}}$. Therefore, there should be an
additional local symmetry which allows one to reduce the number of
independent components in
$(v^{-1})^{~\underline\beta}_{\underline\alpha}$
from 16 to 8.

This symmetry indeed occurs since half of the harmonics,
namely $v^{~~\underline\beta}_{+q'}$, do not enter the relations
defining the
pullback of the target--space supervielbein (\ref{4.63}), (\ref{4.65}).
Hence, we can vary the form of $v^{~\underline\beta}_{+q'}$ in a way
which
keeps the harmonic condition (\ref{4.36}) intact.
The allowed variations have the form of boost transformations $K_i$
\begin{equation}\label{4.66}
\delta v^{~\underline\beta}_{-q}=0, \quad
\delta v^{~\underline\beta}_{+q'}
=K_{++}^i(z)\gamma^i_{qq'}v^{~\underline\beta}_{-q},
\end{equation}
$$
\delta u_{--}^{~\underline a}=0, \quad \delta u_{++}^{~\underline a}=
K_{++}^iu_{i}^{~\underline a}, \quad \delta u_{i}^{~\underline a}=
{1\over 2}K_{++i}u_{--}^{~\underline a},
$$
where $K_{++}^i(z)$ are eight independent parameters.

Therefore, we see, from a somewhat different point of view, that
in the case of the $N=1$, $D=10$ superparticles and superstrings
the commuting (harmonic or twistor--like) spinor variables parametrize
the compact manifold $S^8$--sphere, which is realized as a coset space
\cite{ghs,dghs92}
$$
S^8={{SO(1,9)}\over{[SO(1,1)\times SO(8)]\otimes_S K_i}},
$$
where the stability subgroup is the semidirect product of
$SO(1,1)\times SO(8)$ with the boost transformations $K_i$ (\ref{4.66}).

We conclude that the doubly supersymmetric formulation of $N=1$
superparticles and superstrings, which we have developed in Sections 2
and 3
on dynamical grounds,
completely fits into the general geometrical picture of superembeddings.

Let us now apply the general properties of superembedding to the
description
of M--branes.

\section{M--theory branes}
\setcounter{equation}0

M--theory \cite{m} is a $D=11$ theory whose low energy limit is
$D=11$ supergravity \cite{cjs} and which also contains supermembranes
\cite{bst1,duffst}
and super--5--branes \cite{guv} as part of its nonperturbative spectrum.

The $D=11$ supergravity multiplet consists of the graviton
$e^{\underline a}_{\underline m}(x)$, the gravitino
$\psi^{\underline\alpha}_{\underline m}(x)$ and the antisymmetric
3--rank
gauge
field $A_{\underline{lmn}}(x)$, which are leading components
of the supervielbein $E^{\underline a}(X,\Theta)$,
$E^{\underline\alpha}(X,\Theta)$ and the three--form
$A^{(3)}(X,\Theta)$,
respectively $(\underline a=0,1...,10$; $\underline\alpha=1,...,32)$.

The supermembrane minimally couples to the gauge field $A^{(3)}$,
as superstrings minimally couple to the Neveu--Schwarz field $B^{(2)}$
(\ref{3.7}). The Green--Schwarz--type supermembrane action is
\cite{bst1}
\begin{equation}\label{5.01}
S_{M2}=-\int d^3\xi\sqrt{-\det{g_{mn}}}+\int_{{\cal M}_3}
dZ^{\underline L}dZ^{\underline M}dZ^{\underline
N}A_{\underline{NML}}(Z)
\end{equation}
($g_{mn}(\xi)=E_m^{\underline a}E_{n\underline a}$, and the tension is
put
equal to one).

The super--5--brane minimally couples to a six--form gauge field
$A^{(6)}$
whose seven--form field strength $F^{(7)}=dA^{(6)}-A^{(3)}dA^{(3)}$
is dual to the
four--form field strength
$F^{(4)}=dA^{(3)}$ of $A^{(3)}$.
The 5--brane also (nonminimally) couples to $A^{(3)}$ via the extended
self--dual
field strength
\begin{equation}\label{5.1a}
H^{(3)}(\xi)=db^{(2)}-A^{(3)}(Z(\xi))
\end{equation}
of an antisymmetric field $b_{mn}(\xi)$
living in the worldvolume of the 5--brane \cite{aha}.
In this sense the 5--brane is a dyonic object.
The form of the 5--brane supergravity coupling
is uniquely defined by local symmetries of the 5--brane worldvolume
action
which are responsible for
the self--duality properties of $b_{mn}(\xi)$ \cite{pst3}.
The `Born--Infeld--like' super--5--brane action is \cite{m5,s1}
\begin{eqnarray}\label{5.001}
S_{M5}&=&\int d^6 \xi (-\sqrt{-\det(g_{mn}+{H}^*_{mn})}
+ {\sqrt{-g} \over 4}
H^{*mn}H_{mn})\nonumber \\
&+&
\int_{{\cal M}_6}(A^{(6)}-{1\over 2} d b^{(2)}\wedge A^{(3)}),
\end{eqnarray}
where $g_{mn}(\xi)=E_m^{\underline a}E_{n\underline a}$,
$H_{mn}\equiv H_{mnl}v^l(\xi)$, $H^{*mn}
\equiv {1\over{\sqrt{-g}}}\epsilon^{mnll_1l_2l_3}v_lH_{l_1l_2l_3}$ and
$v_l(\xi)={{\partial_l a(\xi)}\over{\sqrt{-(\partial a)^2}}}$ is the
normalized derivative of the auxiliary scalar field $a(\xi)$ which
ensures
the covariance of the 5--brane action, its construction being based on
a generic method for the covariant
Lagrangian description of duality--symmetric fields proposed in
\cite{pst1,pst2}.

A duality--symmetric component action for $D=11$ supergravity,
which includes both gauge fields,
$A^{(3)}$ and its dual $A^{(6)}$, and which thus couples to both
the membrane and the 5--brane, was constructed in \cite{bbs}. This
action
reduces to the Cremmer--Julia--Scherk action \cite{cjs}
when the field $A^{(6)}$
is eliminated by solving for an algebraic part of the duality relations
between $F^{(7)}$ and $F^{(4)}$.

The $\kappa$--symmetry of the supermembrane and the super--5--brane
worldvolume actions requires
the superbranes to propagate in target--superspace backgrounds
whose supergeometry is constrained to be that of $D=11$ supergravity.

In the case of the superspace description of $D=11$ supergravity the
constraints imposed on components of the torsion
$T^{\underline A}(Z)=\nabla E^{\underline A}=dE^{\underline A}+
E^{\underline B}\Omega^{~\underline A}_{\underline B}
$, curvature
$R(Z)_{\underline A}^{~\underline B}
=d\Omega_{\underline A}^{~\underline B}+\Omega_{\underline
A}^{~\underline C}
\Omega_{\underline C}^{~\underline B}$,
and on the dual field strengths
$F^{(4)}(Z)$ and $F^{(7)}(Z)$
have the following form \cite{11sg,cal}
\begin{equation}\label{5.1}
T^{\underline a}=-i\bar E\Gamma^{\underline a}E,
\end{equation}
\begin{equation}\label{5.2}
T^{\underline\alpha}={1\over{288}}
F_{\underline b_1...\underline b_4}E^{\underline a}
\left(\Gamma_{\underline a}^{~\underline b_1...\underline b_4}-
8\delta_{\underline a}^{[\underline b_1}
\Gamma^{\underline b_2...\underline b_4]}\right)
^{\underline\alpha}_{~\underline\beta}E^{\underline\beta},
\end{equation}
\begin{equation}\label{5.03}
R^{\underline a\underline b}=-{i\over 144}\bar E
\left(\Gamma_{~~\underline{c_1}...\underline{c_4}}^{\underline
a\underline b}
+24\Gamma_{[\underline{c_1}\underline{c_2}}\delta_{\underline{c_3}}
^{\underline a}\delta_{\underline{c_4}]}^{\underline b}\right)E
F^{\underline{c_1}...\underline{c_4}}+
{1\over 2}E^{\underline d}E^{\underline c}
R_{\underline c\underline d}
^{~~~\underline a\underline b},
\end{equation}
\begin{equation}\label{5.3}
F^{(4)}=dA^{(3)}=
{i\over 2}E^{\underline a}E^{\underline b}\bar
E\Gamma_{\underline{ba}}E+
{1\over{4!}}E^{\underline a_4}...E^{\underline a_1}F_{\underline a_1...
\underline a_4},
\end{equation}
\begin{eqnarray}\label{5.4}
F^{(7)}&=&dA^{(6)}-{1\over 2}A^{(3)}dA^{(3)} \nonumber \\
&=&
{i\over 5!}E^{\underline a_1}...E^{\underline a_5}
\bar E\Gamma_{\underline a_5...\underline a_1}E+
{1\over{7!}}E^{\underline a_7}...E^{\underline a_1}F_{\underline a_1...
\underline a_7},
\end{eqnarray}
where the field strength components $F_{\underline a_1...\underline
a_4}$
and $F_{\underline a_1...\underline a_7}$ are related by the Hodge
duality
\begin{equation}\label{5.5}
F_{\underline a_1...\underline a_7}={1\over {4!}}
\epsilon_{\underline a_1...\underline a_7\underline b_1...\underline
b_{4}}
F^{\underline b_1...\underline b_{4}}.
\end{equation}

The constraints (\ref{5.1})--(\ref{5.5}) are {\sl on--shell}
$D=11$ supergravity constraints since they imply the $D=11$ supergravity
equations of motion (which can be alternatively obtained from the
$D=11$ supergravity component actions \cite{cjs,bbs}).

Note that in $D=11$ we deal with 32--component Majorana spinors
$\bar\Psi_{\underline\alpha}=(\Psi^T)^{\underline\beta}
C_{\underline\beta\underline\alpha}$ where $C$ is the charge
conjugation matrix, and we
can use $C_{\underline\alpha\underline\beta}$ and its inverse
$(C^{-1})^{\underline\alpha\underline\beta}
=-C^{\underline\alpha\underline\beta}=-C_{\underline\alpha\underline\beta}$
to lower and raise
the spinor indices. Then the following antisymmetric products of the
$\Gamma$--matrices
\begin{equation}\label{5.6}
\Gamma^{\underline a}_{\underline\alpha\underline\beta}\equiv
C_{\underline\alpha\underline\gamma}
(\Gamma^{\underline a})^{\underline\gamma}_{~\underline\beta}, \quad
(C\Gamma^{\underline a_1\underline
a_2})_{\underline\alpha\underline\beta},
\quad (C\Gamma^{\underline a_1...\underline a_5})_
{\underline\alpha\underline\beta}
\end{equation}
are symmetric in the spinor indices.

We are now in a position to describe the M-branes by applying the
superembedding methods of Section 4.

\subsection{The supermembrane}
The superembedding of the $D=11$ supermembrane was first
analyzed in \cite{bpstv}.

The target superspace ${\underline{\cal M}}_{11,32}$ is that of $D=11$
supergravity and the embedded supersurface ${\cal M}_{3,16}$,
associated with the supermembrane worldvolume, is characterized
by a local supervielbein basis $e^A(z)=(e^a,~e^{\alpha q})$
$(a=0,1,2;$ $\alpha=1,2;$ $q=1,...,8)$ transformed under the
action of $SO(1,2)\times SO(8)$.
The $d=3$ spinors are two--component Majorana spinors.

An appropriate form of the $D=11$ $\Gamma$--matrices (\ref{4.26}),
(\ref{4.27}), which reflects the embedding, and of the charge conjugate
matrix $C^{\underline\alpha\underline\beta}$ to raise indices, is
\begin{equation}\label{5.1.1}
\Gamma^a_{\underline\alpha\underline\beta}=\left(
\begin{array}{cc}
\gamma^a_{\alpha\beta}\delta_{qr}  &  0\\
0  &  \gamma^{a\alpha\beta}\delta_{q'r'}
\end{array}
\right)\,, \quad a=0,1,2\, ,
\end{equation}
\begin{equation}\label{5.1.2}
\Gamma^i_{\underline\alpha\underline\beta}=\left(
\begin{array}{cc}
0  &  \delta^\alpha_\beta\gamma^i_{qq'}\\
\delta_\alpha^\beta\tilde\gamma^i_{q'q}   &  0
\end{array}
\right)\,, \quad i=1,...,8\,,
\end{equation}
\begin{equation}\label{5.1.C}
C^{\underline\alpha\underline\beta}=\left(
\begin{array}{cc}
\epsilon^{\alpha\beta}\delta_{qr}  &  0\\
0  &  -\epsilon_{\alpha\beta}\delta_{q'r'}
\end{array}
\right)\, .
\end{equation}

where  $\tilde\gamma^i_{q'q}\equiv\gamma^i_{qq'}$ are
the $SO(8)$ $\Gamma$--matrices, $\gamma^a_{\alpha\beta}
\equiv \epsilon_{\alpha\gamma}\gamma^{a\gamma}_{~~\beta}$ are symmetric,
and
the $d=3$ $\Gamma$--matrices $\gamma^{a\gamma}_{~~\beta}$ were
defined in eq. (\ref{2.5.9}). The $d=3$ spinor indices are raised and
lowered by the unit antisymmetric matrices $\epsilon_{\alpha\beta}$ and
$\epsilon^{\alpha\beta}$, and the $SO(8)$ indices are raised and lowered
by the unit matrices $\delta_{qr}$, $\delta_{q'r'}$, so that there is no
distinction between the upper and lower $SO(8)$ indices. (Recall that
the
unprimed index $q$ and the primed index $q'$ correspond to different
spinor representations of $SO(8)$).

Induced supergeometry on ${\cal M}_{3,16}$ is described by the general
superembedding conditions (\ref{4.07})--(\ref{4.24}), (\ref{4.29})--
(\ref{4.34}), (\ref{4.038})--(\ref{4.41}) and (\ref{4.46}).

We shall now analyze whether in the case under consideration
their integrability  puts additional
restrictions on the form of the supervielbein pullback (\ref{4.32}) and
on components of the one--form
$\hat\Omega_a^{~i}=e^A\hat\Omega_{A,a}^{~~~i}$ in (\ref{4.39}). Remember
that $\hat\Omega_{A,a}^{~~~i}$ is related to the second fundamental form
(\ref{4.18}), (\ref{4.42}) of the (super)surface, and if its components
satisfy additional conditions, such as eqs. (\ref{4.19}) and
(\ref{4.042}),
these reproduce
(super)brane equations of motion, as eq. (\ref{4.20}) in the bosonic
case.

Consider in more detail the equation (\ref{4.39}). Making use of
the harmonic relations (\ref{4.25}), the form of the
$D=11$ $\Gamma$--matrices (\ref{5.1.1}), (\ref{5.1.2}),
and the equations (\ref{4.29}), (\ref{4.32}),
we can rewrite eq. (\ref{4.39})
in the following form
\begin{eqnarray}\label{5.1.3}
e^a\hat\Omega_a^{~i}&=&e^a{\nabla} E_a^{~\underline b}u_{\underline
b}^{~i}=
e^a(dE_a^{~\underline b}+E_a^{~\underline c}
\Omega_{\underline c}^{~\underline b})u_{\underline b}^{~i} \nonumber \\
&=&-2i e^{\alpha r}e^{\beta q}\gamma^i_{qq'}h^{~~~q'}_{\alpha
r,\beta} +2ie^ae^{\alpha q}\gamma^i_{qq'}E_a^{\underline\beta}
v_{\underline\beta}^{~\beta q'}\epsilon_{\alpha\beta}.
\end{eqnarray}
(Recall that $E_a^{~\underline b}=u_a^{~\underline b}$,
see eq. (\ref{4.014})~).

Comparing the left-- and  right--hand sides of (\ref{5.1.3}) we find
that
\begin{equation}\label{5.1.4}
e^ae^b\hat\Omega_{b,a}^{~~i}=0 \quad \rightarrow \quad
\hat\Omega_{b,a}^{~~i}=\hat\Omega_{a,b}^{~i}
={\nabla}_b E_a^{~\underline b}u_{\underline b}^{~i},
\end{equation}
\begin{equation}\label{5.1.5}
\hat\Omega_{\alpha q, a}^{~~~~i}
={\nabla}_{\alpha q} E_a^{~\underline b}u_{\underline b}^{~i}=
2i\gamma^i_{qq'}E_a^{\underline\beta}
v_{\underline\beta}^{~\beta q'}\epsilon_{\alpha\beta},
\end{equation}
and
\begin{equation}\label{5.1.6}
\gamma^i_{qq'}h^{~~~q'}_{\alpha r,\beta}
=-\gamma^i_{rq'}h^{~~~q'}_{\beta q,\alpha}.
\end{equation}
Let us analyze eq. (\ref{5.1.6}). It implies that
the matrix $\gamma^i_{qq'}h^{~~~q'}_{\alpha r,\beta}$ is antisymmetric
with respect to the pair of the indices $(\alpha r)$ and $(\beta q)$.

A general matrix $h_{\alpha r,\beta q'}$ is decomposed
in the basis of the elements of the
$SO(8)$ and $SO(1,2)$ Clifford algebras as follows
\begin{equation}\label{5.1.7}
h_{\alpha r,\beta q'}=\epsilon_{\alpha\beta}(h_j\tilde\gamma^i_{q'r}
+h_{j_1j_2j_3}
\gamma^{j_1j_2j_3}_{q'r})+\gamma^a_{\alpha\beta}(h_{aj}\tilde\gamma^j_{q'r}+
h_{aj_1j_2j_3}
\gamma^{j_1j_2j_3}_{q'r}),
\end{equation}
where $\gamma^{j_1j_2j_3}_{q'r}$ denotes the antisymmetric product of
the $SO(8)$ $\Gamma$--matrices
$\tilde\gamma^{[j_1}_{q'q}\gamma^{j_2}_{qr'}
\tilde\gamma^{j_3]}_{r'r}$.

Substituting (\ref{5.1.7}) into (\ref{5.1.6}) we get
\begin{equation}\label{5.1.8}
\gamma^i_{qq'}(h_j\tilde\gamma^j_{q'r}+h_{j_1j_2j_3}\gamma^{j_1j_2j_3}_{q'r})
- (q~\leftrightarrow ~ r)=0,
\end{equation}
and
\begin{equation}\label{5.1.9}
\gamma^i_{qq'}(h_{aj}\tilde\gamma^j_{q'r}+h_{aj_1j_2j_3}
\gamma^{j_1j_2j_3}_{q'r})
+(q~\leftrightarrow ~r)=0.
\end{equation}

Using the defining properties of the matrices $\gamma^i_{qq'}$ and
$\tilde\gamma^i_{q'q}\equiv\gamma^i_{qq'}$
$$
\gamma^i_{qq'}\tilde\gamma^j_{q'r}+\gamma^j_{qq'}\tilde\gamma^i_{q'r}=
2\delta_{qr},
$$
we find that eqs. (\ref{5.1.8}) and (\ref{5.1.9}) imply that all
components
of the matrix (\ref{5.1.7}) vanish
\begin{equation}\label{5.1.10}
h_{\alpha r,\beta q'}=0.
\end{equation}
Hence, in the case of the supermembrane
eq. (\ref{4.34}) for the spinor components
$E^{\underline\beta}_{\alpha q}$ of the $\underline{\cal M}_{11,32}$
supervielbein pullback reduces to
\begin{equation}\label{5.1.11}
E^{\underline\beta}_{\alpha q}\equiv
e_{\alpha q}^M\partial_MZ^{\underline M}E_{\underline
M}^{\underline\beta}=
v^{~\underline\beta}_{\alpha q}(z).
\end{equation}
This, in particular, means that the induced superworldvolume
torsion (\ref{4.38})
obeys the standard $d=3$, $n=8$ supergravity constraint
\begin{equation}\label{5.1.12}
T^a_{\alpha q,\beta r}=-2i\delta_{qr}\gamma^a_{\alpha\beta},
\end{equation}
which follows from (\ref{4.38}), (\ref{4.25}), (\ref{5.1.1})
and (\ref{5.1.11}).

The vanishing of the $h$--matrix (\ref{5.1.10}) also implies that, in
view of the $D=11$ supergravity torsion constraint (\ref{5.2}),
the purely spinor components of the superworldvolume torsion $T^{\alpha
q}$
defined in (\ref{4.044}) are zero
\begin{equation}\label{5.1.13}
T^{\alpha q}_{\beta r,\gamma s}=0.
\end{equation}

In turn, the torsion constraints (\ref{5.1.12}) and (\ref{5.1.13})
imply that the  covariant spinor derivatives ${\cal D}_{A}\equiv
e_{A}^M\partial_M+\omega_A+A_A$ of the superworldvolume
satisfy the algebra
\begin{equation}\label{5.1.14}
\{{\cal D}_{\alpha q},{\cal D}_{\beta r}\}
=2i\delta_{qr}\gamma^a_{\alpha\beta}{\cal D}_a+R_{\alpha q\beta r}(z),
\end{equation}
where $R_{\alpha q\beta r,A}^{~~~~~~B}(z)$ are components of the
superworldvolume curvature (\ref{curva2}), (\ref{4.47}).

\subsubsection{The fermionic equation}
Applying the covariant derivative ${\cal D}_{\beta r}$ to eq.
(\ref{5.1.11}),
symmetrizing the result
with respect to ${\beta r}$ and ${\alpha q}$, and
making use of the basic superembedding condition (\ref{4.24}), the
torsion
constraint $T^{\underline\beta}_{\underline\alpha\underline\gamma}=0$
(\ref{5.2})
and the superalgebra (\ref{5.1.14}) we get
\begin{eqnarray}\label{5.1.15}
{\cal D}_{\beta r}E^{\underline\beta}_{\alpha q}+
{\cal D}_{\alpha q}E^{\underline\beta}_{\beta r}&=&2i\delta_{qr}
\gamma^a_{\alpha\beta}E^{\underline\beta}_a
-v^{~\underline\gamma}_{\alpha q}\Omega_{\beta r\underline\gamma}
^{~~~~\underline\beta}
-v^{~\underline\gamma}_{\beta r}\Omega_{\alpha q\underline\gamma}
^{~~~~\underline\beta}\nonumber \\
&=&{\cal D}_{\beta r}v^{~\underline\beta}_{\alpha q}+
{\cal D}_{\alpha q}v^{~\underline\beta}_{\beta r}.
\end{eqnarray}
Then multiplying eq. (\ref{5.1.15}) by $v_{\underline\beta}^{~\gamma
q'}$
and using the orthogonality properties of the spinor harmonics we
obtain
\begin{equation}\label{5.1.16}
2i\delta_{qr}
\gamma^a_{\alpha\beta}E^{\underline\beta}_av_{\underline\beta}^{~\gamma
q'}
=
(\nabla_{\beta r}v^{~\underline\beta}_{\alpha q})
v_{\underline\beta}^{~\gamma q'}~+~(\alpha q~\leftrightarrow~\beta r).
\end{equation}
(Remember that $\nabla_{\beta r}$ is the spinor component of the
pullback of the external target--space differential $\nabla=d+\Omega$
and $\Omega_{\underline\alpha}^{~\underline\beta}={1\over
4}\Omega^{\underline
{ab}}(\Gamma_{\underline{ab}})_{\underline\alpha}^{~\underline\beta}$).

Making use of the harmonic relations (\ref{4.25}), (\ref{4.36}) and
the representation (\ref{5.1.1}), (\ref{5.1.2}) of the
$\Gamma$--matrices,
one can show that
\begin{equation}\label{5.1.17}
(\nabla_{\beta r}v^{~\underline\beta}_{\alpha q})
v_{\underline\beta}^{~\gamma q'}=
{1\over 4}(\nabla_{\beta r}u_a^{~\underline a})u_{\underline a}^{~i}
(\gamma^a)^{~\gamma}_\alpha(\gamma_{i})^{~q'}_q
\equiv{1\over 4}\hat\Omega_{\beta r,a}^{~~~~i}
(\gamma^a)^{~\gamma}_\alpha(\gamma_{i})^{~q'}_q.
\end{equation}

Substituting (\ref{5.1.17}) into (\ref{5.1.16}) and taking the trace
of its left and right hand side with respect to the indices $(q,r)$ and
$({}^\gamma_\beta)$ we get
\begin{equation}\label{5.1.18}
16i\gamma^a_{\alpha\beta}E_a^{\underline\beta}v_{\underline\beta}^{~\beta
q'}
={1\over 4}\hat\Omega_{\beta r,a}^{~~~~i}
(\gamma^a)^{~\beta}_\alpha(\gamma_{i})^{rq'}.
\end{equation}

On the other hand, let us take the earlier found relation (\ref{5.1.5})
and
multiply it by $(\gamma^a)^{~\beta}_\alpha$ and $\gamma_{i}^{~q'r}$,
the result is
\begin{equation}\label{5.1.19}
16i\gamma^a_{\alpha\beta}E_a^{\underline\beta}v_{\underline\beta}^{~\beta
q'}
=-\hat\Omega_{\beta r,a}^{~~~~i}
(\gamma^a)^{~\beta}_\alpha\gamma_{i}^{~rq'}.
\end{equation}
Comparing (\ref{5.1.19}) with (\ref{5.1.18}) we see that their right
hand
sides have different coefficients and, hence, are zero. We thus get the
additional restriction on components $\hat\Omega_{\beta r,a}^{~~~~i}$
of the second fundamental form (\ref{4.42})
\begin{equation}\label{5.1.20}
\hat\Omega_{\beta r,a}^{~~~~i}
(\gamma^a)^{\alpha\beta}(\gamma_{i})^{rq'}
=(\gamma^{a})^{\alpha}_{~\beta}E_a^{\underline\beta}
v_{\underline\beta}^{~\beta q'}=0,
\end{equation}
or, in view of (\ref{5.1.5}),
\begin{equation}\label{5.1.0020}
\hat\Omega_{\beta r,a}^{~~~~i}
(\gamma^a)^{\alpha\beta}=
\nabla_{\beta r}E^{\underline b}_au^{~i}_{\underline b}
(\gamma^a)^{\alpha\beta}=0.
\end{equation}

Eq. (\ref{5.1.20}) is nothing but the {\bf 16}--component Dirac--type
fermionic equation of motion of the $D=11$ supermembrane \cite{bst1}
written in the Lorentz--harmonic form  \cite{zb0} and promoted to
the worldvolume superspace \cite{bpstv}.
To see this let us rewrite (\ref{5.1.20}) in
a form similar to the fermionic equation which one derives from the
supermembrane action (\ref{5.01}).
First of all notice that, with the use of
the harmonic relations (\ref{4.36}), the form of the
$\Gamma$--matrices (\ref{5.1.1}), and the embedding conditions
$E_a^{\underline\beta}v_{\underline\beta}^{~\alpha q}=0$ (\ref{4.30})
and
$E_a^{\underline a}=u_a^{~\underline a}$ (\ref{4.014}), eq.
(\ref{5.1.20})
can be presented in the following form
\begin{equation}\label{5.1.020}
(\gamma^{a})_{\alpha}^{~\beta}E_a^{\underline\beta}
v_{\underline\beta,\beta q'}=
E_a^{\underline\beta}(\eta^{ab}E_b^{\underline a}
\Gamma_{\underline a})_{\underline\beta\underline\gamma}
v_{\alpha q'}^{~~~\underline\gamma}=0
\end{equation}
Multiplying eq. (\ref{5.1.020}) by
$v^{~\alpha q'}_{\underline\alpha}$ we get
\begin{equation}\label{5.1.22}
E_a^{\underline\beta}(\eta^{ab}E_b^{\underline a}
\Gamma_{\underline a})_{\underline\beta\underline\gamma}
v_{\alpha q'}^{~~~\underline\gamma}
v^{~\alpha q'}_{\underline\alpha}
\equiv {1\over 2}
\eta^{ab}E_a^{\underline\beta}\left[E_b^{\underline a}
\Gamma_{\underline a}(1-\bar\Gamma)
\right]_{\underline\beta\underline\alpha}
=0
\end{equation}
where, because of the projector
$$
{1\over 2}(1-\bar\Gamma)^{\underline\gamma}_{~\underline\alpha}
\equiv
v^{~\alpha q'}_{\underline\alpha}v_{\alpha q'}^{~~~\underline\gamma},
$$
the l.h.s. of (\ref{5.1.22}) identically vanishes under the action
of the projector
\begin{equation}\label{5.1.022}
{1\over 2}(1+\bar\Gamma)_{~\underline\beta}^{\underline\alpha}\equiv
v_{\underline\beta}^{~\alpha q}v_{\alpha q}^{~\underline\alpha}
\end{equation}
in which one can recognize the $\kappa$--symmetry projector of eq.
(\ref{4.57}). Thus $\kappa$--symmetry insures that among the
{\bf 32} equations in (\ref{5.1.22}), only
{\bf 16} are independent. And these {\bf 16}
equations reduce the number of the
fermionic physical modes of the supermembrane
down to {\bf 8}.

Using the eqs. (\ref{4.014}), (\ref{4.25}), (\ref{4.36}), (\ref{5.1.1})
and (\ref{5.1.2})
it is not hard to verify that the matrix $\bar\Gamma$ defined in
(\ref{5.1.022}) has exactly the same form as in the $\kappa$--symmetry
projector and in the equations of motion of the Green--Schwarz
formulation
of the supermembrane \cite{bst1}
\begin{equation}\label{5.1.23}
\bar\Gamma
={1\over{3!}}\epsilon^{abc}\Gamma_{abc}, \quad {\rm where} \quad
\Gamma_a=E^{\underline a}_a\Gamma_{\underline a}
\end{equation}
(Note that the Lorentz harmonics have disappeared from eq.
(\ref{5.1.23}).)
Thus, at ($\eta^\mu=0)$ the worldvolume superfield equation
(\ref{5.1.22})
(which is the consequence of (\ref{5.1.20})~)
reduces to the standard fermionic component equations of the
supermembrane.
To show this one should choose the Wess--Zumino gauge
(see, for instance, \cite{wb}) for the components of the inverse
supervielbein
$e^{~M}_A(z)|_{\eta=0}$ on the supersurface such that
\begin{equation}\label{5.1.023}
e^{~M}_A(z)|_{\eta=0}=
\left(
\begin{array}{cc}
e^{~m}_a(\xi)  &  -\psi_a^{\mu r}(\xi)\\
0  &  \delta^{~\mu}_\alpha\delta^{~r}_q
\end{array}
\right)\, ,
\end{equation}
and take into account the superembedding conditions
$E_{\alpha q}^{\underline a}=0$ (\ref{4.24}) and
$E_{\alpha q}^{\underline\alpha}=v_{\alpha q}^{~~\underline\alpha}$
(\ref{5.1.11}). Then,
\begin{equation}\label{wz1}
E_a^{\underline a}|_{\eta=0}=e_a^m(\xi)\partial_mZ^{\underline M}
E_{\underline M}^{\underline a}|_{\eta=0}\equiv
e_a^m(\xi)E_{m}^{\underline a}(Z(\xi)),
\end{equation}
\begin{equation}\label{wz2}
E_a^{\underline\alpha}v_{\underline\alpha,\beta q'}|_{\eta=0}=
e_a^m(\xi)\partial_mZ^{\underline M}
E_{\underline M}^{\underline\alpha}v_{\underline\alpha,\beta
q'}|_{\eta=0}
\equiv e_a^m(\xi)E_{m}^{\underline\alpha}(Z(\xi))
v_{\underline\alpha,\beta q'}(\xi).
\end{equation}
And the worldvolume gravitino field $\psi^{\mu q}_a(\xi)$
(\ref{5.1.023}) is express in terms of
$E_{m}^{\underline\alpha}(Z(\xi))$
\begin{equation}\label{psi}
\psi^{\mu q}_a(\xi)=e(\xi)_a^{~m}E_{m}^{\underline\alpha}
v_{\underline\alpha}^{~\mu q}(\xi),
\end{equation}
due to the superembedding condition  (\ref{4.30}).

Then the component equation of motion for $\theta^{\underline\mu}(\xi)$
takes the form
\begin{equation}\label{5.1.0022}
g^{mn}E_m^{\underline\beta}\left[E_n^{\underline a}
\Gamma_{\underline a}(1-\bar\Gamma)
\right]_{\underline\beta\underline\alpha}
=0, \quad g_{mn}(\xi)=e^a_me_{na}(\xi)
=E_m^{\underline a}E_n^{\underline b}\eta_{\underline{ab}}.
\end{equation}

\subsubsection{The bosonic equation}

Once the supermembrane fermionic equations are implied by
superembedding,
we can expect that (because of supersymmetry) also the bosonic equations
should appear. As we have shown, the bosonic field equations
(\ref{4.20})
of a  bosonic brane are encoded in the minimal embedding condition
(\ref{4.19}). We shall now find an inhomogeneous generalization of the
minimal embedding condition, the inhomogeneity reflecting
the interaction of the
supermembrane with the supergravity gauge field $A^{(3)}(Z)$.

It is a simple exercise to show that the bosonic field equations
must hold provided
that the fermionic equations (\ref{5.1.20}) hold.
This can be verified in different ways, one of which is to analyze
the pullback
components $T_{a,\alpha
q}^{\underline\alpha}v_{\underline\alpha}^{~\beta r'}$
of the torsion constraint (\ref{5.2}). We, however, find it here
simpler to look at the
$e^{\alpha q}e^{\beta r}$ components of the  Codazzi equation
(\ref{4.48})
\begin{equation}\label{5.1.24}
-2i\delta_{qr}\gamma^b_{\alpha\beta}\hat\Omega_{b,a}^{~~~i}-
{\cal D}_{\alpha q}\hat\Omega_{\beta r, a}^{~~~~i}-
{\cal D}_{\beta r}\hat\Omega_{\alpha q, a}^{~~~~i}=
R_{\alpha q\beta r,a}^{~~~~~~i}.
\end{equation}
Multiplying (\ref{5.1.24}) by $\gamma^{a\alpha\beta}$ and $\delta^{qr}$,
and
taking into account the fermionic equation (\ref{5.1.20}) we get
\begin{equation}\label{5.1.25}
\eta^{ab}\hat\Omega_{b,a}^{~~~i}\equiv \eta^{ab}
(\nabla_aE_b^{\underline a})u_{\underline a}^{~i}=
-{i\over{32}}\gamma^{a\alpha\beta}\delta^{qr}R_{\alpha q\beta
r,a}^{~~~~~~i}.
\end{equation}
We observe that, in comparison with the bosonic minimal embedding
condition
(\ref{4.19}), the bosonic equation (\ref{5.1.25}) acquires the r.h.s. If
we now look at the supergravity curvature constraint (\ref{5.03}) we
realize that the r.h.s. of (\ref{5.1.25}) contains components of
the field strength $F^{(4)}$ of the $D=11$
supergravity gauge field $A^{(3)}$.
Hence, the r.h.s. of (\ref{5.1.25})
describes the coupling of the supermembrane to $A^{(3)}$. Its
form is the same as that obtained from the Wess--Zumino term of the
supermembrane worldvolume action (\ref{5.01}). To see this we should
take the corresponding $e^{\alpha q}e^{\beta r}$ components of the
curvature pullback (\ref{5.03}) and
make use of the basic superembedding condition (\ref{4.24}), the
relation
(\ref{5.1.11}), the form of the $\Gamma$--matrices (\ref{4.26}),
(\ref{4.27})
and the harmonic relations (\ref{4.25}), (\ref{4.36}).
If we do this, and take the leading $(\eta=0$)
component of eq. (\ref{5.1.25})
in the Wess--Zumino gauge (\ref{5.1.023})--(\ref{psi}), the r.h.s. of
(\ref{5.1.25}) takes the form
\begin{eqnarray}\label{5.1.026}
-{i\over{32}}\gamma^{a\alpha\beta}\delta^{qr}
R_{\alpha q\beta r,a}^{~~~~~~i}|_{\eta=0}
&=&-{1\over 3!}
\epsilon^{abc}
E_a^{\underline d}E_b^{\underline c}E_c^{\underline b}
F_{\underline{bcd}}^{~~~\underline a}u_{\underline a}^{~i}\nonumber\\
&=&
-{1\over{3!\sqrt{-g}}}
\epsilon^{lmn}
E_l^{\underline d}E_n^{\underline c}E_n^{\underline b}
F_{\underline{bcd}}^{~~~\underline a}u_{\underline a}^{~i} ,
\end{eqnarray}
where
$$E_m^{\underline A}\equiv
\partial_mZ^{\underline M}E_{\underline M}^{\underline A}|_{\eta=0},
\qquad
g_{mn}(\xi)=e^a_me_{na}(\xi)
=E_m^{\underline a}E_n^{\underline b}\eta_{\underline{ab}}.
$$

The $\eta=0$ component of the left--hand side of eq. (\ref{5.1.25}) is
rewritten as follows
\begin{eqnarray}\label{5.1.0026}
&& \eta^{ab}(\nabla_aE_b^{\underline a})u_{\underline a}^{~i}|_{\eta=0}=
\eta^{ab}e_a^m(\xi)(\nabla_me_b^nE_n^{\underline a})
u_{\underline a}^{~i}(\xi)-\eta^{ab}\psi_a^{\alpha q}(\xi)
(\nabla_{\alpha q}E_b^{\underline a})u_{\underline
a}^{~i}(\xi)\nonumber\\
&=&{1\over{\sqrt{-g}}}\nabla_m\left({\sqrt{-g}}g^{mn}
E_n^{\underline a}
\right)
u_{\underline a}^{~i}-2ig^{mn}E^{\underline\alpha}_m
v_{\underline\alpha}^{~\alpha q}E^{\underline\beta}_n
v_{\underline\beta,\alpha r'}(\gamma^i)_q^{~r'}\\
&=&{1\over{\sqrt{-g}}}\nabla_m\left({\sqrt{-g}}g^{mn}
E_n^{\underline a}
\right)
u_{\underline a}^{~i}-2iE^{\underline\alpha}_m
v_{\underline\alpha}^{~\alpha q}(\gamma^{[m}\gamma^{n]})_{\alpha\beta}
E^{\underline\beta}_n
v_{\underline\beta}^{~\beta r'}(\gamma^i)_{qr'}\nonumber\\
&=&{1\over{\sqrt{-g}}}\nabla_m\left({\sqrt{-g}}g^{mn}
E_n^{\underline a}
\right)
u_{\underline a}^{~i}-{{2i}\over{\sqrt{-g}}}\epsilon^{lmn}
E^{\underline\alpha}_m
v_{\underline\alpha}^{~\alpha q}(\gamma^l)_{\alpha\beta}
E^{\underline\beta}_n
v_{\underline\beta}^{~\beta r'}(\gamma^i)_{qr'}\nonumber\\
&=&\left[{1\over{\sqrt{-g}}}\partial_m\left({\sqrt{-g}}g^{mn}
E_n^{\underline a}
\right)+g^{mn}E^{\underline A}_mE^{\underline b}_n
\Omega_{\underline A,{\underline b}}^{~~~~\underline a}\right]
u_{\underline a}^{~i}-{i\over{\sqrt{-g}}}\epsilon^{lmn}E_l^{\underline
b}
\bar E_m\Gamma_{\underline {b}}^{~~\underline a}E_n
u_{\underline a}^{~i}. \nonumber
\end{eqnarray}
When developing eq. (\ref{5.1.0026}) we used the relations
(\ref{5.1.023})--(\ref{psi}) and (\ref{5.1.5}), the harmonic relations
(\ref{4.25}), the $\Gamma$--matrices (\ref{5.1.1}) and  (\ref{5.1.2}),
and the fermionic equation (\ref{5.1.20}).

Putting together (\ref{5.1.026}) and (\ref{5.1.0026}) we finally get
\begin{equation}\label{5.1.26}
\left[{1\over{\sqrt{-g}}}\partial_m\left({\sqrt{-g}}g^{mn}
E_n^{\underline a}
\right)+g^{mn}E^{\underline A}_mE^{\underline b}_n
\Omega_{\underline A,{\underline b}}^{~~~~\underline
a}+{1\over{3!\sqrt{-g}}}
\epsilon^{lmn}
E_l^{\underline D}E_n^{\underline C}E_n^{\underline B}
F_{\underline{BCD}}^{~~~~~\underline a} \right]
u_{\underline a}^{~i}
=0,
\end{equation}
where
$$
\epsilon^{lmn}E_l^{\underline D}E_m^{\underline C}E_n^{\underline B}
F_{\underline{BCD}}^{~~~~~\underline a}=\epsilon^{lmn}[
E_l^{\underline d}E_m^{\underline c}E_n^{\underline b}
F_{\underline{bcd}}^{~~~\underline a} -3!iE_l^{\underline b}
\bar E_m\Gamma_{\underline {b}}^{~~\underline a}E_n]
$$
is the pullback onto the bosonic worldvolume of the constrained field
strength (\ref{5.3}).

In the square brackets of (\ref{5.1.26}) one can recognize the
conventional
scalar field equations of the supermembrane \cite{bst1} obtained from
the
action (\ref{5.01}). If we multiply
them by $E_{m\underline a}$ we may check that they are identically zero
(modulo the fermionic equations (\ref{5.1.0022})~), which reflects the
$d=3$
worldvolume reparametrization invariance of the theory. Hence, indeed,
only
eight worldvolume scalar field equations are independent.

To conclude this subsection we repeat the main expressions which
describe
the $D=11$ supermembrane in the superembedding approach.

\subsubsection{Main superembedding equations for the M2--brane}

The superembedding conditions are
\begin{equation}\label{4.22s}
E^{\underline a}(X^{\underline m}(z),\Theta^{\underline\mu}(z))=
e^a u_a^{~\underline a}(z)
\quad \Rightarrow \quad E_{\alpha q}^{\underline a}=0,
\end{equation}
\begin{equation}\label{4.23s}
E^{\underline\alpha}(X^{\underline m}(z),\Theta^{\underline\mu}(z))=
{1\over 2}(1-\bar\Gamma)^{\underline\alpha}_{~\underline\beta}\,
e^a E_a^{\underline\beta}
+e^{\alpha q} v_{\alpha q}^{~\underline\alpha}(z),
\end{equation}
$$
{1\over 2}(1-\bar\Gamma)_{~\underline\beta}
^{\underline\alpha}\equiv v^{~~\alpha q'}_{\underline\beta}
v_{\alpha q'}^{~~\underline\alpha},
\quad \bar\Gamma={1\over{3!}}\epsilon^{abc}\Gamma_{abc}, \quad
\Gamma_a=E^{\underline a}_a\Gamma_{\underline a}.
$$
This form of the projector was introduced in \cite{bzm,banquant}.

The supermembrane equations of motion (encoded in (\ref{4.22s}) and
(\ref{4.23s})~) are
\begin{equation}\label{5.1.20s}
(\nabla_{\beta r}E_a^{\underline a})u_{\underline a}^{~i}
(\gamma^a)^{\alpha\beta}=0
=(\gamma^{a})^{\alpha}_{~\beta}E_a^{\underline\beta}
v_{\underline\beta}^{~\beta q'},
\end{equation}
\begin{equation}\label{5.1.25s}
 \left(\eta^{ab}\nabla_aE_b^{\underline a}+{1\over 3!}
\epsilon^{abc}
E_a^{\underline d}E_b^{\underline c}E_c^{\underline b}
F_{\underline{bcd}}^{~~~\underline a}\right)u_{\underline a}^{~i}=0.
\end{equation}

In the static gauge $X^a=\xi^a$,
$\Theta^{\underline\alpha}v_{\underline\alpha}^{~\beta r}=\eta^{\beta
r}$,
in flat target superspace and in the flat (linearized) limit of
the worldvolume supergeometry, where $E_{\alpha q}^{\underline a}=
D_{\alpha q}X^{\underline a}-iD_{\alpha q}\Theta\Gamma^{\underline a}
\Theta$, the superembedding condition reduces to the superfield
constraint
on the transversal oscillations of the supermembrane
\begin{equation}\label{5.1.26s}
D_{\alpha q}X^i=i(\gamma^i)_q^{~q'}\theta_{\alpha q'},
\quad \theta_{\alpha q'}
\equiv \Theta^{\underline\alpha}v_{\underline\alpha,\alpha q'}.
\end{equation}
This constraint describes an $n=8$, $d=3$ on--shell scalar
supermultiplet.

Note that acting on $\Theta^{\underline\alpha}$ by the harmonics
$v_{\underline\alpha}^{~\beta r}$ and $v_{\underline\alpha,\alpha q'}$
we transform the worldvolume scalar fermions $\Theta^{\underline\alpha}$
into the worldvolume spinors $\theta^{\beta r}$ and $\theta_{\alpha
q'}$.
When this is done, eq. (\ref{5.1.20s}) becomes the standard $d=3$
Dirac equation for $\theta_{\alpha q'}$. Indeed, using the
superembedding relations, it is a simple exercise to varify that in flat
target superspace, where
$E^{\underline\alpha}=d\Theta^{\underline\alpha}$,
eq. (\ref{5.1.20s}) can be rewritten (modulo (\ref{5.1.25})
or (\ref{5.1.25s})) in the Dirac form
\begin{equation}\label{5.1.25ss}
(\gamma^a)^{\alpha\beta}{\cal D}_a\theta_{\beta q'}=0,
\end{equation}
where
${\cal
D}_a=e_a^{~M}\partial_M+\omega_{a\beta}^{~~\gamma}+A_{aq'}^{~~r'}$
is the worldvolume covariant derivative.

Thus, the supersymmetric field theory in the worldvolume of the
supermembrane governed by eqs. (\ref{5.1.20s}) and (\ref{5.1.25s}) is
that of the  $n=8$, $d=3$ scalar supermultiplet having
eight bosonic and eight fermionic physical modes.

\subsection{The super--5--brane}
We now turn to the most impressive example of the power
of superembedding which has allowed one to get the on--shell
description of the M--theory 5--brane without using the action principle
\cite{hs2,hsw1}.

The supersurface ${\cal M}_{6,16}$ associated with the super--5--brane
worldvolume is characterized by a local supervielbein frame
$e^A(z)=(e^a,~e^{\alpha q})$
$(a=0,1,...,5$, $\alpha=1,...,4;$ $q=1,...,4)$ whose components form the
vector and spinor representations of the group
$SO(1,5)\times SO(5)$. Namely, the index $\alpha$ stands for a spinor
representation of $SU^*(4)\sim Spin(1,5)$ and the index $q$ is that of
the
spinor representation of $USp(4)\sim Spin(5)$.

The $d=6$ spinors are $USp(4)$ simplectic Majorana--Weyl spinors
\cite{kugo,west}.
They are
defined in the same way as the $SU(2)$ simplectic Majorana--Weyl
spinors,
which we introduced in Subsection 2.5.4,
eqs. (\ref{d6mw})--(\ref{d6gamma}).
\begin{equation}\label{d6mws}
\overline{e^{\alpha q}}:= {\bar e}^{\dot\alpha}_{q}
=B^{\dot\alpha}_{~\beta}e^{\beta r}C_{rq},
\end{equation}
where $C_{qr}$ is the antisymmetric $USp(4)$ invariant tensor, its
inverse
being $(C^{-1})^{qr}=-C^{qr}\equiv -C_{qr}$. The matrices
$C^{qr}$ and $C_{qr}$
can be used to raise and lower the $USp(4)$ spinor indices using the
rules
$e_r=e^qC_{qr}$, $e^q=C^{qr}e_r$.

The matrix $B$ is defined by the conditions
$$
B\gamma^{a}B^{-1}=(\gamma^{a})^*, \quad B^*B=-1
$$
and * denotes complex conjugation.

The $4\times 4$ matrices $(\gamma^{a})_{\alpha\beta}$ are antisymmetric.

We note that the matrix $B^{\dot\alpha}_{~\beta}$ can be used to convert
the
dotted indices (of the complex conjugate representation)
into undotted ones, so that one can always deal with
only undotted indices, but there is no $SU^*(4)$ invariant tensor for
lowering the $SU^*(4)$ spinor indices. Thus, the spinors $\psi^{\alpha
q}$
and $\psi_{\alpha}^q$ have different
$SU^*(4)$ chiralities,
one of them is chiral (Weyl) and another one is antichiral.

The supersurface ${\cal M}_{6,16}$ under consideration
is therefore a $d=6$, $(2,0)$ chiral superspace (where 2 stands for
$USp(4)$)
\footnote{The $N=1$, $D=6$ superspace with one $SU(2)$ simplectic
Majorana--Weyl
coordinate (discussed in Subsection 2.5.4) is also called  $D=6$, (1,0)
chiral superspace, since the number of its Grassmann coordinates is half
the
number of those in the $(2,0)$  superspace.}.

An appropriate form of the $D=11$ $\Gamma$--matrices (\ref{4.26}),
(\ref{4.27}) and of the charge conjugation matrices
$C_{\underline\alpha\underline\beta}=C^{\underline\alpha\underline\beta}$
, which reflects the embedding of ${\cal M}_{6,16}$ into
$\underline{\cal M}_{11,32}$, is
\begin{equation}\label{5.2.1}
\Gamma^a_{\underline\alpha\underline\beta}=\left(
\begin{array}{cc}
-\gamma^a_{\alpha\beta}C_{qr}  &  0\\
0  &  \tilde\gamma^{a\alpha\beta}C^{qr}
\end{array}
\right)\,, \quad a=0,1,...,5\, ,
\end{equation}
\begin{equation}\label{5.2.2}
\Gamma^i_{\underline\alpha\underline\beta}=\left(
\begin{array}{cc}
0  &  \delta^\alpha_\beta(\gamma^{i})_q^{~r}\\
-\delta_\alpha^\beta(\gamma^i)^q_{~r}   &  0
\end{array}
\right)\,, \quad i=1,...,5\,,
\end{equation}
\begin{equation}\label{5.2.C}
C_{\underline\alpha\underline\beta}=C^{\underline\alpha\underline\beta}
=\left(
\begin{array}{cc}
0  &  \delta^\alpha_\beta \delta_q^{r}\\
-\delta_\alpha^\beta\delta^q_{r}   &  0
\end{array}
\right)\,.
\end{equation}
In (\ref{5.2.2}) $(\gamma^i)^q_{~r}=C^{qs}(\gamma^{i})_s^{~t}C_{tr}$ are
$USp(4)\sim SO(5)$ gamma--matrices. The matrices
$(\gamma^{i})_{qr}=(\gamma^{i})_q^{~t}C_{tr}$ are antisymmetric.

In (\ref{5.2.1}) the $SU^*(4)$ matrices
$\gamma^{a}_{\alpha\beta}$ and $\tilde\gamma^{a\alpha\beta}$
are antisymmetric and defined by
the following relations
\begin{equation}\label{5.2.03}
\gamma^a_{\alpha\gamma}\tilde\gamma^{b\gamma\beta}+
\gamma^b_{\alpha\gamma}\tilde\gamma^{a\gamma\beta}
=2\delta^{\beta}_\alpha\eta^{ab},
\qquad {\rm tr}(\gamma^a\tilde\gamma^b)=4\eta^{ab}, \quad
\gamma_{a\alpha\beta}\gamma^a_{\gamma\delta}
=-2\epsilon_{\alpha\beta\gamma\delta}.
\end{equation}
\begin{equation}\label{5.2.003}
(\gamma^{[a}\tilde\gamma^{b}
\gamma^{c]})_{\alpha\beta}\equiv \gamma^{abc}_{\alpha\beta}=
\gamma^{abc}_{\beta\alpha}
=-{1\over 6}\epsilon^{abcdef}(\gamma_{def})_{\alpha\beta}.
\end{equation}
\begin{equation}\label{5.2.0003}
(\tilde\gamma^{[a}\gamma^{b}
\tilde\gamma^{c]})^{\alpha\beta}={1\over
6}\epsilon^{abcdef}(\tilde\gamma_d
\gamma_e\tilde\gamma_f)^{\alpha\beta}.
\end{equation}

Eqs. (\ref{5.2.003}) and (\ref{5.2.0003})
imply that the antisymmetric product of three
$SO(1,5)$ $\gamma$--matrices is symmetric in spinor indices and
(anti)--self--dual in vector indices.

The $SU^*(4)\times USp(4)$ splitting of the
$D=11$ Lorentz--spinor harmonics
$v_{\underline\alpha}^{~\underline\beta}$
and their inverse is as follows
\begin{equation}\label{5.2.04}
v_{\underline\alpha}^{~\underline\beta}=
(v_{\underline\alpha}^{~\beta r},v_{\underline\alpha,\beta r}), \quad
(v^{-1})^{~\underline\alpha}_{\underline\beta}=
(v^{~~\underline\alpha}_{\beta r},v^{\beta r,\underline\alpha}).
\end{equation}

The induced supergeometry on ${\cal M}_{6,16}$ is described by the
general
superembedding conditions (\ref{4.07})--(\ref{4.24}),
(\ref{4.29})--(\ref{4.34}), (\ref{4.038})--(\ref{4.41}), (\ref{4.46}).
As in the case of the supermembrane, our goal is to find further
restrictions on components of
$\hat\Omega_a^{~i}=e^A\hat\Omega_{A,a}^{~~~i}$
required by the integrability of the superembedding conditions and to
identify these restrictions with the 5--brane equations of motion.
So we repeat the steps made in the previous subsection.

We again analyze the condition (\ref{4.39}).
Making use of
the harmonic relations (\ref{4.25}) and (\ref{5.2.04}), the form of the
$D=11$ $\Gamma$--matrices (\ref{5.2.1}) and (\ref{5.2.2}),
and the equations (\ref{4.29}) and (\ref{4.32}),
we rewrite eq. (\ref{4.39})
in the following form
\begin{eqnarray}\label{5.2.3}
e^a\hat\Omega_a^{~i}&=&e^a{\nabla} E_a^{~\underline b}u_{\underline
b}^{~i}
=e^a(du_a^{~\underline b}+u_a^{~\underline c}
\Omega_{\underline c}^{~\underline b}) u_{\underline b}^{~i}
\nonumber \\
&=&
-2i e^{\alpha q}e^{\beta r}(\gamma^i)_q^{~s}h_{\alpha r,\beta s}
+2ie^ae^{\alpha q}(\gamma^i)_{q}^{~r}E_a^{\underline\beta}
v_{\underline\beta,\alpha r}.
\end{eqnarray}
Comparing the left-- and the right--hand side of (\ref{5.2.3}) we find
that
\begin{equation}\label{5.2.4}
e^ae^b\hat\Omega_{b,a}^{~~i}=0 \quad \rightarrow \quad
\hat\Omega_{b,a}^{~~~i}=\hat\Omega_{a,b}^{~i}
={\nabla}_b E_a^{~\underline b}u_{\underline b}^{~i},
\end{equation}
\begin{equation}\label{5.2.5}
\hat\Omega_{\alpha q, a}^{~~~~i}
={\nabla}_{\alpha q} E_a^{~\underline b}u_{\underline b}^{~i}=
2i(\gamma^i)_q^{~r}E_a^{\underline\beta}
v_{\underline\beta,\alpha r},
\end{equation}
and
\begin{equation}\label{5.2.6}
(\gamma^i)_{q}^{~s}h_{\alpha r,\beta s}
=-(\gamma^i)_{r}^{~s}h_{\beta q,\alpha s}.
\end{equation}

In the basis of the $SO(1,5)\times SO(5)$ $\gamma$--matrices an
arbitrary matrix $h_{\alpha q,\beta r}$ has the following decomposition
\begin{equation}\label{5.2.7}
h_{\alpha q,\beta r}
=C_{qr}[h_a\gamma^a_{\alpha\beta}
+{1\over 6}h_{abc}\gamma^{abc}_{\alpha\beta}]+
\gamma^{ij}_{qr}[h_{[ij]a}\gamma^a_{\alpha\beta}
+h_{[ij]abc}\gamma^{abc}_{\alpha\beta}].
\end{equation}
Substituting eq. (\ref{5.2.7}) into (\ref{5.2.6}) we find that the only
nonzero component of $h_{\alpha q,\beta r}$ is
\begin{equation}\label{5.2.8}
h_{\alpha q,\beta r}=C_{qr}h_{\alpha\beta}
={1\over 6}C_{qr}h_{abc}\gamma^{abc}_{\alpha\beta}, \quad
h_{abc}={1\over 6}\epsilon_{abcdef}h^{def},
\end{equation}
where $h_{abc}(z)$ is a self--dual tensor due to the $\gamma$--matrix
relations (\ref{5.2.003}).

The matrix $h_{\alpha\beta}$ satisfies the identities
\begin{equation}\label{hh1}
h_{\alpha\gamma}\tilde\gamma^{a\gamma\delta}h_{\delta\beta}
=-2h^{abc}h_{bcd}\gamma^d_{\alpha\beta},
\end{equation}
\begin{equation}\label{hh2}
\tilde\gamma_a^{\alpha\gamma}h_{\gamma\delta}\tilde\gamma^{a\delta\beta}=0,
\end{equation}
\begin{equation}\label{hh3}
{\rm tr}({h\tilde\gamma}_{[a}\gamma_{b}
\tilde\gamma_{c]})=-8h_{abc},
\end{equation}
which we shall use below.

We have thus observed that the integrability of the
superembedding conditions reveals
the presence of a self--dual tensor field on the embedded supersurface.
Later on this self--dual tensor will be related to
the field strength $H^{(3)}=db^{(2)}-A^{(3)}$ of the 5--brane tensor
gauge
field which appeared in the M5--brane action (\ref{5.001}).

In view of eq. (\ref{5.2.8}) the general embedding condition
(\ref{4.34})
reduces to
\begin{equation}\label{5.2.9}
E^{\underline\beta}_{\alpha q}\equiv
e_{\alpha q}^M\partial_MZ^{\underline M}E_{\underline
M}^{\underline\beta}=
v^{~~\underline\beta}_{\alpha q}(z)+h_{\alpha\beta}C_{qr}
v^{\beta r,\underline\beta}.
\end{equation}

We may now derive the $e^{\alpha q}e^{\beta r}$ components of the
supersurface torsion (\ref{4.38}) and
(\ref{4.044}).

Using eqs. (\ref{4.38}) and (\ref{5.2.9}), the harmonic relations
(\ref{4.25}), the $\Gamma$--matrix decomposition (\ref{5.2.1}) and
the identity (\ref{hh1})
we get
\begin{equation}\label{5.2.10}
T^a_{\alpha q,\beta r}=-2iC_{qr}\gamma^b_{\alpha\beta}m_b^{~a},
\quad m_{b}^{~a} \equiv \delta_{b}^{~a}-2h_{bcd}h^{acd}.
\end{equation}

We observe that eq. (\ref{5.2.10})
differs from the standard supergravity torsion constraint
(\ref{5.1.12}) by the non--unit matrix $m_{b}^{~a}$.
The reason
for this is that (as in the case of the bosonic surface)
we have chosen the bosonic embedding condition
(\ref{4.07}) such that the induced metric of the embedded surface
is of the standard `Green--Schwarz' type (\ref{4.014}) which is used
in the construction of the M5--brane action (\ref{5.001}).
If we wanted to have the standard constraint on the torsion of
the supersurface ${\cal M}_{6,16}$ (i.e. with $\delta_{b}^{~a}$ instead
of $m_{b}^{~a}$ in (\ref{5.2.10})), we should redefine the
${\cal M}_{6,16}$ supervielbein $e^a$ in (\ref{4.07}) as follows
$$
E^a\equiv E^{\underline b}u^{~a}_{\underline b}(z)
=e^a(z)=\hat e^bm_b^{~a},
$$
The price for this would be that in the new frame $\hat e^a$ the metric
$\hat g_{mn}=\hat e^a_m\hat e_{na}$ on ${\cal M}_{6,16}$ does not
coincide
with the induced metric of the Green--Schwarz--type.
So we prefer to work in the standard (induced)
frame on ${\cal M}_{6,16}$ \footnote{The $\hat e^a$--frame has been
used for the
description of the M5--brane superembedding in the reference \cite{hs2},
and
the transition to the Green--Schwarz frame has been discussed in
\cite{hsw1}.}.

We now look at the $e^{\alpha q}e^{\beta r}$ components of the torsion
$T^{\gamma s}$ (\ref{4.044}). Because of our choice of the supersurface
connection (\ref{4.43}), (\ref{4.46}) and the presence of the tensor
$h(z)$,
they are nonzero and have the following form
\begin{equation}\label{5.2.11}
T^{\gamma s}_{\alpha q,\beta r}=
-C_{qt}h_{\alpha\delta}(\nabla_{\beta r} v^{\delta t,\underline\alpha})
v_{\underline\alpha}^{~\gamma s}+ (\alpha q)~\leftrightarrow~(\beta r) .
\end{equation}
In view of (\ref{5.2.10}) and (\ref{5.2.11})
the anticommutator of the supersurface covariant derivatives
${\cal D}_{\alpha q}$ gets modified and acquires an additional term
in comparison with that
of the supermembrane (\ref{5.1.14})
\begin{equation}\label{5.2.12}
\{{\cal D}_{\alpha q},{\cal D}_{\beta r}\}
=2iC_{qr}\gamma^a_{\alpha\beta}m_a^{~b}{\cal D}_b
-T^{\gamma s}_{\alpha q,\beta r}{\cal D}_{\gamma s}
+R_{\alpha q,\beta r}(z).
\end{equation}

\subsubsection{The tensor field equation}

We are now in a position to relate $h_{abc}(z)$ to the field strength
$H_{lmn}(\xi)$ of the worldvolume gauge field $b_{mn}(\xi)$ of
the M5--brane action.
By definition, in the bosonic worldvolume (\ref{5.1a})
\begin{equation}\label{t1}
H^{(3)}=db-A^{(3)}.
\end{equation}
The Bianchi identity for (\ref{t1}) is
\begin{equation}\label{t2}
dH^{(3)}=-dA^{(3)}=F^{(4)},
\end{equation}
where $A^{(3)}$ and $F^{(4)}$ are the pullbacks onto the bosonic
worldvolume
of the $D=11$ gauge field potential and of its field strength,
respectively.

To find the relation between $H^{(3)}$ and $h_{abc}(\xi,\eta)$ we should
promote the equations (\ref{t1}) and (\ref{t2}) to the supersurface
${\cal M}_{6,16}$, i.e. to consider $b^{(2)}(z)$ and $H^{(3)}(z)$
as ${\cal M}_{6,16}$ superfields.

Since $h_{abc}(z)$ carries only vector indices it is natural to assume
that the superfield
$$
H_{ABC}=e^{~L}_Ae^{~M}_Be^{~N}_CH_{LMN}(Z),
$$
also has only nonzero components with three vector indices $H_{abc}(z)$,
i.e.
\begin{equation}\label{t3}
H_{\alpha q,BC}(z)=0.
\end{equation}

The assumption (\ref{t3}) about the structure of the
superform $H^{(3)}$ is consistent with the Bianchi identity (\ref{t2})
extended to the supersurface,
which allows one to express $H_{abc}(z)$ in terms of $h_{abc}(z)$ and to
find the equation of motion of $h_{abc}(z)$ (or $H_{abc}(z)$), as we
shall
do in a moment.

We should stress in advance that eq. (\ref{t3}) is {\it not}
an additional constraint on the superembedding since, as the analysis
has shown \cite{hs2,hsw1}, the equations for $h_{abc}(z)$ which follow
from
the Bianchi identity (\ref{t2}) are equivalent to those which are
contained
in the basic superembedding condition $E^{\underline a}_{\alpha q}=0$
(\ref{4.24}).

Let us analyze the Bianchi identity (\ref{t2}).
To this end we rewrite it in the following form
\begin{eqnarray}\label{t4}
&&dH^{(3)}={1\over 2}({\cal D}e^C)e^Be^AH_{ABC}
+ {1\over 6}e^De^Ce^Be^A{\cal D}_AH_{BCD} \nonumber \\
&=&
{1\over 2}T^Ce^Be^AH_{ABC}
+ {1\over 6}e^De^Ce^Be^A{\cal D}_DH_{ABC}
=-{1\over 4!}e^De^Ce^Be^AF_{ABCD}.
\end{eqnarray}
In view of (\ref{t3}), of the torsion constraint (\ref{5.2.10}) and of
the
$D=11$ field strength constraint (\ref{5.3}), the first nonzero
component of eq. (\ref{t4}) is that of $e^{\alpha q}e^{\beta r}e^ce^d$
\begin{equation}\label{t5}
iC_{qr}\gamma^a_{\alpha\beta}m_a^{~b}H_{bcd}=-iE^{\underline\alpha}_{\alpha
q}
(\Gamma_{\underline{ab}})_{\underline\alpha\underline\beta}
E^{\underline\beta}_{\beta r}
E^{\underline a}_cE^{\underline b}_d.
\end{equation}
Making use of the harmonic relations (\ref{4.25}) and (\ref{5.2.04}),
the form
(\ref{5.2.1})--(\ref{5.2.3}) of the $\Gamma$--matrices, the
relation (\ref{5.2.9}) for $E^{\underline\alpha}_{\alpha q}$,
and that $E^{\underline a}_c=u^{\underline a}_c$, we reduce (\ref{t5})
to
\begin{equation}\label{t6}
\gamma^a_{\alpha\beta}m_a^{~d}H_{dbc}
=-(\gamma_{[b}\tilde\gamma_{c]})_\alpha^{~\gamma}
h_{\gamma\beta}-h_{\alpha\gamma}
(\tilde\gamma_{[b}\gamma_{c]})_{~\beta}^{\gamma}.
\end{equation}
Then, multiplying (\ref{t6}) by $\tilde\gamma_a^{\beta\alpha}$ and
making
use of the identities (\ref{5.2.03}) and (\ref{hh3}), we finally
arrive at the relation between $H_{abc}$ and $h_{abc}$
\begin{equation}\label{t7}
m_a^{~d}H_{dbc}=4h_{abc}~\Leftrightarrow~
H_{abc}=4(m^{-1})_a^{~d}h_{dbc}.
\end{equation}

Using the self--duality
of $h_{abc}$ and the definition (\ref{5.2.10}) of
$m_a^{~b}=\delta_{a}^{~b}-2h_{acd}h^{bcd}$ one can show that
\begin{equation}\label{t8}
(m^{-1})_a^{~b}={1\over{1-{{2\over 3}k^2}}}(2\delta_{a}^{~b}-m_a^{~b}),
\quad k_a^{~b}\equiv h_{acd}h^{bcd}, \quad k^2\equiv{1\over 6}{\rm
tr}({kk}).
\end{equation}
It is then not hard to check that
the r.h.s. of (\ref{t7}) is indeed totally
antisymmetric because of the self--duality of $h_{abc}$. For this one
should notice that $k_a^{~d}h_{dbc}$ is anti--self--dual
\begin{equation}\label{t9}
k_a^{~d}h_{dbc}=-{1\over 6}\epsilon_{abcdef}k^{dd_1}h_{d_1}^{~~ef}.
\end{equation}

The form of the expressions (\ref{t7}) implies that though $h_{abc}$
obeys ordinary Hodge self--duality (\ref{5.2.8}),
the field strength $H_{abc}$ satisfies
a much more complicated nonlinear self--duality condition. Its explicit
form can be derived from the M5--brane action (\ref{5.001})
as an equation of motion of $b_{mn}$ \cite{perry,pst1,m5,s1}, or
directly from
(\ref{t7}) \cite{hsw2}. We refer the reader to these papers for details
on the
different forms of the generalized self--duality condition.

Note that at the linearized level, when $m_{b}^{~a}$
is replaced with $\delta_{b}^{~a}$, $H_{abc}$ becomes proportional
to $h_{abc}$ and satisfies the ordinary Hodge self--duality condition.

When the field strength of a gauge field is self--dual, its Bianchi
identities
are equivalent to the gauge field equations of motion. We shall now show
how the Bianchi identities (\ref{t2}) produce the tensor field equations
in the case under consideration.

Consider the $e^ae^be^ce^d$ component of the Bianchi identity
(\ref{t4}).
It has the form
\begin{equation}\label{t10}
{1\over 6}\epsilon^{abcdef}\left({\cal
D}_cH_{def}+3T^{f_1}_{cd}H_{eff_1}
\right)=-{1\over{4!}}\epsilon^{abcdef}E^{\underline D}_cE^{\underline
C}_d
E^{\underline B}_eE^{\underline A}_fF_{\underline {ABCD}},
\end{equation}
or
\begin{equation}\label{t11}
{1\over 6}\epsilon^{abcdef}\hat{\cal D}_cH_{def}=
-{1\over{4!}}\epsilon^{abcdef}E^{\underline D}_cE^{\underline C}_d
E^{\underline B}_eE^{\underline A}_fF_{\underline{ABCD}},
\end{equation}
where, for convenience,
we have introduced the covariant derivative $\hat{\cal D}_c$
\begin{equation}\label{t12}
\hat{\cal D}_c=e_c^{~M}\partial_M+\hat\omega_{c,ba}=
e_c^{~M}\partial_M+
\left(\omega_{c,ba}-{1\over 2}(T_{cb,a}-T_{ca,b}-T_{ba,c})\right).
\end{equation}
Note that in the pure bosonic limit the connection $\hat\omega_{ba}$ is
torsion free, i.e. for such a connection the components $\hat T^a_{bc}$
of the torsion $\hat T^a=\hat{\cal D}e^a$ are zero.

We now substitute $H_{def}$ in (\ref{t11}) with
$(m^{-1})_d^{~f_1}h_{f_1ef}$
(\ref{t7}),
and use the relations (\ref{t8}), (\ref{5.2.10}) and (\ref{t9}) to get
\begin{equation}\label{t13}
4\hat{\cal D}_c({1\over{1-{{2\over 3}k^2}}}m^c_{d}h^{dab})=
-{1\over{4!}}\epsilon^{abcdef}E^{\underline D}_cE^{\underline C}_d
E^{\underline B}_eE^{\underline A}_fF_{\underline{ABCD}}.
\end{equation}
Eq. (\ref{t13}) is the equation of motion for the self--dual
tensor field $h_{abc}$.

We conclude that the Bianchi identity (\ref{t2}) with $H^{(3)}$
defined by the relation (\ref{t7}) is equivalent to the field equations
of the self--dual worldvolume tensor field.

Let us now proceed with deriving

\subsubsection{The fermionic equation}
To get the fermionic field $\Theta^{\underline\mu}(z)$
equation of the $M5$--brane
we hit the left-- and the right--hand side of
eq. (\ref{5.2.9}) with the covariant derivative ${\cal D}_{\beta r}$,
symmetrize with respect to the pairs of indices $(\alpha q)$ and
$(\beta r)$, and multiply the resulting expression by
$v_{\underline\beta,\gamma s}$. We thus obtain
\begin{eqnarray}\label{5.2.13}
&&iC_{qr}\gamma^a_{\alpha\beta}m_a^{~b}E_b^{\underline\beta}
v_{\underline\beta,\gamma s}
+C_{qt}h_{\alpha\delta}(\nabla_{\beta r} v^{\delta t,\underline\alpha})
v_{\underline\alpha}^{~\sigma t_1}
E_{\sigma t_1}^{\underline\beta}v_{\underline\beta,\gamma s}-
E_{\alpha q}^{\underline\alpha}
\Omega_{\beta r,\underline\alpha}^{~~~~\underline\beta}
v_{\underline\beta,\gamma s}+ (\alpha q)~\leftrightarrow ~(\beta r)
\nonumber \\
&&={\cal D}_{\beta r}v^{~\underline\beta}_{\alpha q}
v_{\underline\beta,\gamma s}+({\cal D}_{\beta r}h_{\alpha\delta}C_{qt}
v^{\delta t,\underline \beta})v_{\underline\beta,\gamma s}+
 (\alpha q)~\leftrightarrow ~(\beta r).
\end{eqnarray}

To simplify eq. (\ref{5.2.13}) we should make use of the expression
(\ref{5.2.9}), the identity (\ref{hh1}), the relations
(\ref{4.43}) and (\ref{4.46}) between the
supersurface connection $\omega_A^{~B}$ and the target superspace
connection
$\Omega_{\underline A}^{~\underline B}$, and to recall that
due to the harmonic relations (\ref{4.41}) and (\ref{4.53}), and the
form of the $\Gamma$--matrices (\ref{5.2.1}) and (\ref{5.2.2})
\begin{eqnarray}\label{5.2.0014}
&&(\nabla_{\beta r}v^{~\underline\beta}_{\alpha q})
v_{\underline\beta,\gamma s}=
{1\over 4}(\nabla_{\beta r}u_a^{~\underline a})u_{\underline a}^{~i}
\gamma^a_{\alpha\gamma}(\gamma_{i})_{qs}
\equiv{1\over 4}\hat\Omega_{\beta r,a}^{~~~~i}
\gamma^{a}_{\alpha\gamma}(\gamma_{i})_{qs}, \nonumber \\
&&
(\nabla_{\beta r}v^{\alpha q,\underline\beta})
v_{\underline\beta}^{~\gamma s}=
{1\over 4}(\nabla_{\beta r}u_a^{~\underline a})u_{\underline a}^{~i}
\gamma^{a\alpha\gamma}(\gamma_{i})^{qs}
\equiv{1\over 4}\hat\Omega_{\beta r,a}^{~~~~i}
\gamma^{a\alpha\gamma}(\gamma_{i})^{qs}.
\end{eqnarray}
We thus reduce (\ref{5.2.13}) to
\begin{equation}\label{5.2.14}
2iC_{qr}\gamma^a_{\alpha\beta}m_a^{~b}E_b^{\underline\beta}
v_{\underline\beta,\gamma s}=
{1\over 4}\gamma^b_{\alpha\gamma}m_b^{~a}\hat\Omega_{\beta r,a}^{~~~~i}
(\gamma_{i})_{qs}+{\cal D}_{\beta r}h_{\alpha\gamma}C_{qs}+
(\alpha q)~\leftrightarrow ~(\beta r).
\end{equation}
Let us multiply eq. (\ref{5.2.14}) by $C^{qr}$ and
$\tilde\gamma_c^{\beta\alpha}\tilde\gamma^{c\delta\gamma}$. Because of
the
identity (\ref{hh2}) the term with the covariant derivative
of $h_{\alpha\gamma}$ vanishes and we get
\begin{equation}\label{5.2.014}
16i\tilde \gamma_b^{\delta\gamma}m^{ba}E_a^{\underline\beta}
v_{\underline\beta,\gamma s}
=-\tilde\gamma_b^{\delta\gamma}m^{ba}\hat\Omega_{\gamma q,a}^{~~~~i}
(\gamma_{i})^q_{~s}.
\end{equation}
Now take the relation (\ref{5.2.5}) and multiply it by
$(\gamma_i)^q_{~s}$
and $\tilde\gamma_b^{\delta\alpha}m^{ba}$. We have
\begin{equation}\label{5.2.15}
10i\tilde \gamma_b^{\delta\gamma}m^{ba}E_a^{\underline\beta}
v_{\underline\beta,\gamma s}
=-\tilde\gamma_b^{\delta\gamma}m^{ba}\hat\Omega_{\gamma q,a}^{~~~~i}
(\gamma_{i})^q_{~s}.
\end{equation}
Comparing (\ref{5.2.014}) with (\ref{5.2.15}) we see that the
coefficients
on their left--hand sides do not match, and hence (also in view of
(\ref{5.2.5}))
\begin{equation}\label{5.2.16}
\tilde \gamma_b^{\alpha\beta}m^{ba}E_a^{\underline\beta}
v_{\underline\beta,\beta q}=0=
\tilde\gamma_b^{\alpha\beta}m^{ba}\hat\Omega_{\beta q,a}^{~~~~i}.
\end{equation}

We have thus arrived at the {\bf 16}--component
fermionic field equation of motion of the
$M5$--brane similar to that of the supermembrane (\ref{5.1.20}), the
only difference being in the presence of the worldvolume self--dual
tensor $h_{abc}(z)$ in the $M5$--brane equation via the matrix
$m_b^{~a}=\delta_b^{~a}-h_{bcd}h^{acd}$.


Let us now rewrite eq. (\ref{5.2.16}) in the Green--Schwarz form similar
to the supermembrane equation (\ref{5.1.22}).

First of all we use the
harmonic relations (\ref{4.25}) and (\ref{5.2.04}),  the form of the
$\Gamma$--matrices (\ref{5.2.1}), and the embedding conditions
$E_a^{\underline\beta}v_{\underline\beta}^{~\alpha q}=0$ (\ref{4.30})
and
$E_a^{\underline a}=u_a^{~\underline a}$ (\ref{4.014}) to get
\begin{equation}\label{5.2.17}
m^{ba}E_a^{\underline\beta}
(E_b^{\underline a}\Gamma_{\underline
a})_{\underline\beta\underline\gamma}
v^{\alpha q,\underline\gamma}=0.
\end{equation}
Then, to `hide' the supersurface spinor indices,
we multiply (\ref{5.2.17}) by
\begin{equation}\label{5.2.018}
E_{\underline\alpha,\alpha q}\equiv v_{\underline\alpha\alpha q}
-v_{\underline\alpha}^{~\beta r}h_{\beta\alpha}C_{rq}.
\end{equation}
We get
\begin{equation}\label{5.2.18}
m^{ba}E_a^{\underline\beta}
(E_b^{\underline a}\Gamma_{\underline
a})_{\underline\beta\underline\gamma}
v^{\alpha q,\underline\gamma}E_{\underline\alpha,\alpha q}
\equiv
{1\over 2}m^{ba}E_a^{\underline\beta}
\left[E_b^{\underline a}\Gamma_{\underline a}(1-\bar\Gamma)\right]
_{\underline\beta\underline\alpha}
=0.
\end{equation}
The choice of the matrix $E_{\underline\alpha,\alpha q}$ is prompted
by the requirement that the resulting fermionic equation is invariant
under the $\kappa$--symmetry transformations (as the superbrane
dynamical
equations must always be). This means that this equation should be
annihilated
by the $\kappa$--symmetry projector (\ref{4.57}), (\ref{4.057})
whose form
is
\begin{equation}\label{5.2.0019}
{1\over 2}(1+\bar\Gamma)_{~~\underline\beta}^{\underline\alpha}\equiv
v_{\underline\beta}^{~\beta r}E_{\beta r}^{~\underline\alpha}=
v_{\underline\beta}^{~\beta r}(v_{\beta r}^{~\underline\alpha}
+C_{rs}h_{\beta\gamma}v^{\gamma s,\underline\alpha}).
\end{equation}
So we have
chosen $E_{\underline\alpha,\alpha q}$ in such a way that
$E_{\beta r}^{~\underline\alpha}E_{\underline\alpha,\alpha q}\equiv 0$.

Finally we use the form (\ref{5.2.8}) of $h_{\beta\alpha}$ and the
harmonic relations (\ref{4.25}) to find the explicit form of the matrix
$\bar\Gamma$ of eq. (\ref{5.2.18}) in terms of antisymmetric products of
the
$D=11$ $\Gamma$--matrices (\ref{5.2.1}),
(\ref{5.2.2})
\begin{equation}\label{5.2.19}
\bar\Gamma={1\over {6!}}\epsilon^{a_1...a_6}
\Gamma_{a_1...a_6}+{1\over 3}h^{abc}\Gamma_{abc},
\quad \Gamma_a=E_a^{\underline a}\Gamma_{\underline a}.
\end{equation}

Thus we have derived the Green--Schwarz--type fermionic equation of
motion
(\ref{5.2.18}) of the super--5--brane. In the Wess--Zumino gauge
(\ref{5.1.023}) its $\eta=0$ component has the form
\begin{equation}\label{5.2.019}
{1\over 2}m^{ba}e_b^me^n_aE_m^{\underline\beta}
\left[E_n^{\underline a}\Gamma_{\underline a}(1-\bar\Gamma)\right]
_{\underline\beta\underline\alpha}
=0,
\end{equation}
where $E_m^{\underline A}=\partial_mZ^{\underline M}E_{\underline M}
^{\underline A}(Z(\xi))$,
$({\underline A}={\underline \beta},{\underline a})$.

\subsubsection{The scalar equation}
To identify the equations of motion of the superworldvolume
scalar fields $X^{\underline m}(z)$ we
analyze the $e^ae^{\alpha q}$ components of the pullback of the
target superspace torsion
$T^{\underline\alpha}$ (\ref{5.2}) multiplied by
$E_{\underline\alpha,\beta r}$
defined in (\ref{5.2.018}).
\begin{eqnarray}\label{5.2.20}
T^{\underline\alpha}E_{\underline\alpha,\beta r}
&\equiv&(\nabla E^{\underline\alpha})E_{\underline\alpha,\beta
r}\nonumber \\
&&=T^AE^{\underline\alpha}_AE_{\underline\alpha,\beta r}+
e^A(\nabla E^{\underline\alpha}_A-\omega_A^{~B}E^{\underline\alpha}_B)
E_{\underline\alpha,\beta r} \nonumber \\
&&= T^bE^{\underline\alpha}_bE_{\underline\alpha,\beta r}+
e^A(\nabla E^{\underline\alpha}_A-\omega_A^{~B}E^{\underline\alpha}_B)
E_{\underline\alpha,\beta r}
\end{eqnarray}
where $T^A=(T^b,T^{\alpha q})$
is the supersurface induced torsion defined in (\ref{4.38}) and
(\ref{4.044}). Note that the term with $T^{\alpha q}$ has disappeared
from
(\ref{5.2.20}) since by definition
$E^{\underline\alpha}_{\alpha q}E_{\underline\alpha,\beta r}\equiv 0$
(see
eqs. (\ref{5.2.018}), (\ref{5.2.0019})~).
The $e^ae^{\alpha q}$ components of (\ref{5.2.20}) are
\begin{equation}\label{5.2.21}
T_{a,\alpha q}^{\underline\alpha}E_{\underline\alpha,\beta r}=
T_{a,\alpha q}^bE^{\underline\alpha}_bE_{\underline\alpha,\beta r}
-(\nabla_{\alpha q}E^{\underline\alpha}_a
+\omega_{\alpha q,a}^{~b}E^{\underline\alpha}_b)
E_{\underline\alpha,\beta r}-(\nabla_aE^{\underline\alpha}_{\alpha q})
E_{\underline\alpha,\beta r}.
\end{equation}
Using the definition of $E^{\underline\alpha}_{\alpha q}$ (\ref{5.2.9})
and
$E_{\underline\alpha,\beta r}$ (\ref{5.2.018}), i.e.
$$
E^{\underline\alpha}_{\alpha q}=
v^{~\underline\alpha}_{\alpha q}(z)+h_{\alpha\gamma}C_{qs}
v^{\gamma s,\underline\alpha}, \quad
E_{\underline\alpha,\beta r}=
v_{\underline\alpha\beta r}
-v_{\underline\alpha}^{~\gamma s}h_{\gamma\beta}C_{sr},
$$
the orthogonality properties of the harmonics (\ref{5.2.04}) and
the relations (\ref{4.43}), (\ref{4.45}) and (\ref{5.2.0014}) for the
supersurface and target superspace connections, the last term of
(\ref{5.2.21}) can be rewritten as
\begin{equation}\label{5.2.22}
(\nabla_aE^{\underline\alpha}_{\alpha q})
E_{\underline\alpha,\beta r}=
{1\over
4}\hat\Omega_{a,b}^{~~i}m_b^{~c}\gamma_{c\alpha\beta}(\gamma_i)_{qr}+
{\cal D}_ah_{\alpha\beta}C_{qr}.
\end{equation}
Thus eq. (\ref{5.2.21}) takes the following form
\begin{equation}\label{5.2.23}
{1\over
4}\hat\Omega_{a,b}^{~~~i}m_b^{~c}\gamma_{c\alpha\beta}(\gamma_i)_{qr}+
{\cal D}_ah_{\alpha\beta}C_{qr}=
(T_{a,\alpha q}^bE^{\underline\alpha}_b-T_{a,\alpha
q}^{\underline\alpha})
E_{\underline\alpha,\beta r}-K_{\alpha q,a,\beta r},
\end{equation}
where
$$
K_{\alpha q,a,\beta r}\equiv(\nabla_{\alpha q}E^{\underline\alpha}_a
+\omega_{\alpha q,a}^{~~~~b}E^{\underline\alpha}_b)
E_{\underline\alpha,\beta r}
$$
are components of the second fundamental form introduced in
(\ref{4.42}).

The equation of motion of the worldvolume scalar fields $X^{\underline
m}(z)$
is obtained from (\ref{5.2.23}) by multiplying the latter by
$m^{ad}\tilde\gamma_d^{\beta\alpha}(\gamma_i)^{rq}$
\begin{eqnarray}\label{5.2.24}
&&m^{bc}m_c^{~a}\hat\Omega_{a,b}^{~~i}\equiv
m^{bc}m_c^{~a}(\nabla_aE^{\underline a}_b)u^{~i}_{\underline a}
\nonumber \\
&&= {1\over 4} m^{ad}\tilde\gamma_d^{\beta\alpha}(\gamma^i)^{qr}\left[
(T_{a,\alpha q}^bE^{\underline\alpha}_b-T_{a,\alpha
q}^{\underline\alpha})
E_{\underline\alpha,\beta r}-K_{\alpha q,a,\beta r}
\right].
\end{eqnarray}

We observe that the l.h.s. of (\ref{5.2.24})
has the form similar to the supermembrane bosonic
equation (\ref{5.1.25})
but with the matrix $m^{bc}m_c^{~a}$ replacing the Minkowski metric
$\eta^{ab}$. The complicated r.h.s. of eq. (\ref{5.2.24})
describes the interaction of the super--5--brane with the
$D=11$ gauge field strength
$F^{(4)}=dA^{(3)}$.

Note that since (\ref{5.2.23}) contains the vector derivative
of $h_{\alpha\beta}$, from (\ref{5.2.23}) we can also derive another
form of
the equation of motion of the tensor field $h_{abc}$.
To this end we  multiply (\ref{5.1.24}) by $C^{qr}$ and
$m^{ad}(\tilde\gamma_{[d}\gamma_b\tilde\gamma_{c]})^{\alpha\beta}$.
Using
the identity (\ref{hh3}) we get
\begin{equation}\label{5.2.25}
m^{ad}{\cal D}_ah_{dbc}=-{1\over{32}}C^{qr}
m^{ad}(\tilde\gamma_{[d}\gamma_b\tilde\gamma_{c]})^{\alpha\beta}
\left[
(T_{a,\alpha q}^bE^{\underline\alpha}_b-T_{a,\alpha
q}^{\underline\alpha})
E_{\underline\alpha,\beta r}-K_{\alpha q,a,\beta r}
\right].
\end{equation}

The form of eq. (\ref{5.2.25}) differs from that of eq. (\ref{t3}), but
a somewhat tedious analysis shows that the two equations
are in fact equivalent \cite{hs2,hsw1}
modulo the fermionic equation (\ref{5.2.18}).

The five scalar field equations (\ref{5.2.24}) have a rather complicated
structure of the right--hand side. Cumbersome manipulations using
the form of the torsion constraints (\ref{5.1}) and (\ref{5.2}), the
superembedding conditions $E_{\alpha q}^{\underline a}=0$ and
(\ref{5.2.9}),
the harmonic relations (\ref{4.25}), the form of the $D=11$
$\Gamma$--matrices
(\ref{5.2.1})--(\ref{5.2.C}) and the identities (\ref{t7})--(\ref{t9})
allow one to disentangle eqs. (\ref{5.2.24})
and to present their leading ($\eta=0$) components (which are the proper
equations for $x^{\underline m}(\xi)$) in the following
form
\begin{eqnarray}\label{5.2.26}
&&\left[G^{mn}({\cal D}_nE^{\underline a}_m
+E^{\underline b}_m\Omega_{n\underline b}^{~~\underline a})\right.
\nonumber \\
&&\left.-
{{\varepsilon^{m_1\cdots m_6}}\over{\sqrt{-g}(1-{{2\over 3}k^2})}}
\left({1\over 6!}
F^{\underline{a}}{}_{m_6\cdots m_1} -{1\over (3!)^2}(
F^{\underline{a}}{}_{m_6 m_5 m_4}
-E_{n}^{\underline{a}}
F^{n}{}_{m_6 m_5 m_4})H_{m_3 m_2 m_1}\right)\right]u^{~i}_{\underline a}
=0, \nonumber \\
&&
\end{eqnarray}
where all quantities depend only on the bosonic worldvolume coordinate
$\xi^m$,
$$
G^{mn}(\xi)=m^{bc}m_c^{~a}e_b^me^n_a(\xi),
$$
$$
E^{\underline A}_m=\partial_mZ^{\underline M}
E^{\underline A}_{\underline M}(Z(\xi)), \qquad
F_{m_p\cdots m_1}=E^{\underline A_p}_{m_p}\cdots E^{\underline
A_1}_{m_1}
F_{\underline{A_1}\cdots \underline{A_p}}(Z(\xi)), \quad (p=4,7)
$$
and
$$
{\cal D}_m=\partial_m+\Gamma_{mn}^{l}(\xi)
$$
is the covariant derivative with the Christoffel symbol
$\Gamma_{mn}^{l}(\xi)$ determined by the worldvolume induced metric
$g_{mn}(\xi)=E^{\underline a}_mE^{\underline b}_n\eta_{\underline{ab}}$.

The expression in the square brackets of (\ref{5.2.26}) is the same as
the
one obtained from the M5--brane action (\ref{5.001}) as the
$x^{\underline m}(\xi)$ equation of motion.

The equivalence of the M5--brane equations of motion (\ref{5.2.019})
and (\ref{5.2.26}), and of the `self--duality' relation (\ref{t7}) to
the
equations of motion yielded by the M5--brane action (\ref{5.001}) has
been demonstrated in \cite{equ}. The proof is not straightforward, since
in the action one deals directly with the field strength $H_{mnl}$
and not with $h_{abc}$, and
a projector ${1\over 2}(1+\Gamma)$, which appears in an natural way in
the
$\kappa$--symmetry transformations of the action, differs from the
$\kappa$--symmetry projector (\ref{5.2.0019}), (\ref{5.2.19}) of the
superembedding approach. The two projectors are related to each other
by the identities
$$
{1 + \Gamma
\over 2}~~{1 + \bar\Gamma \over 2} = {1 + \bar\Gamma
\over 2}, \quad {1 + \bar\Gamma
\over 2}~~{1 + \Gamma \over 2} = {1 + \Gamma
\over 2}.
$$
For details we refer the reader to the papers \cite{m5,equ}, and to
\cite{lw99}
where the equivalence of the covariant energy--momentum tensors of the
M5--brane in both approaches has been discussed.

The two formulations of the M5--brane have been applied to studying
various aspects of M--theory and its duals in
\cite{kallosh,cherkis}--\cite{br,bbs,bhoo}--\cite{sato,t1}--\cite{bv}
and \cite{open,hlwstri}--\cite{lwmono}.

To summarize, in the superembedding approach the M5--brane is described
by
the following main relations.

\subsubsection{Main superembedding equations for the M5--brane}

The superembedding conditions are
\begin{equation}\label{4.22ss}
E^{\underline a}(X^{\underline m}(z),\Theta^{\underline\mu}(z))=
e^a u_a^{~\underline a}(z)
\quad \Rightarrow \quad E_{\alpha q}^{\underline a}=0,
\end{equation}
\begin{equation}\label{4.23ss}
E^{\underline\alpha}(X^{\underline m}(z),\Theta^{\underline\mu}(z))=
{1\over 2}(1-\Gamma_0)^{\underline\alpha}_{~\underline\beta}\,
e^a E_a^{\underline\beta}
+e^{\alpha q} (v_{\alpha q}^{~\underline\alpha}(z)+h_{\alpha\beta}C_{qr}
v^{\beta r,\underline\beta}),
\end{equation}
$$
h_{\alpha\beta}={1\over 6}\gamma^{abc}_
{\alpha\beta}h_{abc},
$$
$$
{1\over 2}(1-\Gamma_0)_{~~\underline\beta}^{\underline\alpha}=
v_{\underline\beta,\beta r}v^{\beta r,\underline\alpha}, \quad
\Gamma_0={1\over{6!}}\epsilon^{a_1\cdots a_6}\Gamma_{a_1\cdots a_6},
\quad
\Gamma_a=E^{\underline a}_a\Gamma_{\underline a}.
$$
The M5--brane equations of motion encoded in (\ref{4.22ss}) and
(\ref{4.23ss})
are the fermionic field equation
\begin{equation}\label{5.2.16s}
\tilde \gamma_b^{\alpha\beta}m^{ba}E_a^{\underline\beta}
v_{\underline\beta,\beta q}=0=
\tilde\gamma_b^{\alpha\beta}m^{ba}
\nabla_{\beta q}E^{\underline b}_au^{~i}_{\underline b},
\end{equation}
$$
m_b^{~a}=\delta_b^{~a}-2k_b^{~a}, \quad k_b^{~a}=h_{acd}h^{bcd},
$$
the worldvolume  scalar field equation
\begin{equation}\label{5.2.24s}
m^{bc}m_c^{~a}(\nabla_aE^{\underline a}_b)u^{~i}_{\underline a}=
{{\varepsilon^{a_1\cdots a_6}}
\over{1-{{2\over 3}k^2}}}
\left({1\over 6!}
F^{\underline{a}}{}_{a_6\cdots a_1} -{1\over (3!)^2}
F^{\underline{a}}{}_{a_6 a_5 a_4}H_{a_3a_2a_1}\right)u_{\underline
a}^{~i},
\end{equation}
where $F^{\underline{a}}{}_{a_p\cdots a_1}=F^{\underline{a}}{}
_{\underline a_p\cdots \underline a_1}E^{\underline a_p}_{a_p}\cdots
E^{\underline a_1}_{a_1}$, $(p=3,6)$;\\
and the self--dual tensor field equation
\begin{equation}\label{t13s}
dH^{(3)}=-{1\over{4!}}E^{\underline D}E^{\underline C}E^{\underline B}
E^{\underline A}F_{\underline{ABCD}}, \quad
H_{abc}=4(m^{-1})_a^{~d}h_{dbc},
\quad h_{abc}={1\over 6}\epsilon_{abcdef}h^{def}.
\end{equation}

As in the case of the supermembrane (see eq. (\ref{5.1.26s}))
in the linearized limit
the superembedding condition (\ref{4.22ss}) reduces to the constraint on
the superfields
corresponding to the transverse oscillations of the super--5--brane
$$
D_{\alpha q}X^i=i(\gamma^i)_q^{~r}\theta_{\alpha r}, \quad
\theta_{\alpha r}
\equiv \Theta^{\underline\alpha}v_{\underline\alpha,\alpha r},
\quad i=0,1\cdots 5, \quad q,r=1,\cdots , 4\, .
$$
This constraint describes an $n=(2,0)$, $d=6$ on--shell tensor
supermultiplet.

Thus, the supersymmetric field theory in the worldvolume of the
M5--brane
governed by eqs. (\ref{5.2.16s})--(\ref{t13s}) is that of the $n=(2,0)$,
$d=6$
tensor supermultiplet which on the mass shell has five scalar, three
(self--dual) tensor and eight fermionic physical modes.

\section{Other developments and applications}
\setcounter{equation}0
\subsection{The generalized action principle}
In Sections 2 and 3 we have constructed doubly supersymmetric
superfield actions for superparticles and superstrings which
produce the basic superembedding condition
$E^{\underline a}_{\alpha q}=0$ dynamically. This has been possible
because in these cases the superembedding condition  does not
contain dynamical field equations, and the latter should be obtained
by imposing the minimal embedding conditions, or from an action.

This method of constructing worldvolume superfield actions can
be used in other cases where the superembedding condition does
not put the superbrane on the mass shell, as, for example,
$N=2$, $D=3,4$ and 6 superparticles \cite{gs2,bms}, the $N=2$, $D=3$
superstring \cite{gs2}, the $N=1$, $D=4$ supermembrane \cite{hrs}
and an $N=2$, $D=4$ space--filling D3--brane.

As we have seen, in the case of the M--theory branes the
superembedding condition implies the dynamical equations of motion.
Therefore, if one tried to construct a worldvolume superfield action
for these branes using the prescription of Sections 2 and 3,
i.e. by introducing the superembedding condition into the action
with a Lagrange multiplier, this Lagrange multiplier superfield
would acquire dynamical equations and thus would contain redundant
degrees of freedom, which are absent from the spectrum of the
brane theories of interest.
This is analogous, for instance, to the superfield formulations of
ten--dimensional Super--Yang--Mills theory and $D=10$ and $D=11$
 supergravities whose superfield constraints
put the theories on the mass shell, and the way of constructing
superfield actions for them has not been found.

In such cases one should either deal with component actions,
as Green--Schwarz--type actions for the superbranes,
or consider so called generalized actions, whose construction
is based on a group manifold (rheonomic) approach which
has been developed in application to supersymmetric field theories
in \cite{gma} and in application to superbranes in \cite{bsv} and,
from somewhat different perspective, in \cite{hrs}.

The main principles of the construction of the generalized
actions (which we enumerate for superbranes \cite{bsv,bpst}) are

i) In the superworldvolume of the superbrane with the number of
bosonic dimensions $d=p+1$ embedded into
a D--dimensional target superspace one
constructs a $p+1$--superform
$L^{(p+1)}$ which is closed, $dL^{(p+1)}=0$, modulo the
superembedding condition $E^{\underline a}_{\alpha q}=0$ (or the
one which replaces it in the case of the space--filling branes
\cite{akub}).
The superform is constructed from the pullbacks of the
target--space supervielbeins, the harmonic variables (which
are auxiliary worldvolume superfields) and
from the field strengths of the worldvolume gauge fields,
when present. It contains both the kinetic and the Wess--Zumino
term of the superbrane Lagrangian.
A systematic way to get this form (which can always be found)
has been proposed in \cite{hrs}.
An example of such a form has been considered
in the case of the heterotic string in Subsections 3.2.2
and 3.2.3 (eqs. (\ref{3.4.1}) and (\ref{3.5.1})).

ii) The integral of this superform over an arbitrary $p+1$--dimensional
bosonic submanifold ${\cal M}_{p+1}=(\xi^m, \eta^\mu(\xi))$ of the
superworldvolume is the generalized
action
\begin{equation}\label{6.1}
S=\int_{{\cal M}_{p+1}}L^{(p+1)}.
\end{equation}
So the generalized action is not a fully fledged worldvolume
superfield action.

iii) But, in spite of the fact that the integration
is performed over
a bosonic submanifold, the action (\ref{6.1})
is superdiffeomorphism invariant in the whole superworldvolume
modulo the superembedding condition
\begin{equation}\label{6.2}
\delta S=\int_{{\cal M}_{p+1}}\delta L^{(p+1)}=
\int_{{\cal M}_{p+1}}d(i_\delta L^{(p+1)})
+\int_{{\cal M}_{p+1}}i_\delta dL^{(p+1)},
\end{equation}
where $i_\delta L^{(p+1)}:=
{1\over{p!}}dz^{L_p}\cdots dz^{L_1}
\delta z^ML_{ML_1\cdots L_p}.
$
We see that the first term in the variation (\ref{6.2}) is
a total derivative and the second term vanishes because of
the condition i).

iv) The variation of the surface ${\cal M}_{p+1}$ in the
action functional (\ref{6.1}) is equivalent to the
superdiffeomorphisms $\delta\eta(\xi)$ of the superworldvolume
in the odd directions orthogonal to ${\cal M}_{p+1}$, which
leads to the condition $dL^{(p+1)}=0$.
As a result,
because of iii) the variation of the action (\ref{6.2})
with respect to
the worldvolume superfields yields the superembedding conditions
and the superfield equations of motion in the {\sl whole}
worldvolume superspace. When contained in the superembedding
conditions, the equations of motion do not produce any new
information in addition to the superembedding conditions.

v) When the integration surface is chosen to be
$(\xi^m;\eta^\mu=0)$ and we take the integral of
the leading component $L^{(p+1)}|_{\eta=0=d\eta}$ of the
superform, the generalized action (\ref{6.1}) is reduced to a
component action, which can be then rewritten in the
Green--Schwarz form.

An example related to the generalized action taken at
$\eta=0$ is the property of the Weil triviality of the term
(\ref{3.5.6}) of the
doubly supersymmetric string action considered
in Subsection 3.2.3, with the two--form (\ref{3.5.1}) being
closed up to the superembedding condition.

Because of a solid geometrical ground the generalized action
formalism can be useful, for example, for deriving  component
actions of supersymmetric models for which actions have been
unknown.

The generalized actions have been constructed for the ordinary
super--p--branes \cite{bsv,banur}, the super--D--branes \cite{bst,akub}
and
(implicitly)
for so called L--branes \cite{Lbranes} whose physical modes form
linear supermultiplets.
For the M5--brane the generalized action which
extends the Green--Schwarz-type
action (\ref{5.001}) to the worldvolume superspace and which
produces the superembedding conditions is still unknown
because of problems caused by the presence of the self--dual field.

For further details on the generalized action approach to
superfield theories we refer the reader to original
literature \cite{gma,bsv,bst,bpst,banur,akub,hrs}.

\subsection{D--branes, L--branes and branes ending on branes}
We have already mentioned that the superembedding approach is
applicable to the description of all known superbranes, including
the Dirichlet branes \cite{hs1,bst,akub}, and it has also been used to
derive equations of motion and actions for a class of
branes called L--branes \cite{hs1,Lbranes} which contain on their
worldvolumes antisymmetric gauge fields dual to worldvolume scalars
or vectors.

As in the case of the M5--brane the worldvolume gauge fields
of the D--branes and L--branes (or more precisely their field strengths)
show up in the superembedding condition (\ref{4.32}) as
the spin--tensor field $h_{\alpha q}^{~\alpha'}(z)$.
The analysis of the superembedding
integrability conditions involving $h^{~\alpha'}_{\alpha q}$
is made along the lines explained with the example of the M5-brane and
reveals the Born--Infeld structure of equations of motion of the gauge
fields.

L--branes can be obtained, by dualizing vector fields
of Dp--brane actions \cite{dual,pst3,abou,bg3}, and
from standard super--p--branes by a
direct dimensional reduction of the target superspace
and the dualization of worldvolume scalars, corresponding
to compactified dimensions, into
rank $p-1$ antisymmetric fields, which enter the linear
supermultiplet of the supersymmetric worldvolume field theory.

In some cases, when, for example, there are eight supersymmetries
in the worldvolume, the scalar supermultiplets are on the mass shell,
while the corresponding dual linear supermultiplets are off--shell, and
the L--branes admit the off-shell worldvolume superfield description
\cite{Lbranes}.

In the superembedding approach L--branes naturally appear when
for a given supersurface and target superspace the basic
superembedding condition yields the superfield constraint
of the linear supermultiplet \cite{hs1}.
The L--brane duals of D-branes also occur when one considers
Dp--branes ending on D(p+2)--branes \cite{open}.

The consideration of open branes ending on another (host)
branes from the point of view of superembedding is
an interesting problem
of its own, as, for instance, studying an M2--brane ending on
an M5--brane. The investigation of such brane configurations has
been carried out in \cite{open},
where it has been demonstrated that the basic principles of
superembedding work perfectly well also in these cases.
It has been shown that if the constraints on the superworldvolume
of the open brane are imposed, the superembedding conditions
determine the superworldvolume constraints for the host branes,
and provide one with information about the dynamics of the
boundary of the open brane in the host brane.

For further details we refer the reader to the original literature
cited above.

\subsection{Nonlinear realizations and superembeddings}

We have already discussed that the presence of the superbranes
in the target superspaces (partially) breaks supersymmetry of
the background vacuum. This supersymmetry breaking is spontaneous,
since the worldvolume equations of motion of the superbrane are
manifestly invariant under target--space supersymmetry, while
their (classical vacuum) solutions preserve only a
fraction of the supersymmetry transformations.

The superbrane configurations whose worldvolume actions
possess $\kappa$--symmetry,
with the number of independent parameters being half the number of
target--space supersymmetries, include BPS states which preserve
half the supersymmetry.

An effective group--theoretical and geometrical method
to describe theories with spontaneously broken symmetries
is the method of nonlinear realizations of symmetries
(or the coset space approach) which
is based on the Cartan theory of group manifolds and coset spaces.

For the first time this method was
applied in physics to the construction
of phenomenological Lagrangians of particle interactions
by Callan, Coleman, Wess and Zumino \cite{ccwz},
and independently by Volkov
\cite{volkov}. It was then used
for the construction of the first globally supersymmetric field
theory \cite{vak} and supergravity \cite{vs}.

Since branes naturally provide us with a geometrical mechanism of
partial (super)symmetry breaking
(which is nothing but the Goldstone--Higgs mechanism),
it is natural to apply
to the description of the field theory on the worldvolume of the
brane the method of nonlinear realizations.

This has been done in a number of papers.

A  super--3--brane
in $N=1$, $D=6$ superspace as a model of partial spontaneous
$n=2$ supersymmetry breaking in $d=4$ was considered in
\cite{hlp,bg1}.

An $N=1$, $D=4$ supermembrane and  $N=2$, $D=2$ superparticles
were discussed in \cite{gaunt5}.

In the static gauge
a worldvolume $n=1$ superfield Born--Infeld--type
action for the D3--brane in $N=2$, $D=4$ superspace
was first constructed in \cite{bg2} as the nonlinear action
of the Goldstone--Maxwell supermultiplet for partially broken
$N=2$ supersymmetry. And the gauge fixed superfield action for
its dual L3--brane was derived in \cite{bg3}.

The formalism of partial supersymmetry breaking has been
further developed in application to branes in superstring
and M--theory in \cite{bik,rt,ik1,ketov}.

Because of its nature, this method
is applicable to the description of (super)branes propagating
in the backgrounds with isometries,
which have the geometry of coset (super)spaces,
such as flat superspace or anti--de--Sitter superspaces.
For instance, in the framework of the AdS/CFT (superconformal field
theory)
correspondence \cite{ads}, the coset space approach has been used to get
the
explicit form of the supervielbeins of $AdS_{p+2}\times S^{D-p-2}$
superspaces and to construct gauge fixed superconformal actions
for superstrings \cite{ckktp,2bs,pst4}, a D3--brane
\cite{ckktp,2bd3,pst4}
and M--branes \cite{ckktp,mbra,pst4}
propagating in these superbackgrounds.

As a simple example of the use of the method of nonlinear realizations
let us consider a  supermembrane
in $N=1$, $D=4$ flat superspace parametrized by coordinates
$X^{\underline a}$ and $\Theta^{\underline\alpha}$
(${\underline a}=0,1,2,3$; ${\underline\alpha}=1,2,3,4$).
From the point of view of the worldvolume field theory this
is the model of spontaneous breaking $n=2$, $d=3$ supersymmetry
down to $n=1$ \cite{gaunt5,ik1}.

The flat superspace is associated with the coset
(actually supergroup) manifold  of the $N=1$, $D=4$ translations
whose element can be exponentially parametrized as
\begin{equation}\label{6.3}
K(X,\Theta)=e^{i(X^{\underline a}P_{\underline a}
+\Theta^{\underline\alpha}\underline Q_{\underline\alpha})},
\end{equation}
where $P_{\underline a}$ and ${\underline Q}_{\underline\alpha}$
are the supertranslation generators
\begin{equation}\label{6.4}
\{\underline Q_{\underline\alpha},\underline Q_{\underline\beta}\}
=2iP_{\underline a}
(C\Gamma^{\underline a})_{\underline\alpha\underline\beta},
\quad [\underline Q_{\underline\alpha},P_{\underline a}]=0.
\end{equation}
The coset element (\ref{6.3}) can be multiplied by the
Lorentz--group coset matrix $U={{SO(1,3)}\over{SO(1,2)}}$
\cite{bik,ik1},
which corresponds to the harmonic matrices (\ref{4.9}) and (\ref{4.028})
of the superembedding approach.

The Cartan one--form $K^{-1}dK$ takes its values in the
superalgebra (\ref{6.4}) and yields the supervielbeins
of the superspace under consideration
\begin{equation}\label{6.5}
{1\over i}K^{-1}dK={\cal E}^{\underline a}P_{\underline a}+
{\cal E}^{\underline\alpha}\underline Q_{\underline\alpha},
\end{equation}
where
\begin{equation}\label{6.6}
{\cal E}^{\underline a}=dX^{\underline a}-
id\bar\Theta\Gamma^{\underline a}\Theta, \quad
{\cal E}^{\underline\alpha}=d\Theta^{\underline\alpha}.
\end{equation}

In the superembedding approach we have considered the pullbacks
${\cal E}^{\underline A}(Z(z))$ of the
supervielbeins (\ref{6.6}) onto the supermembrane
worldvolume parametrized by coordinates $z=(\xi^a,\eta^\alpha)$
($a=0,1,2$; $\alpha=1,2$), and imposed the superembedding condition
${\cal E}^{\underline a}_\alpha=0$ to specify the embedding
corresponding to the dynamics of the brane.

In the method of nonlinear realizations one, from the beginning,
(i.e. already in (\ref{6.3}))
identifies part of the superspace coordinates $X^{\underline a}$
and $\Theta^{\underline\alpha}$ with the superworldvolume
coordinates $\xi^a$ and $\eta^\alpha$
\begin{equation}\label{6.7}
\xi^a=X^a~(a=0,1,2), \quad \eta=(1+\Gamma^{012})\Theta.
\end{equation}
The corresponding supertranslations $P_a$ and
$Q={1\over 2}(1+\Gamma^{012})\underline Q$ generate
$n=1$, $d=3$ supersymmetry
which remains unbroken in the superworldvolume of the brane.

The coordinates $X^3(\xi,\eta)$ and
$\theta(\xi,\eta)=(1-\Gamma^{012})\Theta$ transverse
to the brane are associated with the Goldstone superfields
of spontaneously broken supertranslations generated by
$P_3$ and $S={1\over 2}(1-\Gamma^{012})\underline Q$.

Thus, from the perspective of the $n=2$, $d=3$ worldvolume
field theory the $N=1$, $D=4$ superalgebra (\ref{6.4}) looks
as
\begin{equation}\label{6.8}
\{Q_\alpha,Q_\beta\}=2P_a\gamma^a_{\alpha\beta},
\quad
\{S_\alpha,S_\beta\}=2P_a\gamma^a_{\alpha\beta},
\quad \{Q_\alpha,S_\beta\}=2\epsilon_{\alpha\beta}P_3,
\end{equation}
where $P_3$ plays the role of the central charge.

The Goldstone superfields $X^3(\xi,\eta)$ and $\theta(\xi,\eta)$
are then subject to constraints which should reduce the
number of their components to a suitable irreducible
supermultiplet. (Such constraints can be found, for instance,
with the help of a so called ``inverse Higgs'' effect \cite{ivanov}).
In the case under consideration the
constraint is
\begin{equation}\label{6.08}
D_\alpha X^3=\theta_\alpha,
\end{equation}
which singles out
a scalar $n=1$, $d=3$ supermultiplet describing physical degrees
of freedom of the supermembrane.

In some cases \cite{bg1}--\cite{ketov} the constraints can be solved
in terms of superfields which can be used to construct nonlinear
actions.

Comparing this formalism with the superembedding approach we
see that the choice (\ref{6.7}) of the superbrane coordinates
is nothing but the static gauge, which can be chosen in the
superembedding approach to fix worldvolume superdiffeomorphisms,
and the constraint (\ref{6.08}) imposed on the Goldstone superfields
is similar to the superembedding condition (\ref{4.22s}),
(\ref{5.1.26s}).

Thus, the method of the nonlinear realizations can be regarded as
a gauge fixed version of the covariant superembedding description
of superbranes, which can provide us with
a way of explicit solving for the superembedding constraints and
with an alternative method to construct (gauge fixed) superbrane
actions.

A detailed analysis of the relation between the two approaches
in different cases is an interesting subject for future study.

 \section{Concluding remarks and outlook}
We have given an introduction to generic features of the
geometrical approach to the description of the theory of superbranes.

This powerful approach unifies on the grounds of supersurface theory
various formulations of supersymmetric extended objects, such
as the Green--Schwarz, twistor and Lorentz--harmonic formulation,
and the method of nonlinear realizations.

Being manifestly supersymmetric in the worldvolume and in the
target space, this approach also
establishes (at the classical level) the link between spinning
particles and superparticles, and between
Neveu--Schwarz--Ramond and the Green--Schwarz superstrings.

Superembedding explains the nature of the $\kappa$--symmetry of the
Green--Schwarz--type actions as odd superdiffeomorphisms (local
supersymmetry) of the superbrane worldvolume.

These properties of superembeddings
allow one, in certain cases, to overcome the covariant
quantization problem of superparticles and superstrings.

As we have seen, in many cases the basic superembedding condition
contains the full information about the dynamics of the superbrane,
i.e. it produces the superbrane equations of motion. This is
of particular importance for the description of new objects for
which the use of other methods may encounter problems.

Depending on whether or not the superembedding condition puts the
superbrane on the mass shell, the approach gives a recipe
for the construction of component, generalized or superfield actions
of the superbranes by the use of a closed $(p+1)$--superform
which exists in the brane superworldvolume.

As further applications of the superembedding approach, one may
use it to search for new types of (dual) families of  branes,
study superembeddings which correspond to brane configurations
preserving less than half target--space supersymmetry, as intersecting
branes,
carry out more detailed analysis of the relation between the
superembedding approach and the method of nonlinear realizations,
and to use it for studying the dynamics of branes and
gauge fixed brane actions in AdS
superbackgrounds in connection with the AdS/CFT
correspondence conjecture.

\section*{Acknowledgements}
The author is grateful to J. Bagger for the suggestion to write
down this article and to N. Berkovits, S. Krivonos, P. Pasti, C.
Preitschopf, H. Skarke and M. Tonin for interest to this work,
useful discussions and comments, and especially to I. Bandos, F.
Bettella and D. Smith for reading the manuscript and numerous
discussions which helped the author to make some points more
transparent. The author acknowledges the financial support from
the Alexander von Humboldt Foundation. This work was also
partially supported by the European Commission TMR Programme
ERBFMPX-CT96-0045 to which the author is associated, by the INTAS
Grant 96-308  and under the project N 2.5.1/52 of the Ukrainian
State Committee on Science and Technology.

\end{document}